\documentclass[fleqn,usenatbib]{mnras}
\usepackage[T1]{fontenc}
\usepackage{ae,aecompl}
\usepackage{newtxtext,newtxmath}
\usepackage{graphicx}
\usepackage{epsfig}
\usepackage{color}
\usepackage{subfigure}
\usepackage{caption,booktabs}
\usepackage{footmisc}
\usepackage{multirow}
\usepackage{float}

\usepackage{graphicx}	
\usepackage{amsmath}	

\date{Accepted XXX. Received YYY, in original form ZZZ}
\pubyear{2020}

\date{Accepted XXX. Received YYY, in original form ZZZ}
\pubyear{2020}

\title[Multiwavelength analysis of low surface brightness galaxies]
{Multiwavelength analysis of low surface brightness galaxies to study possible dark matter signature}

\author[P. Bhattacharjee et al.]{
Pooja Bhattacharjee$^{1,2}$ \thanks{Email address: pooja.bhattacharjee@jcbose.ac.in},
Pratik Majumdar$^{3}$,
Mousumi Das$^{4}$,
Subinoy Das$^{4}$,
Partha S. Joarder$^{5}$
\newauthor 
and Sayan Biswas$^{6}$ \\ 
   $^{1}$ Department of Physics, Bose Institute, 93/1 A.P.C. Road, Kolkata 700009, India, \\ 
   $^{2}$ Centre for Astroparticle Physics and Space Science, Bose Institute, Block EN, Sector V, Salt Lake, Kolkata 700091, India, \\
   $^{3}$ Saha Institute of Nuclear Physics, HBNI, 1/AF, Bidhannagar, Kolkata 700064, India, \\
   $^{4}$ Indian Institute of Astrophysics, Koramangala, Bangalore, Karnataka 560034, India, \\
   $^{5}$ Sister Nivedita University,  Newtown, Kolkata 700156, India, \\
   $^{6}$ Raman Research Institute, C.V. Raman Avenue, Sadashivanagar, Bangalore, Karnataka  560080, India.\\
  }

\hypersetup{draft}
\begin{document}
\label{firstpage}
\pagerange{\pageref{firstpage}--\pageref{lastpage}}
\maketitle
\begin{abstract}
\noindent Low Surface Brightness (LSB) galaxies have very diffuse, low surface density stellar disks which appear faint in optical images. They are very rich in neutral hydrogen (HI) gas, which extends well beyond the stellar disks. Their extended HI rotation curves and stellar disks indicate that they have very massive dark matter (DM) halos compared to normal bright galaxies. Hence, LSB galaxies may represent valuable laboratories for the indirect detection of DM. In this paper, we search for WIMP annihilation signatures in four LSB galaxies and present an analysis of nearly nine years of data from the Fermi Large Area Telescope (LAT). Above 500 MeV, no excess emission was detected from the LSB galaxies. We obtain constraints on the DM cross-section for different annihilation channels, for both individual and stacked targets. In addition to this, we use radio data from the Very Large Array (VLA) radio telescope in order to derive DM constraints, following a multiwavelength approach. The constraints obtained from the four considered LSB galaxies are nearly 3 orders of magnitude weaker than the predicted limits for the thermal relic abundances and the combined limits achieved from Fermi-LAT observations of dwarf spheroidal galaxies. Finally, we discuss the possibility of detecting emission from LSB galaxies using the upcoming ground-based $\gamma$-ray and radio observatories, namely the Cherenkov Telescope Array (CTA) and the Square Kilometre Array (SKA).
\end{abstract}




\section{Introduction}
\label{Section1}
\noindent The data from several astrophysical and cosmological observations \citep{kom, ade} reveal that a large portion of the total mass density of our Universe ($\sim$85\%) is made of non-luminous and non-baryonic matter, known as Dark Matter (DM). Cosmological N-body simulations \citep{die08, spring08} and theoretical predictions mainly favour the cold dark matter (CDM) model to explain the formation of the large scale structure of the Universe. Moreover, the physics beyond the Standard Model (SM) anticipates that CDM consists of some form of massive, non-baryonic and neutrally charged particles. In one of the most considered theoretical scenarios, DM is composed of weakly interacting massive particles (WIMPs). In the past decade, theoretical studies have shown that WIMPs are among the most promising DM candidates and their pair annihilation can lead to the production of high energy $\gamma$ rays \citep{ste, jung, evan, bertone, abdo, Bonnivard}. This $\gamma$-ray emission can be detected by space-based telescopes such as the Fermi Large Area Telescope (or Fermi-LAT) \citep{abdo, acknew, acknew1, ack, geringernew1, hoopernew, albert17}. \\

\noindent The dwarf spheroidal galaxies (dSphs) have traditionally been considered among the most promising candidates for WIMP detection, as they are thought to be one of the most DM dominated galaxies in our local Universe and the closest to our Galaxy \citep{die, spr, kuh, abdo}. Although such studies have not yet detected $\gamma$ rays from WIMP annihilation, they have put strong constraints on the $\gamma$-ray fluxes \citep{abdo, acknew, acknew1, ack, geringernew1, hoopernew, poo, albert17, poo2019}, which has led to significant constraints on the theoretical models of DM.\\

\noindent The reason why the dSphs or the satellite galaxies are considered to be DM dominated is that their velocity dispersion is relatively high and indicates that their dynamical masses are much larger than their baryonic masses \citep{strigari2008}. Furthermore, they are relatively close, being satellite galaxies of the Milky Way. However, the larger velocity dispersion could be due to other physical processes such as tidal shocking by the Milky Way or interactions with other nearby satellite galaxies \citep{hammer2018}. In fact, recently several ultra-diffuse dwarf galaxies (UDGs) have been discovered that are similar in many respects to the dSphs but are found to be practically DM free \citep{vandokkum2019}. In addition, several gas rich UDGs are thought to contain very little DM \citep{guo2019}. A diversification of the classes of targets represents, therefore, a well-motivated approach for indirect DM searches.\\

\noindent Another class of interesting targets with significant DM content are the low surface brightness (LSB) galaxies. They contain extended disks of neutral hydrogen (HI) gas \citep{burkholder2001, oneil2004, du2015}, which extend out to 2 to 3 times beyond their stellar disks \citep{blok1996, mishra2017}. This enables one to trace their DM content out to large radii using their HI rotation curves  \citep{das2019}. Also, although LSB galaxies are gas rich, they have diffuse, metal poor and practically dust free stellar disks \citep{impey1997}, with very little molecular hydrogen gas \citep{das2006}. Hence, it is not surprising that they have very low star formation rates (SFRs). Star formation can lead to $\gamma$-ray emission, which can interfere with indirect searches for DM signatures. Another source of $\gamma$-rays is represented by accreting supermassive black holes or active galactic nuclei (AGN). But AGNs are rarely found in LSB galaxies and have never been detected in dwarf LSB galaxies \citep{das2013}.\\

\noindent Another reason for considering LSB galaxies is that they are very massive systems. In fact, LSB galaxies lie in the same baryonic mass range as high brightness galaxies (HSB), the main difference being that the HI gas mass-to-luminosity ratio of LSB galaxies is much higher than that for HSB galaxies of similar baryonic mass  \citep{oneil2004, honey2018}. Rotation curve studies also indicate the presence of massive DM halos in LSB galaxies \citep{blok1997}. Thus LSB galaxies satisfy two of the main criteria that candidates for indirect searches of DM should have i.e. (i)~they have high DM masses and (ii)~they usually do not contain the obvious sources of $\gamma$ radiation such as star-forming regions and AGN.\\

\noindent Within the class of LSB galaxies, the smaller ones termed as LSB dwarfs or irregulars have very little star formation and no nuclear activity \citep{das2013}. They are thus interesting targets to search for $\gamma$ rays from WIMP annihilation. Moreover, the HI rotation curves and gas kinematics of LSB dwarfs may be used to determine the central halo density profiles. This has been used to resolve the `cusp-core' problem in the CDM theory of galaxy formation \citep{bosch, naray2009}. But such studies can also put constraints on the DM mass in the galaxies.\\

\noindent Despite being possible targets for indirect DM searches, the $\gamma$-ray signature from LSB dwarf galaxies have been analyzed in very few cases \citep{gammaldi2017, cadena2018, hasimoto}. The reason why they have not been widely studied in indirect DM searches may be due to the small values of their astrophysical factor (J), which is due to their large distances from the Milky Way compared to dSph galaxies \citep{gammaldi2017, cadena2018}. However, some LSB dwarfs are relatively close by e.g. UGC~12632. Furthermore, their fluxes can be stacked to improve the results. This is particularly important because next generation optical telescopes, like the Large Synoptic Survey Telescope (LSST), are expected to discover large amounts of LSB galaxies, which will provide the possibility to improve significantly the prospects for indirect DM searches with this class of objects. With this in mind, we have done a study of the Fermi-LAT data of four nearby LSB dwarfs to constrain the DM models of WIMP annihilation. Moreover, we have also used a multiwavelength approach to search for the DM signal by exploring whether the DM signal can be detected at radio frequencies.\\

\noindent The paper is organized as follows. We first discuss the properties of the LSB galaxies (section~2) and then describe the analysis of nine years of Fermi-LAT data (section~3). In subsection~3.1 and subsection~4.2, we calculate the upper limits of $\gamma$-ray flux from each LSB galaxy and derive the upper limits of the velocity-averaged pair-annihilation cross-section ($<\sigma v>$) for different WIMP pair-annihilation channels. Next, in subsection~4.3, we perform a joint likelihood analysis of all four LSB galaxies. In subsection~4.4, we predict the possible diffuse radio signal from LSB galaxies with the RX-DMFIT code \citep{alex} and in subsection~4.5, we provide a comparison between three different DM density profiles. In section~5, we compare the flux upper limits of four LSBs obtained from Fermi-LAT data with the Cherenkov Telescope Array (CTA) sensitivity. Finally, in the concluding section, a brief discussion of our results obtained from $\gamma$-ray data with the Fermi-LAT and the prediction from radio emission are presented.

\section{Sample Selection:}
\label{Section2}
\noindent The LSB dwarf galaxies in our sample were selected based on the following criteria. (i)~They should have low luminosities but large DM masses. The galaxies should not show any signs of star formation or active galactic nuclei (AGN) activity in their optical images as such processes can also give rise to $\gamma$ rays \citep{abdo2010, ajello2016}. (ii)~They should be within 15~Mpc so to minimize the impact of the target distance on the value of the astrophysical J-factor. Based on these constraints, we selected a sample of 4 nearby LSB dwarfs from \citet{bosch}. Table~1 lists the sample galaxies, their distances and their properties. The neutral hydrogen observational data is from the Westerbork Synthesis Radio Telescope for the four LSB galaxies and was obtained from \citet{swaters2002}. It should be noted that some LSB dwarfs with smaller distances could have been included, but they showed blue emission in their optical images or signs of UV emission, both of which indicate the presence of star formation.

\begin{itemize}
\item \textbf{UGC 3371 (DDO 039)}: UGC 3371 is an irregular dwarf galaxy (Irr). Observational studies indicate that there is an excellent agreement between the HI and H$\alpha$ flat rotation curve velocities but the initial part of the HI rotation curve rises more steeply than the H$\alpha$ one. This inconsistency in observational data could come from the overcorrection in beam smearing of HI data \citep{swaters2003} and because of this disagreement it is not possible to strongly predict the nature of its rotation curve, it could have either a linear rise or a steep rising. There are some studies such as \citet{swaters1999} which indicate that the best-fit models of UGC 3371 prefer a DM halo with a steep central cusp. The halo profile of UGC~3371 is hence consistent with CDM models \citep{bosch}.\\

\item \textbf{UGC 11707}: UGC 11707 is a small spiral galaxy with loosely bound broken arms made of individual stellar clusters. It also has a very faint central bulge (Sd). Due to the lack of data, the  H$\alpha$ rotation curve of UGC 11707 is poorly sampled but the inner rise in the H$\alpha$ rotation curve appears to have a steeper profile than the HI rotation curve. This galaxy appears to have a very large amount of freedom in its model parameters~\citep{swaters1999}. This is partly due to the relatively large error bars for the inner data points, which is a reflection of the asymmetry between the receding and approaching rotation velocities for radii $\leq$ 7 kpc. The halo profile of UGC 11707 is consistent with CDM models \citep{bosch}. \\

\item \textbf{UGC 12632 (DDO 217)}: UGC 12632 is a weakly barred spiral galaxy (SABm). The observational data shows that the HI gas is uniformly distributed over its disk but has a very high velocity bump on the blue side of the H$\alpha$ rotation curve. The position velocity map indicates a narrow rise of the rotational velocity near the centre and a gradual increase in the outer disk. So the observed rotation curve of UGC 12632 is in excellent agreement with CDM models of halos. Thus it shows that this galaxy is consistent with CDM models \citep{bosch}.\\

\item \textbf{UGC 12732}: UGC 12732 is a weakly barred spiral galaxy (SABm), similar to UGC 12632. Studies suggest that there is a ring of approximate size $0.5^\circ$ $\times$ $0.5^\circ$ in UGC 12732 associated with one long thin arm of very low surface brightness. The H$\alpha$ and HI rotation curves of UGC 12732 are in good agreement and it is also consistent with CDM models \citep{bosch,swaters2009}.\\
\end{itemize}

\begin{table}
\caption{Properties of LSB galaxies. Column~I: Name of LSB galaxies; Column~II: Galactic longitude and latitude of LSB galaxies; Column~III: The adopted distance of the galaxies, based on a Hubble constant ($H_{\circ}$)= 75 $km~s^{-1}~Mpc^{-1}$. We have obtained the value of distance for each LSB galaxies and their corresponding uncertainties from \textit{NASA/IPAC Extragalactic Database}; Column~IV: Observed rotational velocity at last measured point of rotational curve from \citet{bosch}; Column~V: Scale length of stellar disk from \citet{bosch}; Column~VI: B band Luminosity of LSBs from \citet{obrien2012}; Column~VII: Location of the last observed data points of LSB galaxies from \citet{swaters2009}; Column~VIII: Observed HI gas masses of LSB galaxies from \citet{swaters2002}.}
    \centering
    \begin{minipage}{1.0\textwidth}
\begin{tabular}{|p{0.5cm}||p{1.6cm}|p{0.8cm}|p{0.8cm}|p{0.8cm}|p{0.8cm}|p{0.6cm}|p{0.7cm}|}
\hline
\hline
Name & (l,b) & $d$\footnote{\url{https://ned.ipac.caltech.edu/}}  & $V_{last}$ & $R_{d}$ & $L_{B}$ & $R_{last}$ & $M_{HI}$ \\ [0.5ex]
$ $ & [deg],[deg] & (Mpc) & $(km~s^{-1})$ & (Kpc) & $(10^{9}~L_{\odot}^{B})$ & (Kpc) & $(10^{8}~M_{\odot})$ \\ [0.5ex]
\hline
UGC 3371 & 138.43,22.81 & $12.73^{+0.90}_{-0.90}$ & ~~~~~86 & 3.09 & 1.54 & 10.2 & 12.2 \\ [0.5ex]
\hline
UGC 11707 & 74.31,-15.04 & $14.95^{+1.05}_{-1.05}$ & ~~~~~100 & 4.30 & 1.13 & 15.0 & 37.2 \\ [0.5ex]
\hline
UGC 12632 & 106.77,-19.31 & $8.36^{+0.60}_{-0.60}$ & ~~~~~76 & 2.57 & 0.86 & 8.53 & 8.7 \\ [0.5ex]
\hline
UGC 12732 & 103.74,-33.98 & $12.38^{+0.87}_{-0.87}$ & ~~~~~98 & 2.21 & 0.71 & 15.4 & 36.6 \\ [0.5ex]
\hline
\hline
\end{tabular}
\end{minipage}
\end{table}
\newpage
\clearpage
\begin{table}
    \caption{Parameters used for our Fermi-LAT data analysis.}
    \label{table:fermi_lat_parameters}
    \begin{center}
    \begin{tabular}{cccc}
        \hline \hline
        {\bf Parameters needed for Fermi-LAT data extraction} & & &\\
        \hline\hline
        Parameter & Value & &\\
        \hline \hline
        Radius of interest (ROI) &  $10^{\circ}$ & &\\
        Start of observation (TSTART (MET)) & 239557418 (2008-08-04 15:43:37.000 UTC) & &\\
        End of observation (TSTOP (MET)) & 530379822 (2017-10-22 15:43:37.000 UTC) & &\\
        Energy Range & 500 MeV - 300 GeV  & &\\
        Fermitools version & \texttt{1.2.1} & &\\
        \hline \hline
        $~~~~~~~~~~~~~~~~~~~$\texttt{gtselect} for assigning event selection & & &\\
        \hline \hline
        Event class & Source type (128) & & \\
        Event type & Front+Back (3) & &\\
        Zenith angle cut & $90^{\circ}$ & &\\
        \hline \hline
        $~~~~~~~~~~~~~~~~~~~$\texttt{gtmktime} for assigning time selection & & & \\
        \hline \hline
        Filter applied & $\rm{(DATA\_QUAL>0)\&\&(LAT\_CONFIG==1)}$ & &\\
        ROI-based zenith angle cut & No & &\\
        \hline \hline
        $~~~~~~~~~~~~~~~~~~~$\texttt{gtltcube} for generating livetime cube & & &\\
        \hline \hline
        Zenith angle cut ($z_{cut}$) & $90^{\circ}$ & &\\
        Step size in $cos(\theta)$ & 0.025 & & \\
        Pixel size (degrees) & 1 & &\\
        \hline \hline
        $~~~~~~~~~~~~~~~~~~~$\texttt{gtbin} for generating 3-D (binned) counts map & & & \\
        \hline \hline
        Size of the X $\&$ Y axis (pixels) & 140 & &\\
        Image scale (degrees/pixel) & 0.1 & &\\
        Coordinate system & Celestial (CEL) & &\\
        Projection method & AIT & &\\
        Number of logarithmically uniform energy bins & 24 & &\\ 
        \hline \hline
        $~~~~~~~~~~~~~~~~~~~$\texttt{gtexpcube2} for generating exposure map & & &\\
        \hline \hline
        Instrument Response Function (IRF) & $\rm{P8R3\_SOURCE\_V2}$ & & \\
        Size of the X $\&$ Y axis (pixels) & 400 & &\\
        Image scale (degrees/pixel) & 0.1 & &\\
        Coordinate system & Celestial (CEL) & &\\
        Projection method & AIT & &\\
        Number of logarithmically uniform energy bins & 24 & &\\ 
        \hline \hline
        $~~~~~~~~~~~~~~~~~~~$For generating source model in XML format & & &\\
        \hline \hline
        Galactic diffuse emission model & $\rm{gll\_iem\_v07.fits}$ & & \\
        Extragalactic isotropic diffuse emission model & $\rm{iso\_P8R3\_SOURCE\_V2\_v1.txt}$ & &\\
        Source catalog & 4FGL & & \\
        Extra radius of interest &  $5^{\circ}$ & &\\
        Spectral model &  Power law (section 3.1) $\&$ DMFit Function (section 4.2)\citep{jel} & &\\
        \hline \hline
                
    \end{tabular}
    \end{center}
\end{table}
 
\newpage
\clearpage

\section{Fermi-LAT observation and data analysis of LSBs}
\label{Section3}
\noindent The LAT on-board Fermi is a pair-conversion $\gamma$-ray detector capable of measuring $\gamma$ rays in a very wide range of energy from 20 MeV to 500 GeV. It can track the electron and positron resulting from pair conversion of an incident $\gamma$-ray in thin high-Z foils, and can also measure the energy deposition due to the subsequent electromagnetic shower that develops in the calorimeter. The LAT nominally operates in a scanning mode observing the whole sky every 3 hours, resulting in the overall coverage of the sky being fairly uniform. In our paper, we have analyzed almost 9 years of sky survey data (from 2008-08-04 to 2017-10-22) from the direction of each galaxy in our sample. \\ 

\noindent For $\gamma$-ray data-analysis, we have used the latest version of Fermi ScienceTools \textit{v1.2.1}\footnote{\url{https://fermi.gsfc.nasa.gov/ssc/data/analysis/software/}}. This Fermi ScienceTools is processed with an improved PASS 8 instrument response function (IRF) and for our purpose we have used source class IRF\footnote{\url{https://fermi.gsfc.nasa.gov/ssc/data/analysis/documentation/Cicerone/Cicerone_LAT_IRFs/IRF_overview.html}}, $\rm{P8R3\_SOURCE\_V2}$ \footnote{\url{https://fermi.gsfc.nasa.gov/ssc/data/analysis/documentation/Pass8_usage.html}}. For each of our `source of interest,' we have extracted the LAT data within a $10^{\circ}~\times~10^{\circ}$ of the radius of interest (ROI) around each source. The point spread function (PSF) of Fermi-LAT varies with the photon energy and their angle of incidence. At around 500 MeV and 1 GeV, the PSF of LAT is yielding to $4^{\circ}$ and $2.5^{\circ}$, respectively\footnote{\url{https://www.slac.stanford.edu/exp/glast/groups/canda/lat_Performance.htm}}. Thus, in order to reduce the possible uncertainties at low energies and background contamination at high energies, we have put energy limits of $0.5 \le E \le 300$~GeV on the reconstructed energy ($E$). In our analysis, with \texttt{gtmktime} we have extracted the good time interval (GTI) data from the whole dataset and we have also applied zenith-angle cut at $90^{\circ}$ as recommended by the Fermi-LAT team in order to remove the possible contamination from the earth albedo. \\

\noindent Next, with \texttt{gtlike} \citep{cas, matt}, we have performed the binned likelihood analysis\footnote{\url{https://fermi.gsfc.nasa.gov/ssc/data/analysis/scitools/binned_likelihood_tutorial.html}} on the GTI dataset. The spectral model of all the $\gamma$-ray sources lying in our ROIs would vary during the maximum likelihood method. For this course, we have generated a source model file with the inclusion of all the sources from the 4FGL catalog \citep{Abdollahi2020} within a $15^{\circ}~\times~15^{\circ}$ ROI from the location of our `source of interest'. We have extended the spatial coverage up to $15^{\circ}$ ROI to consider all the possible overlapping between the PSF of nearby sources. The angular size of the LSB galaxies is much smaller than the PSF of the LAT instrument in all energy bands, thus for our analysis, we have treated them as the $\gamma$-ray point sources. Furthermore, during the likelihood process, we have included the galactic diffusion emission ($\rm{gll\_iem\_v07.fits}$) model and the corresponding isotropic component ($\rm{iso\_P8R3\_SOURCE\_V2\_v1.txt}$) to the source model for eliminating the possible background photons coming from the galactic and the extragalactic components \footnote{\url{https://fermi.gsfc.nasa.gov/ssc/data/access/lat/BackgroundModels.html}}. During the fitting, the spectral parameters of all the sources within $10^{\circ}~\times~10^{\circ}$ ROI and the normalization value of the two diffuse background components were left free. The remaining all the background sources within the $15^{\circ}~\times~15^{\circ}$ ROI have been kept fixed to the values mentioned in the 4FGL catalog \citep{Abdollahi2020}. In Table~2, we have provided all the necessary information on our Fermi-LAT analysis.

\subsection{Results from the power-law modelling} 
\label{Section3.1}

\begin{figure}
\subfigure[ UGC 3371]
 { \includegraphics[width=1.15\columnwidth]{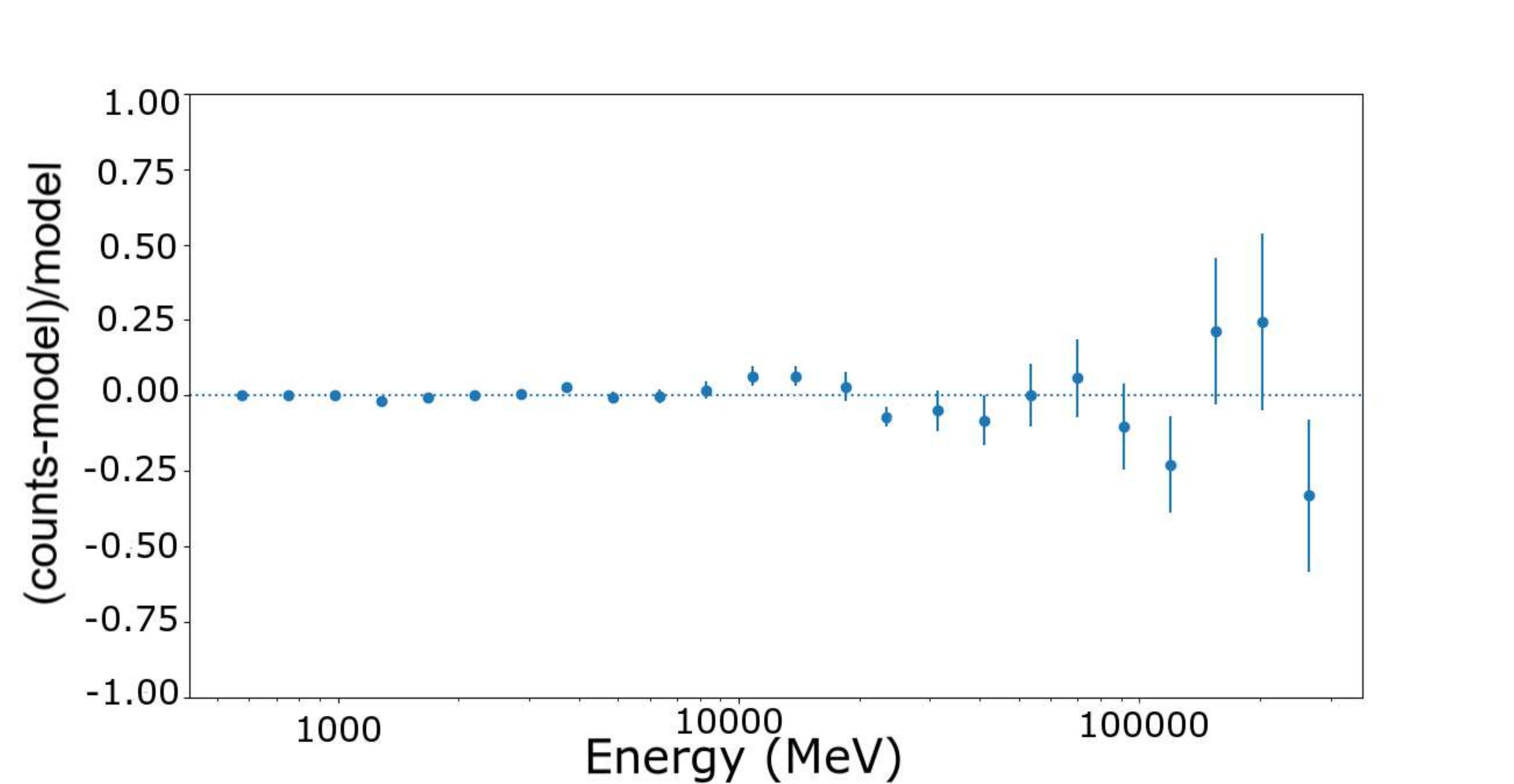}}
\subfigure[ UGC 11707]
 { \includegraphics[width=1.15\columnwidth]{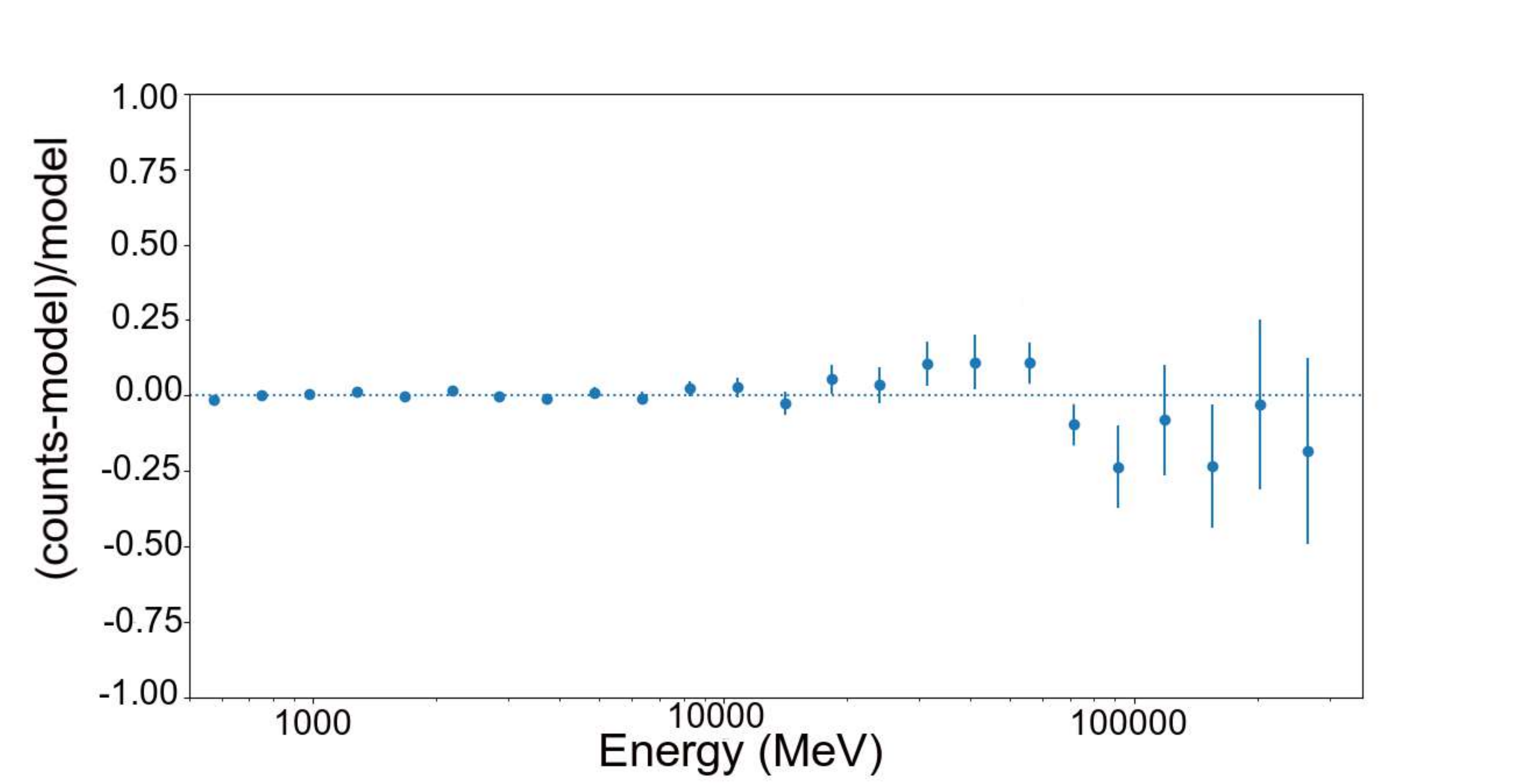}}
\subfigure[ UGC 12632]
 { \includegraphics[width=1.15\columnwidth]{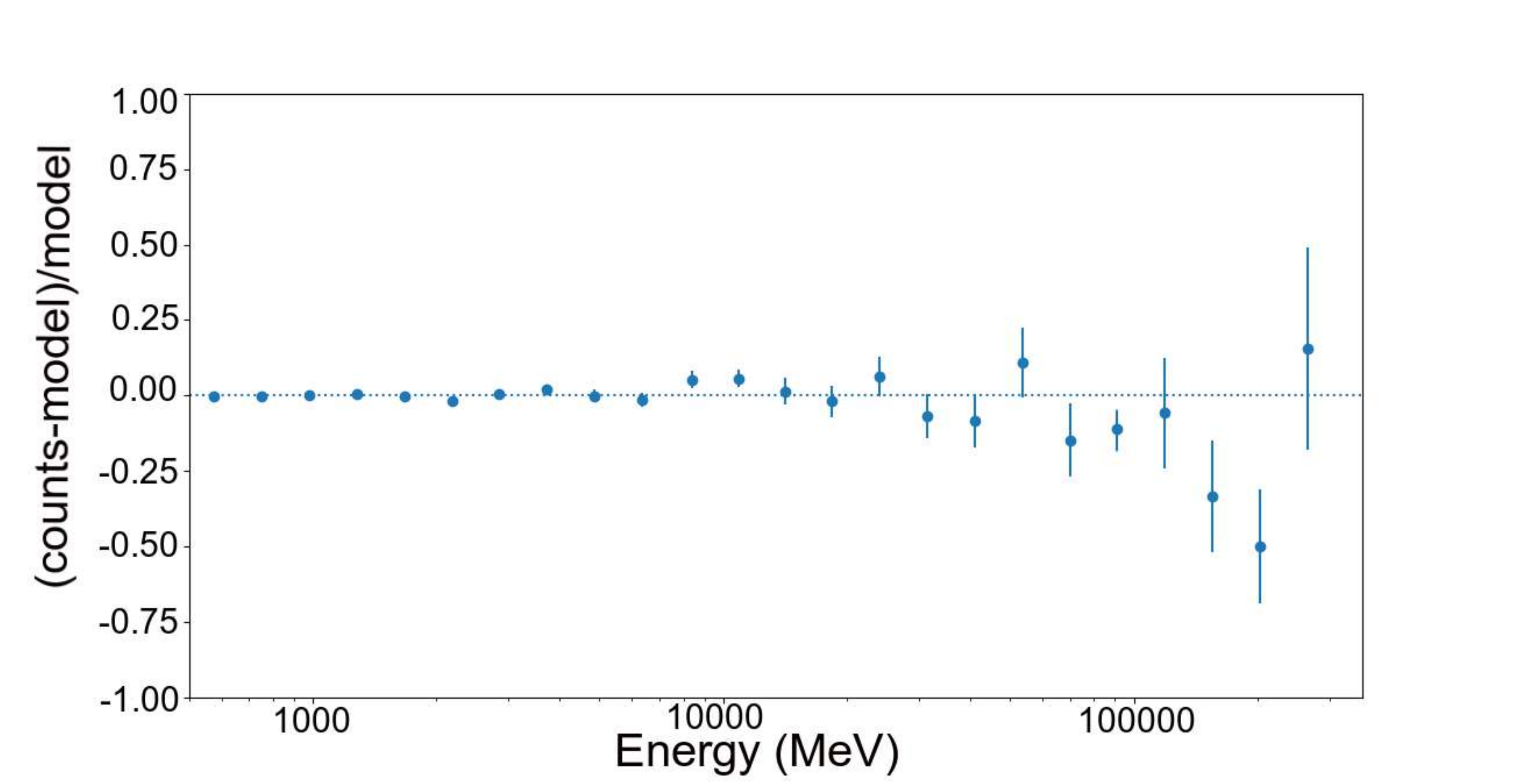}}
\subfigure[ UGC 12732]
 { \includegraphics[width=1.15\columnwidth]{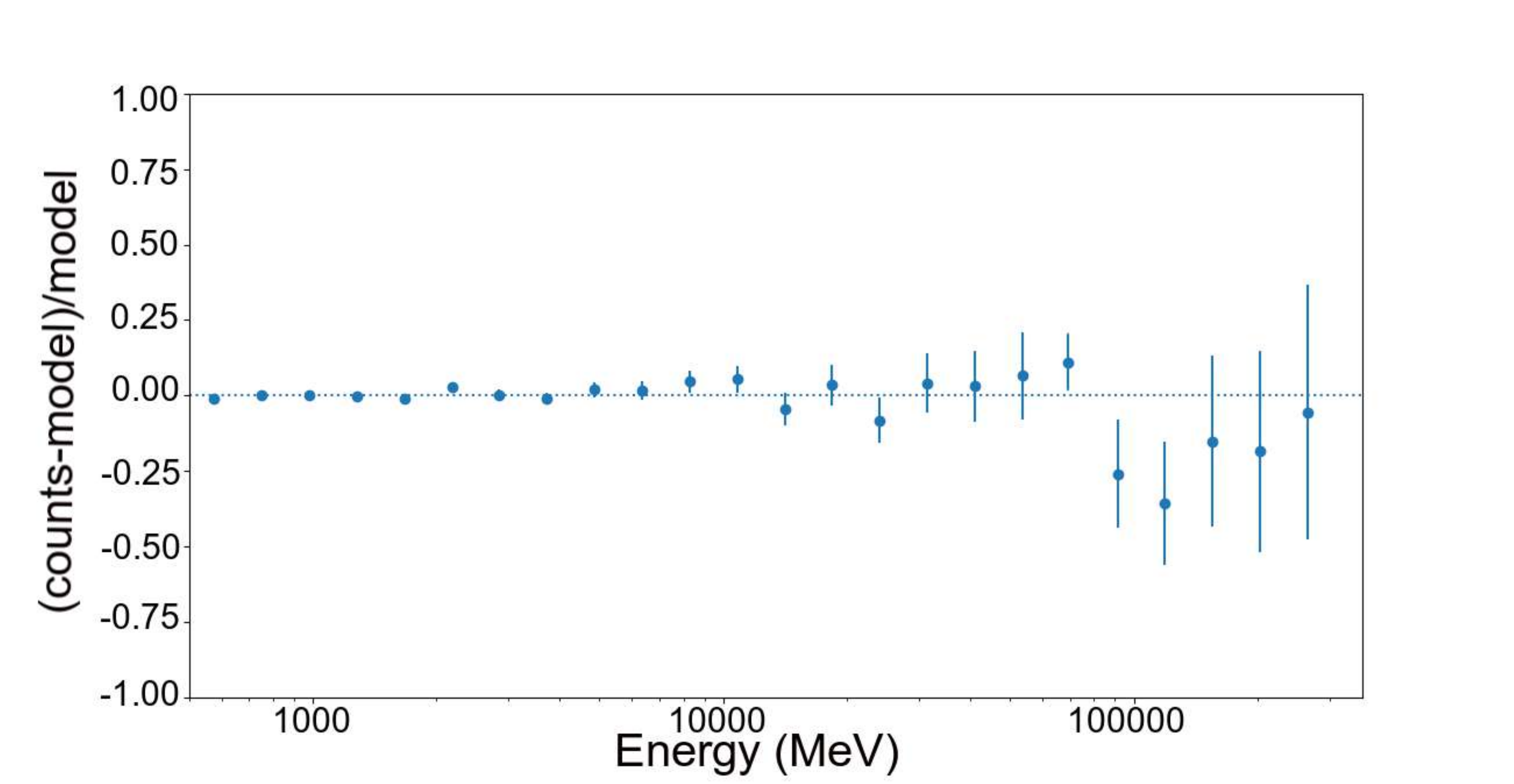}}
\caption{Residual plots of all the four LSB galaxies considered in this work. The spectra of the sources are modeled with a power law with a spectral index, $\Gamma$=2.}
\label{Fig1}
\end{figure}

\noindent In our analysis, we have modelled all the LSB galaxies with power-law spectra to constrain the $\gamma$-ray flux from each source. The expression of differential flux for power-law spectrum is \citep{abdo}:
\begin{equation}
\frac{dN}{dA dE dt} = N_{0} \Big(\frac{E}{E_{0}}\Big)^{-\Gamma},
\end{equation} 
\noindent where, $dN$ denotes the number of photons within energy interval $E$ to $E + dE$ and $dA$ is the elemental area where photons are incident with time interval, $dt$. In Eq.~(1), the reconstructed energy ($E$) is varied between 500 MeV to 300 GeV and  $\Gamma$ and $N_{0}$ are the spectral index and normalisation constants, respectively. We have fixed $E_{0}$ and $\Gamma$ at 500 MeV and 2, respectively. The $\Gamma$=2 is chosen to check the constraints on the standard astrophysical source spectra. During the course of fitting, we have checked the best fit value of the $N_{0}$ along with the isotropic and the galactic diffuse normalization parameters.\\

\noindent In Table~3, we show the results obtained for each LSB galaxy for $\Gamma = 2$, including the best-fitted values of isotropic and the galactic diffuse normalization components along with the $N_{0}$ and its statistical error. In the case of converge-fitting, the ideal value of two diffuse normalization components should be close to $\sim$1 \citep{abdo}. From Table~3, we find that the best-fitted values of the galactic and isotropic components are close to 1 and it provides confidence in our analysis mechanism. It is important to note that for each galaxy the normalization parameter, $N_{0}$ is always lower than the statistical error by at least an order of 1 magnitude. This implies that our analysis did not detect any signal from the direction of the LSB sources. In Table~1, we also report the latitude (b) $\&$ longitude (l) of LSB galaxies. For our source location, the systematic resulting from the IRFs and diffuse background model can only affect the results by $<10\%$ \citep{acknew1, ack}. In Fig.~1(a,b,c,d), we show the residual fit for the four LSB galaxies.\\

\noindent We have then derived the upper limits of $\gamma$-ray flux following the profile likelihood method \citep{rol, bar}. In this procedure, all the normalization parameters, i.e. $N_{0}$ and galactic and isotropic components, were continuously fitted with spectrum model at every step until the logarithmic difference of two likelihood function comes to $1.35$ \citep{abdo}, which corresponds to a one-sided $95\%$ confidence level (C.L.). In Table~4, we have displayed the flux upper limits in the energy range between 500 MeV to 300 GeV, considering $\Gamma = 2$.

\begin{table}
\caption{Fitting values of the three normalization parameters in the chosen ROI for spectral index $\Gamma$ = 2.}
\begin{tabular}{|p{1cm}|p{2cm}|p{2cm}|p{2.4cm}|}
\hline 
\hline
LSB  &  Galactic  & Isotropic  & $N_{0}\times~10^{-5}$ \\ [0.5ex]
Galaxies & Component &  Component &  $ $ \\ [0.5ex]
&  $cm^{-2} s^{-1} MeV^{-1}$ & $cm^{-2} s^{-1} MeV^{-1}$ & $cm^{-2} s^{-1} MeV^{-1}$ \\ [0.5ex]
\hline 
UGC 3371 & $0.95 \pm 0.011$  & $0.95 \pm 0.035$ & $(6.29\pm 21.55)\times10^{-8}$ \\ [0.5ex]
\hline 
UGC 11707 & $0.92 \pm 0.001$  & $1.06 \pm 0.001$ & $(0.1099\pm6.06)\times10^{-7}$ \\ [0.5ex]
\hline 
UGC 12632 & $0.93 \pm 0.011$  & $1.09 \pm 0.05$ & $(0.334\pm 5.82)\times10^{-6}$ \\ [0.5ex]
\hline 
UGC 12732 & $0.97 \pm 0.001$  & $1.004 \pm 0.017$ & $(0.12\pm2.30)\times10^{-8}$  \\ [0.5ex]
\hline 
\hline
\end{tabular}
\end{table}

\begin{table}
\caption{Flux upper limits of the four LSB galaxies at $95\%$ C.L.}
\begin{tabular}{|c|c|} 
\hline 
\hline
LSB galaxies & E~$>$~500 MeV \\ [0.5ex]
$ $ & ($cm^{-2} s^{-1}$) \\ [0.5ex] 
\hline 
UGC 3371 & $2.43\times10^{-10}$ \\ [0.5ex] 
\hline 
UGC 11707 & $3.22\times10^{-10}$ \\ [0.5ex] 
\hline
UGC 12732 & $3.54\times10^{-10}$ \\ [0.5ex]
\hline
UGC 12632 & $3.06\times10^{-10}$ \\ [0.5ex]
\hline
\hline
\end{tabular}
\end{table}

\section{A theoretical framework to estimate $\gamma$-ray flux from pair-annihilation of WIMPs in case of LSB galaxies }
\label{Section4}
\subsection{Dark matter density profile modelling}
\label{Section4.1}
\noindent At given photon energy E, the expression of photon flux originated from WIMP pair annihilation of mass $m_{DM}$, within a solid angle $\Delta \Omega$ is \citep{bal, abdo}:
\begin{equation}
\phi_{DM}(E, \Delta \Omega)~ = ~ \Phi^{pp}(E) \times J(\Delta \Omega),
\end{equation}

\noindent where, $\Phi^{pp}(E)$ and $J(\Delta \Omega)$ are referred as ``particle physics factor" and ``astrophysical factor" (or ``J factor"), respectively.

\subsubsection{\textbf{Particle Physics factor}}
\label{Section4.1.1}
\noindent $\Phi^{pp}(E)$ is defined as the Particle physics factor. It depends on the nature of the particles generating from WIMP annihilation. The expression of $\Phi^{pp}(E)$ is \citep{abdo}
\begin{equation}
\Phi^{pp}(E)~ = ~ \frac{<\sigma v>}{8 \pi ~m^{2}_{DM}} \sum_{f} \frac{dN_{f}}{dE}B_{f}.\\
\end{equation}

\noindent $\Phi^{pp}(E)$ is a function of the DM particle mass ($m_{DM}$) and the thermally averaged WIMP annihilation cross-section times the relative velocity ($<\sigma v>$). Here, $\frac{dN_{f}}{dE}$ and $B_{f}$ denote the $\gamma$-ray spectrum and branching fraction corresponding to the $f^{\rm{th}}$ final state of WIMP annihilation, respectively. In our analysis, we have not considered any effect of the Sommerfeld enhancement \citep{ark, abdo, feng}. Sommerfeld enhancement may change the relative velocity of WIMP and, in that case, the $\gamma$-ray flux can be enhanced by a factor up to $\sim$100 for a wide range of DM masses (i.e. for 100 GeV to 3 TeV).

\subsubsection{\textbf{Astrophysical factor}}
\label{Section4.1.2}
\noindent The astrophysical factor $J$ depends on the DM density distribution in the LSB galaxy and is given by
\begin{eqnarray}
J (\Delta \Omega) &=& \int \int \rho^{2} (r(\lambda)) d\lambda ~ d\Omega  \nonumber \\
                  &=& 2 \pi \int_{\theta_{\rm{min}}}^{\theta_{\rm{max}}} \rm{sin} \theta \int_{\lambda_{\rm{min}}}^{\lambda_{\rm{max}}} \rho^{2}(r(\lambda)) d\lambda ~ d\theta ,
\end{eqnarray}

\noindent where- $\rho(r)$ is the DM mass density profile of the LSB galaxy, whereas $\lambda$ and $r(\lambda)$ denote the line-of-sight and galactocentric distance of the galaxy, respectively. The expression for $r(\lambda)$ is =
\begin{equation}
r(\lambda) = \sqrt{\lambda^{2} + d^{2} - 2~ \lambda ~d~ \rm{cos \theta}} 
\end{equation}

\noindent Here $d$ denotes the distance of LSB galaxy and $\theta$ is the angle between the direction of LAT observation and the centre of LSB galaxy. The minimum and maximum limits of $\lambda$ can be represented as \citep{evan}

\begin{eqnarray}
\lambda_{\rm{max}} = d\rm{cos \theta} + \sqrt{R_{\rm{vir}}^{2} - d^{2} \rm{sin^{2} \theta}} \\
\lambda_{\rm{min}} = d\rm{cos \theta} - \sqrt{R_{\rm{vir}}^{2} - d^{2} \rm{sin^{2} \theta}}
\end{eqnarray}
 
\noindent where, $R_{\rm{vir}}$ is the virial radius of LSB galaxy.\\

\noindent For our analysis, we have considered the Navarro-Frenk-White (NFW) density profile \citep{nav} to model DM mass distribution in LSB galaxies. This profile is motivated by both the numerical simulations and the observed rotation curves of our selected LSB galaxies. The rotational curves of the four selected LSB galaxies are consistent with the prediction for CDM as expected in $\Lambda$CDM cosmology \citep{bosch, bosch2001, swaters2003}. The observational study by \citet{bosch, bosch2001, swaters2003} could not rule out the possibility that the DM distribution in those LSB galaxies may follow the cuspy profile. The data we have used for estimating the J-factor (see Table~5) are taken from \cite{bosch}, where they modelled DM distribution with NFW density profile. The expression of the NFW density profile is \citep{nav, abdo}

\begin{equation}
\rho (r)=\frac{\rho_{s}r_{s}^{3}}{r(r_{s} + r)^{2}}
\end{equation}

\noindent where $\rho_{s}$ and $r_{s}$ are the characteristic density and scale radius, respectively and $r$ is the distance from the center of the LSB galaxy.

\noindent The expression of the $\rho_{s}$ is \citep{lokas, Liddle}:
\begin{equation}
\rho_{s} = \rho_{c}^{0} \delta_{\rm{char}}
\end{equation}

\noindent where $\delta_{\rm{char}}$ is the fitting parameter and $\rho_{c}^{0}$ is the critical density of the Universe. For our calculation, we have adopted the Hubble constant of $H_{0}$=$75~\rm{km~s^{-1}Mpc^{-1}}$ = $100h~\rm{km~s^{-1}Mpc^{-1}}$ from \citet{bosch} and thus $\rho_{c}^{0}$ can be expressed as: $\rho_{c}^{0}$ = $2.78h^{-1}\times10^{11}$ $\frac{M_{\odot}}{(h^{-1}Mpc)^{3}}$.\\

\noindent The expression of the $\delta_{\rm{char}}$ is:
\begin{equation}
\delta_{\rm{char}} = \frac{v c^{3}g(c)}{3}
\end{equation}
\noindent where
\begin{equation}
g(c)=\frac{1}{\ln(1+c)- c/(1+c)}
\end{equation}

\noindent In Eq. 10 and Eq. 11, $c$ is the concentration parameter that defines the shape of the density profile and the value of the virial overdensity, $v$ is assumed to be $\approx$ 178 \citep{bosch}.

\noindent $R_{\rm{vir}}$ (also dubbed as $r_{200}$) is the virial radius at which the mean density is 200 times of the present critical density ($\rho_{c}^{0}$) of our Universe. The circular velocity at $R_{\rm{vir}}$ is defined as \citep{lokas, bosch}
\begin{equation}
V_{200} = \frac{R_{vir}}{h^{-1}}
\end{equation}

\noindent The expression of scale radius is \citep{bosch}:
\begin{equation}
r_{s} = \frac {R_{\rm{vir}}}{c}
\end{equation}

\noindent For this paper, we have taken the values of $c$ and $V_{200}$ from \citet{bosch}. Thus, once we would know the value of these two parameters, using Eqs.~9,10,11,12 $\&$ 13, we can estimate the value of $\rho_{s}$ and $r_{s}$ and can also derive the J-factor from Eq.~4. \\

\noindent For our calculation, we have taken $\theta_{\rm{min}}$~=~$0^{\circ}$ and $\theta_{\rm{max}}$~=~$\sin^{-1}\Big(\frac{R_{vir}}{d}\Big)$. The estimation of the J-factor helps us to rank the LSBs without assuming any specific DM models. Furthermore, it also allows us to estimate any possible detection from LSBs for any theoretical favored DM particle models.\\

\noindent In Table~5, we have displayed some of the necessary parameters for deriving the J-factors from Eq.~4. The distance for each LSB galaxies and their corresponding uncertainties are taken from \textit{NASA/IPAC Extragalactic Database}, while we have adopted the value of $c$ and $V_{200}$ from \citet{bosch}. \\

\noindent The uncertainties in J-factor are calculated from Eq.~4 by using the distribution of distance ($d$) and concentration parameter ($c$). For estimating the uncertainties, we have developed an algorithm that is intended to find limiting values of the J-factor in a 2$\sigma$ limit by a Monte Carlo method, with the input being the values of the variable $d$ and $c$ along with their respective limits. First, we have generated random numbers with a user-defined distribution. For our purpose, we have taken asymmetric normal distribution about the mean with two different values of the standard deviation on each side of the mean, because concentration parameters lie in an asymmetrical limit. The program generates the distribution by the use of Smirnov transform on a set of uniformly distributed random numbers. With this algorithm, we have generated the values of the variables $d$ and $c$ within the tolerance limits (2$\sigma$ or 95$\%$ C.L. for our case) of limiting values. \\

\begin{table}
\centering
\caption{Different parameters needed for the J-factor calculation from Eq.~4 ($h_{0}=0.75$).}
\label{table-1}
\begin{tabular}{|p{0.8cm}|p{1.0cm}|p{0.9cm}|p{1cm}|p{0.8cm}|p{0.7cm}|p{1.5cm}|}
\hline \hline
Galaxy & Distance & ~~~~c & ~~$V_{200}$ & $\theta_{\rm{max}}$ & J~factor\\
name  & ~~Mpc & $ $ & ~~$km~s^{-1}$ & ~~~$^{\circ}$ & $\times10^{16}$~$\frac{GeV^{2}}{cm^{5}}$\\
\hline \hline
UGC 3371  & $12.73^{+0.90}_{-0.90}$ & $14.5^{+14.6}_{-10.2}$ & ~~69.8 & $0.42$ & $0.739^{+2.87}_{-0.63}$  \\
\hline 
UGC 11707 & $14.95^{+1.05}_{-1.05}$ & $14.7^{+14.6}_{-10.3}$ & ~~66.9 & $ 0.34$ & $0.485^{+1.85}_{-0.42}$ \\
\hline 
UGC 12632 & $8.36^{+0.60}_{-0.60}$ & $15.6^{+15.5}_{-10.9}$ & ~~51.4 & $0.47$ & $0.795^{+3.25}_{-0.716}$ \\
\hline 
UGC 12732 & $12.38^{+0.87}_{-0.87}$ & $14.3^{+14.4}_{-10}$ & ~~73.3 & $0.45$ & $0.880^{+3.40}_{-0.75}$ \\
\hline \hline
\end{tabular}
\end{table}

\subsubsection{\textbf{Comparison with Toy Model}}
\label{Section4.1.3}
\noindent In this section, we use the toy model proposed by \citet{char} to cross-check our method of numerical integration to calculate the J-factor for the NFW density profile. In Fig.~2, we have portrayed the toy model for J-factor calculation \citep{char}. In this figure, the vertical hatched region is for contribution from integration and the cross-hatched region is for the toy model.  \\

\begin{figure}
\centering
\includegraphics[width=\columnwidth]{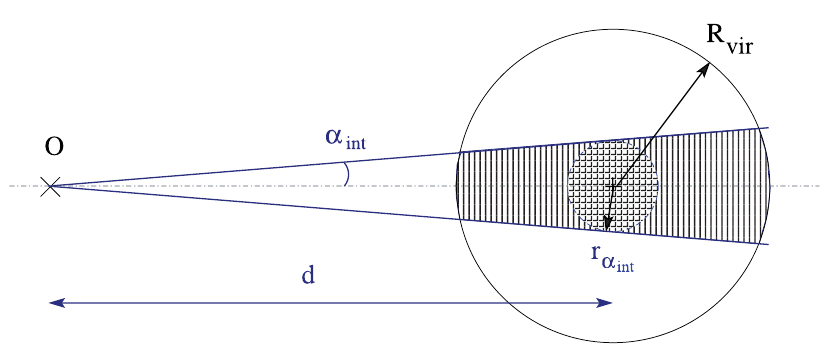}
\caption{Sketch of the toy model used for the J-factor calculation.}
\label{Fig2}
\end{figure}

\noindent From Fig.~2, $d$ is the distance to the center of the galaxy from the observer and $\alpha_{int}$ is the angle for integration where $r_{\alpha_{int}} = d~\rm{sin}\alpha_{int}$. In the toy model, it is assumed that around $90\%$ clump luminosity can be contained in scale radius, $r_{s}$ and it does not depend on any specific density profile. Then the simplified form of Eq.~8 can be written as
\begin{equation}
\rho_{approx} = \rho_{s} r_{s}/r~ \rm{for} ~r_{sat}< r \le  r_{s} 
\end{equation}
where, $r_{sat}$ is the saturation distance. The corresponding approximate form of J-factor is:
\begin{eqnarray}
J_{approx} &=&\frac{4 \pi} {d^{2}} \int_{0}^{\rm{min}[r_{\alpha_{int}}, r_{s}]} \rho_{approx}^{2} r^{2}~ dr \nonumber\\
              &=&  \frac{4 \pi} {d^{2}} \rho_{s}^{2}r_{s}^{2}(\rm{min}[r_{\alpha_{int}}, r_{s}]).
\end{eqnarray}
\noindent If $r_{\alpha_{int}}\gtrsim r_{s}$, the density profile falls faster than $1/r$ for $r \sim r_{s}$. The toy model advised us stop the integration at $r_{x}$ where $\rho_{true} = \frac{\rho_{approx}}{x}$, $x =2$ and $r_{x} = r_{s}[\sqrt2 - 1]$ \citep{char}. Thus we obtain:
\begin{equation}
J_{approx}= \frac{4 \pi} {d^{2}} \rho_{s}^{2}r_{s}^{2}(\rm{min}[r_{x}, r_{\alpha_{int}}]).
\end{equation}
\noindent In Table~6, we have compared our estimated J-factor for all four LSB galaxies with the J-factor obtained from Toy-model calculation proposed by \citet{char}. The J-factor calculated from Eq. 4 is comparable to the one from the Toy model calculation within a factor $\sim$2.
\begin{table}
\centering
\caption{value of J-factor in two different calculation for $h_{0}$=0.75.}
\label{table-1}
\begin{tabular}{|p{1.5cm}|p{2.5cm}|p{2.5cm}|}
\hline \hline
Galaxy & Integration method  & Toy model \\
name & ($\rm{GeV^{2}/cm^{5}}$) & ($\rm{GeV^{2}/cm^{5}}$) \\
\hline \hline
UGC 3371  & $0.739^{+2.87}_{-0.63}\times10^{16}$ & $0.918^{+3.47}_{-0.82}\times10^{16}$  \\
\hline
UGC 11707  & $0.485^{+1.85}_{-0.42}\times10^{16}$ & $0.603^{+2.20}_{-0.54}\times10^{16}$  \\
\hline
UGC 12632  & $0.795^{+3.08}_{-0.68}\times10^{16}$ & $0.987^{+3.84}_{-0.88}\times10^{16}$   \\
\hline
UGC 12732 & $0.880^{+3.40}_{-0.75}\times10^{16}$ & $1.09^{+4.37}_{-0.97}\times10^{16}$   \\
\hline
\end{tabular}
\end{table}
\subsection{Constraints on the annihilation cross-section}
\label{Section4.2}
\begin{figure}
 \begin{center}
\subfigure[]
 { \includegraphics[width=0.75\linewidth]{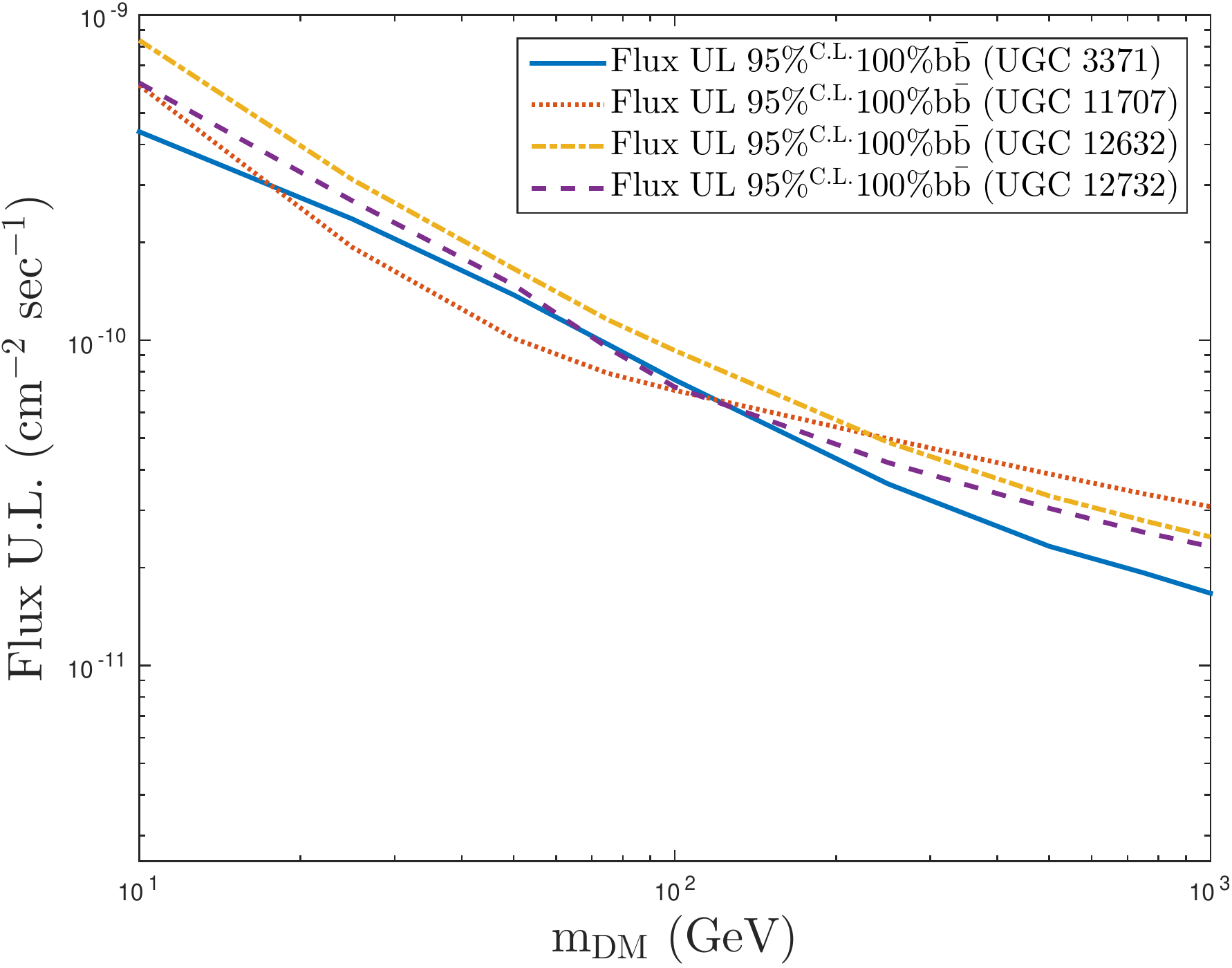}}
\subfigure[]
 { \includegraphics[width=0.75\linewidth]{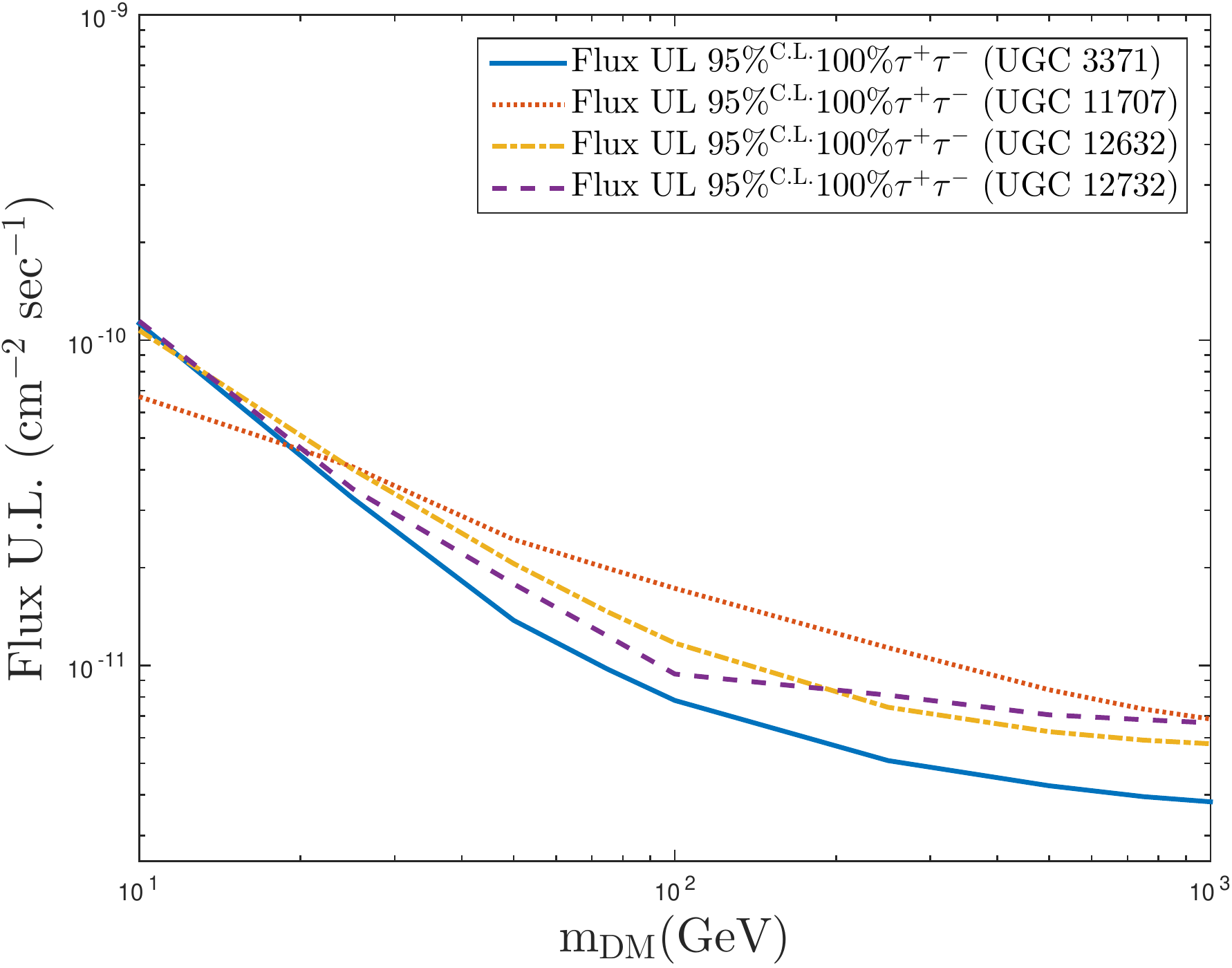}}
 \subfigure[]
 { \includegraphics[width=0.75\linewidth]{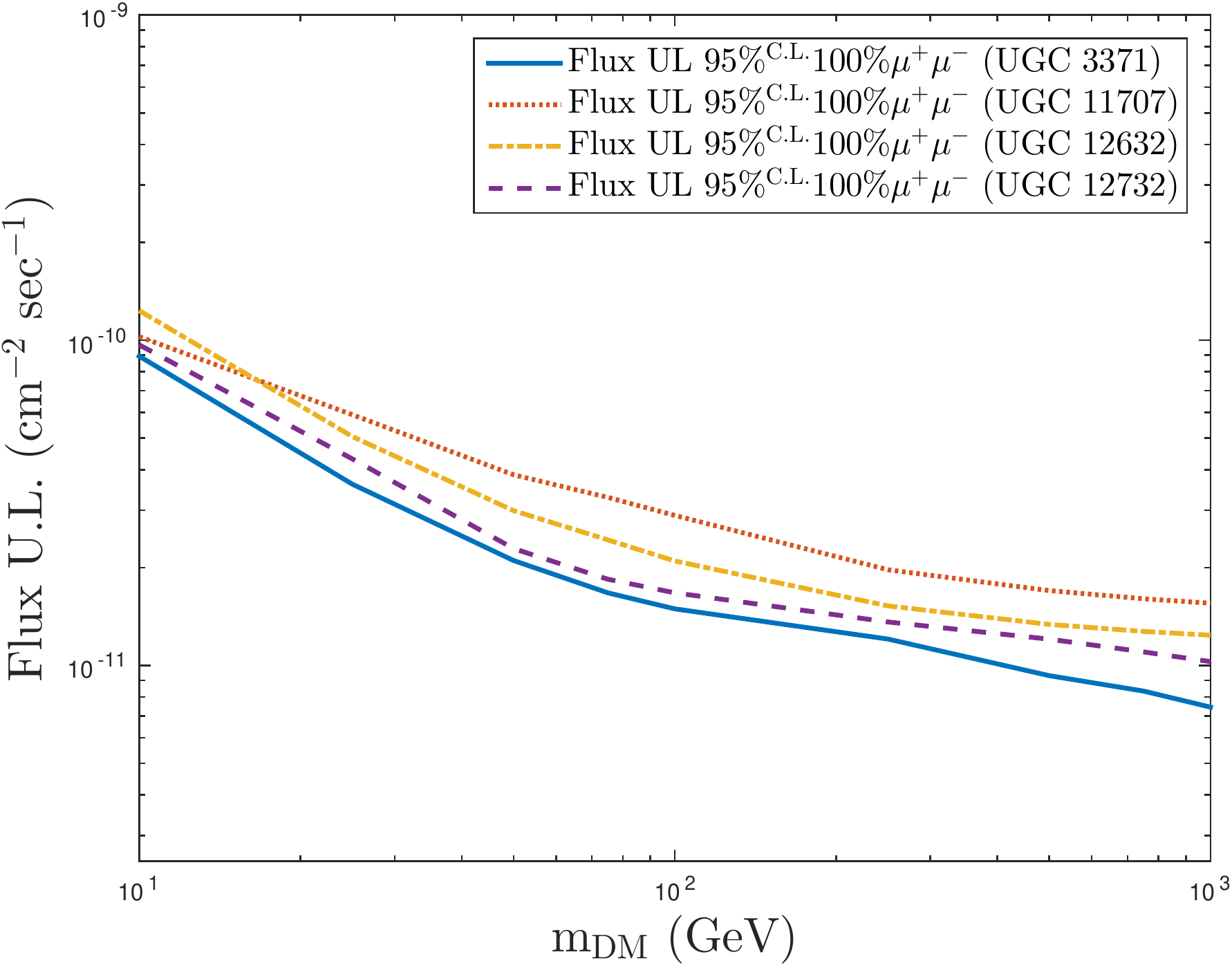}}
\subfigure[]
 { \includegraphics[width=0.75\linewidth]{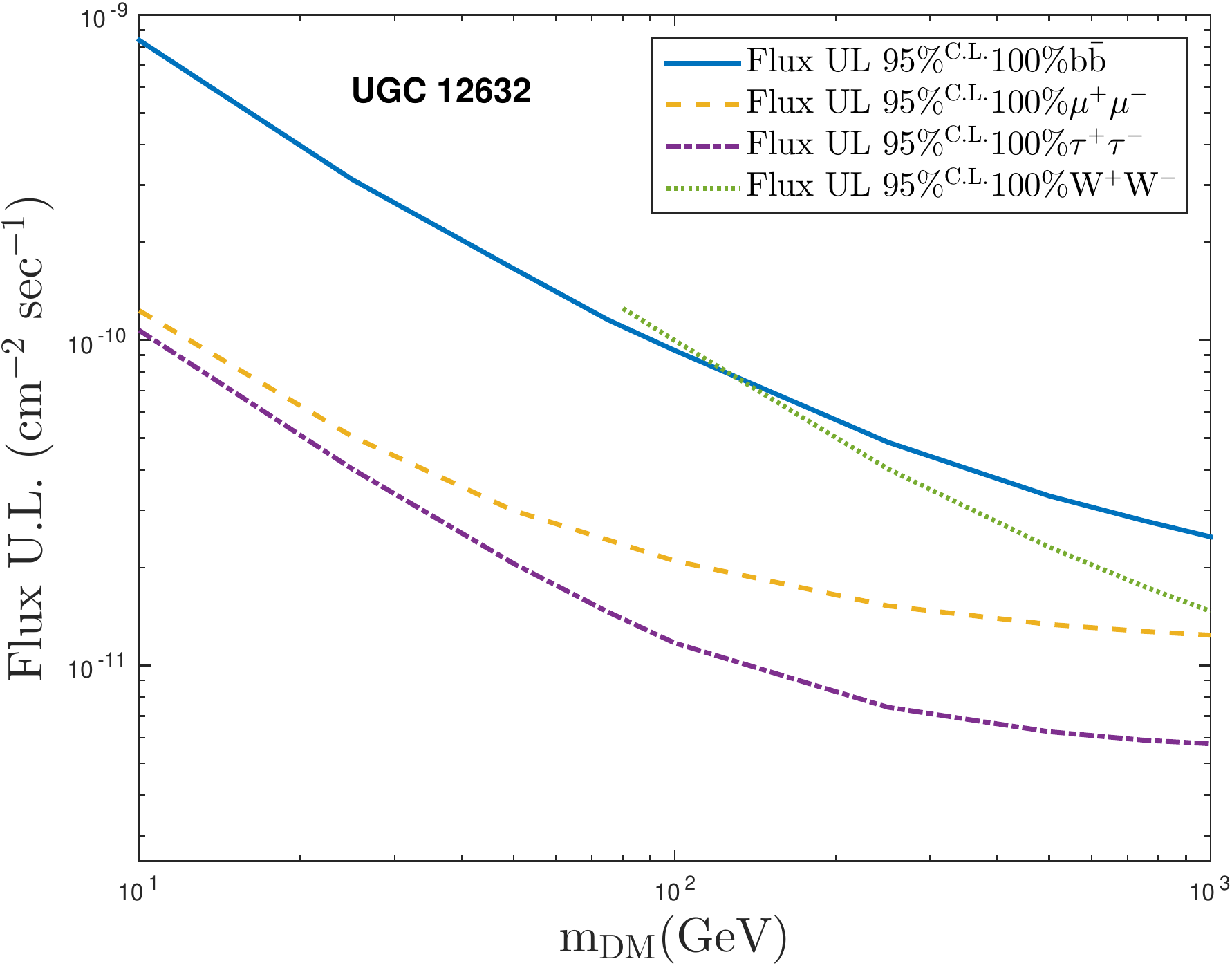}}
\caption{95$\%$ C.L. flux upper limits of the selected LSB galaxies for different pair annihilation final states: (a) $100\%$ $b\overline{b}$, (b) $100\%$ $\rm{\tau^{+} \tau^{-}}$, (c) $100\%$ $\rm{\mu^{+} \mu^{-}}$. Plot (d) shows the flux upper limits of UGC 12632 for all the four selected annihilation final states.}
\end{center}
\end{figure}

\begin{figure}
 \begin{center}
\subfigure[]
 { \includegraphics[width=0.75\linewidth]{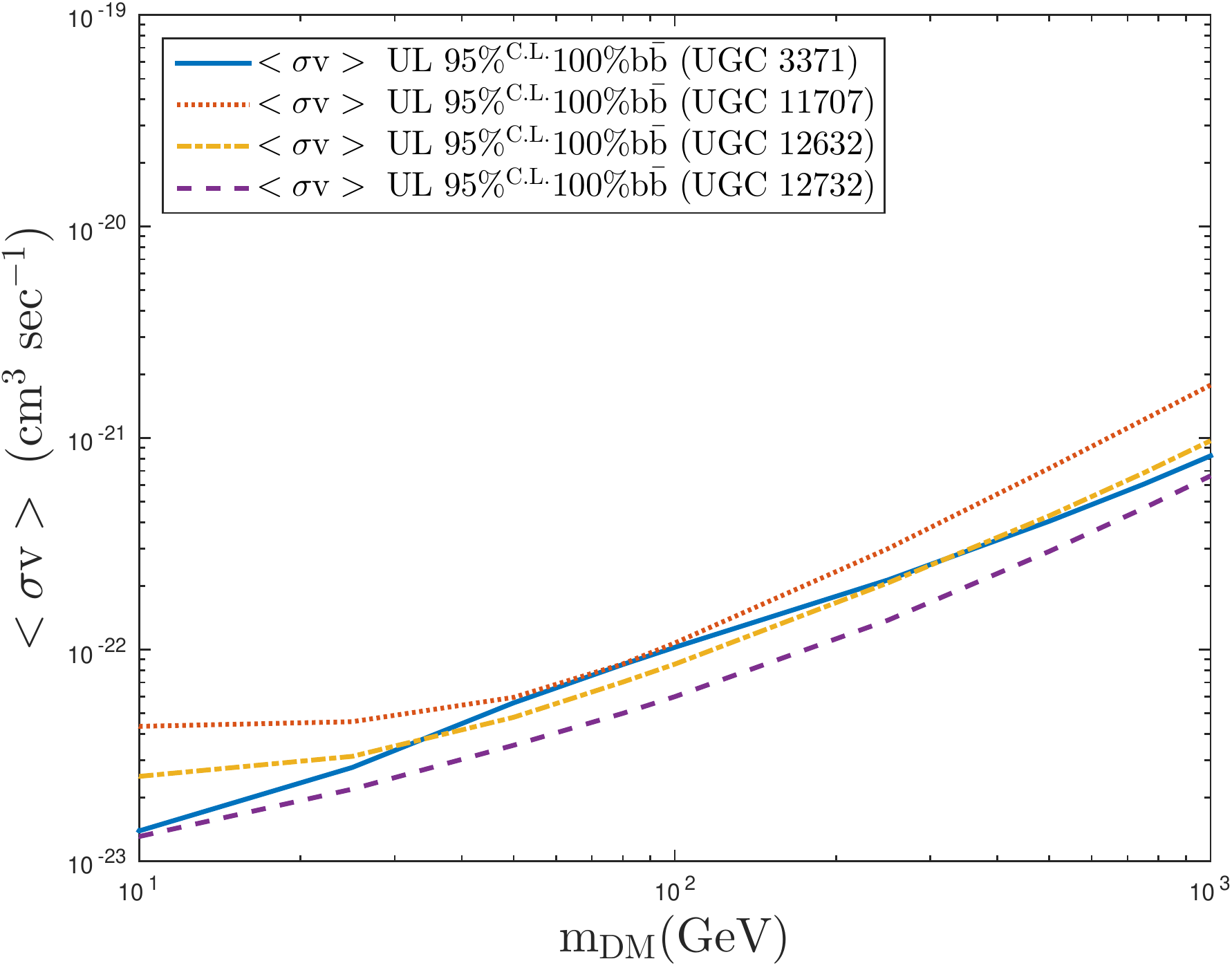}}
\subfigure[]
 { \includegraphics[width=0.75\linewidth]{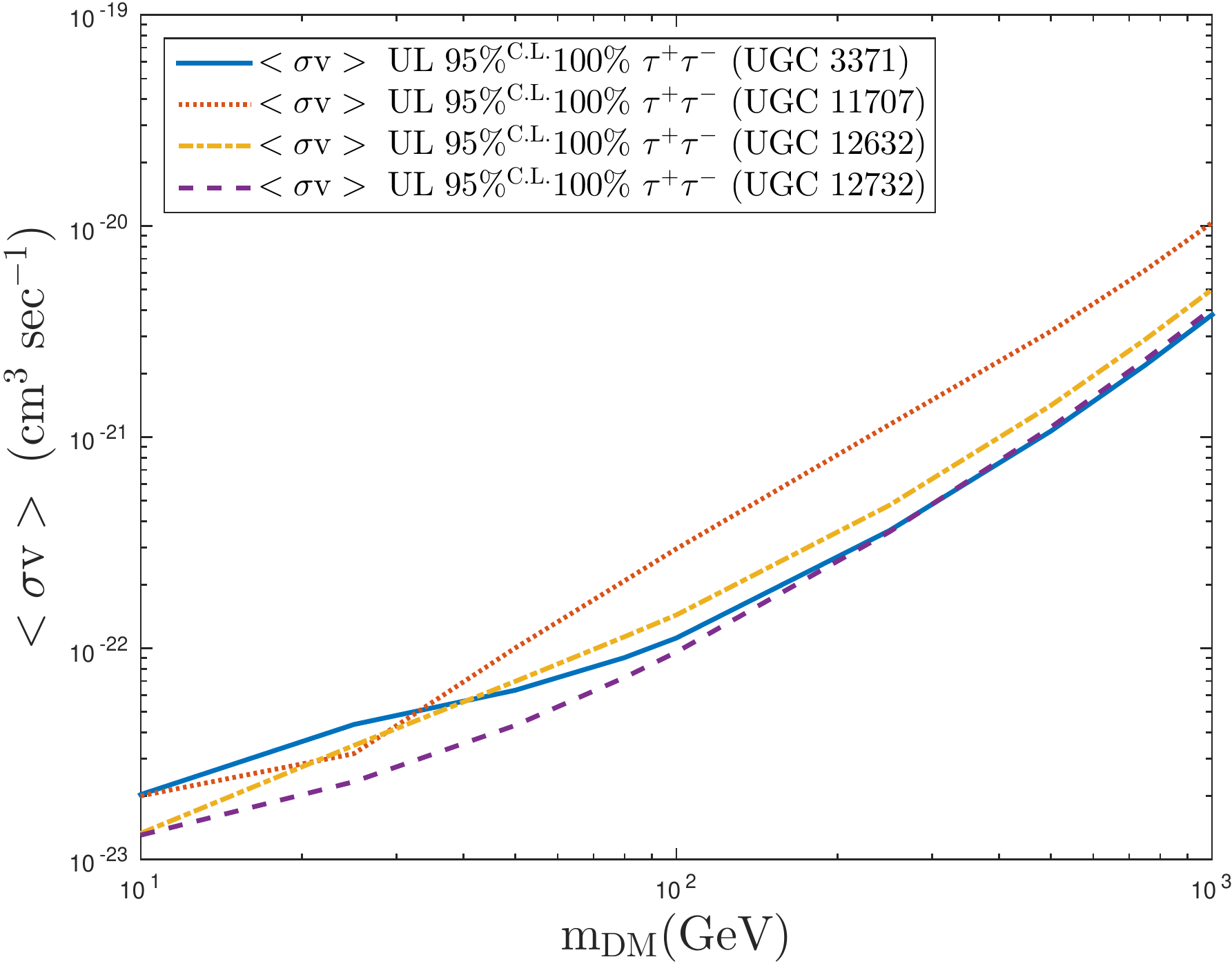}}
 \subfigure[]
 { \includegraphics[width=0.75\linewidth]{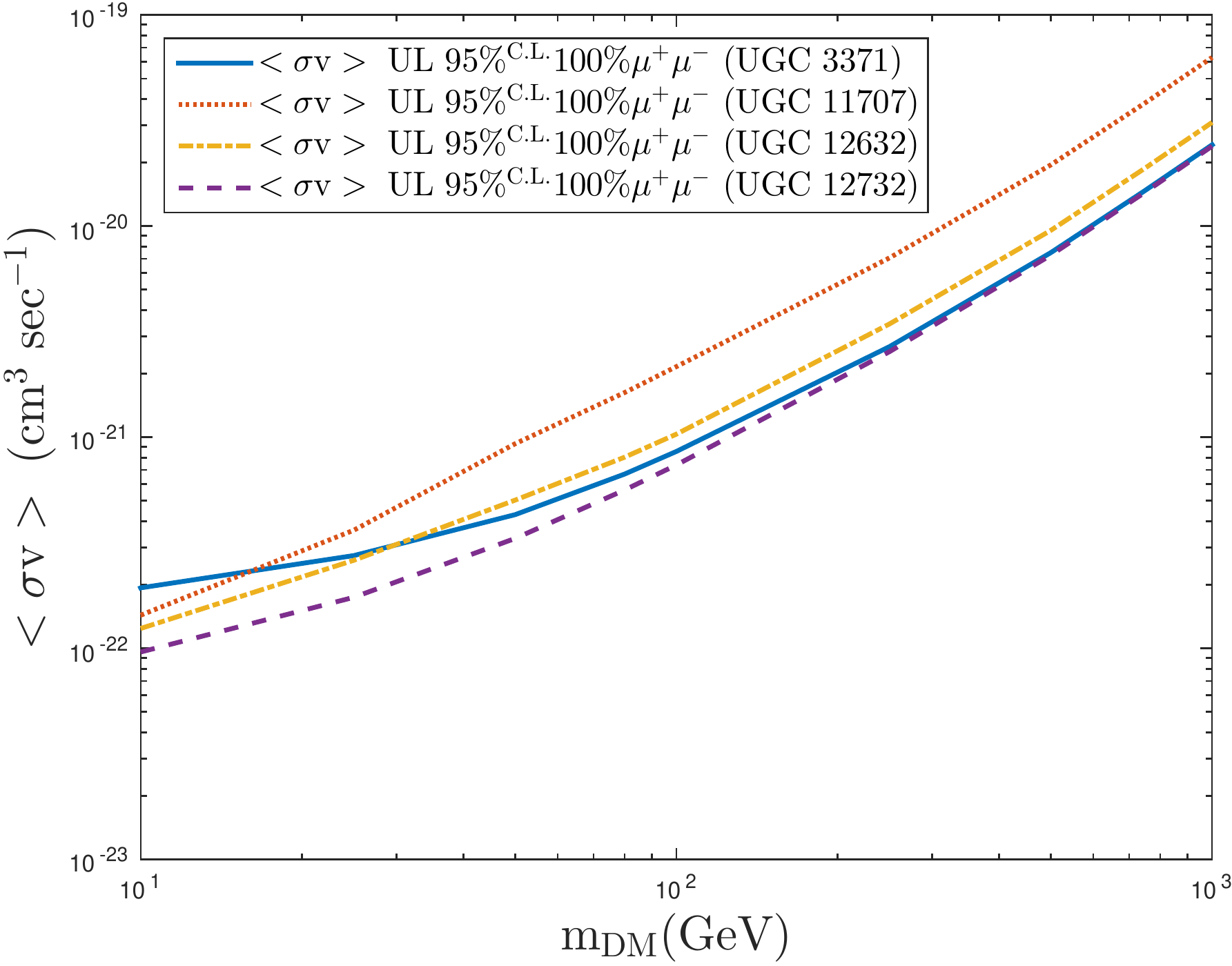}}
\subfigure[]
 { \includegraphics[width=0.75\linewidth]{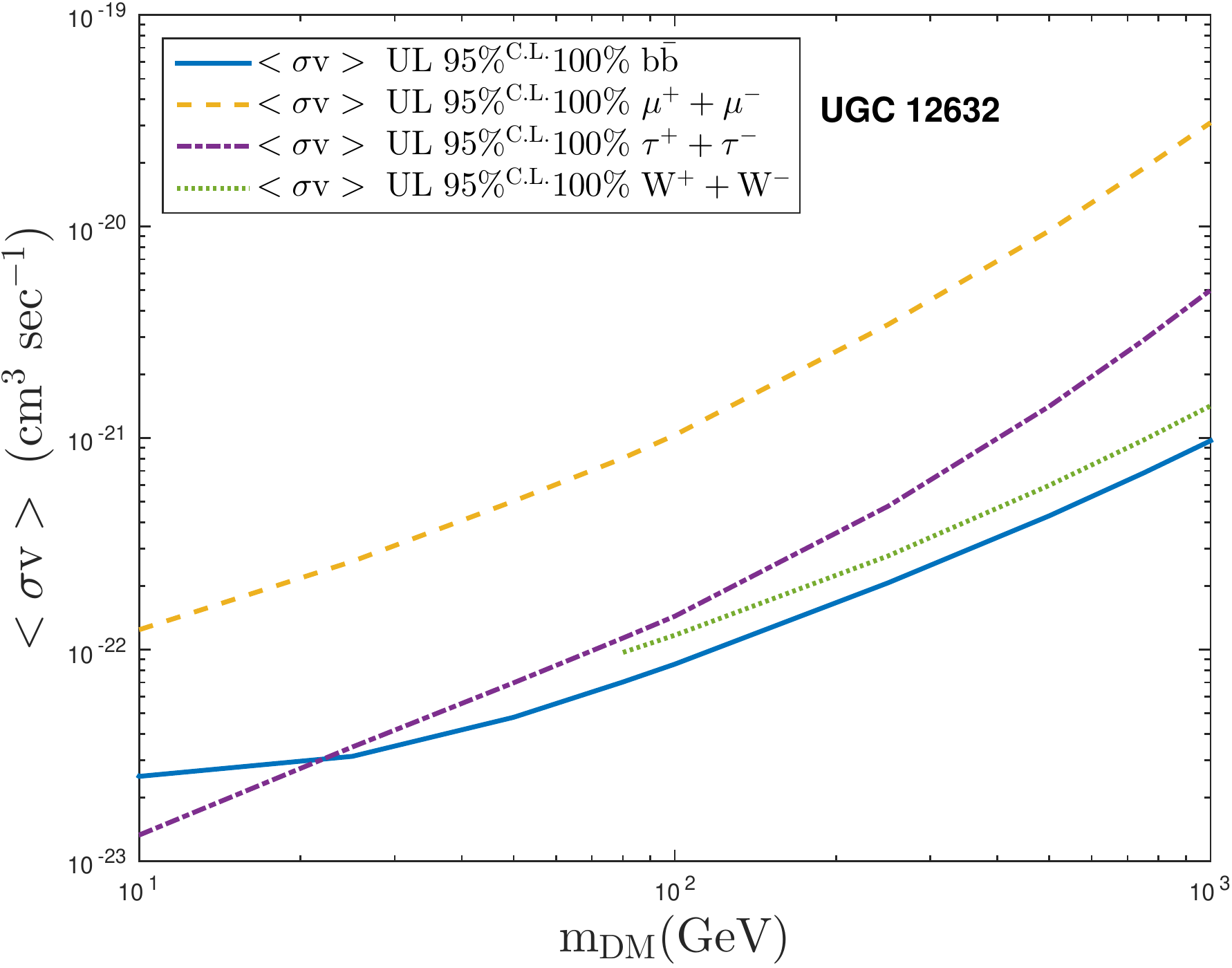}}
\caption{95$\%$ C.L. upper limits on $<\sigma~v>$ for the selected LSB galaxies for different pair annihilation final states: (a) $100\%$ $b\overline{b}$, (b) $100\%$ $\rm{\tau^{+} \tau^{-}}$, (c) $100\%$ $\rm{\mu^{+} \mu^{-}}$. Plot (d) shows the $<\sigma~v>$ upper limits of UGC 12632 for all the four selected annihilation final states.}
\end{center}
\end{figure}

\noindent From gamma-ray flux upper limits, we can derive constraints on the DM annihilation cross-section for each LSB galaxy using the DMFit package \citep{gondo, jel}. For the $\gamma$-ray spectra originating from the WIMP pair annihilation, we have estimated the upper limits of flux at 95$\%$ C.L. from LSB galaxies (already described in subsection 3.1) and the relative thermally averaged pair-annihilation cross-section $<\sigma v>$ as a function of DM mass ($m_{DM}$) and WIMP annihilation final states (f). For our paper, we have chosen four supersymmetry-theorem favored WIMP pair annihilation final states (f): $100\%$ $b\overline{b}$, $100\%$ $\tau^{+}\tau^{-}$, $100\%$ $\rm{\mu^{+} \mu^{-}}$ and $100\%$ $W^{+}W^{-}$ \citep{jung}. These four annihilation channels have been used in several previous studies, particularly when the neutralino is a WIMP candidate. Supersymmetry theory indeed favours neutralino as a CDM candidate, but our results are generic to any WIMP models.\\

\noindent For obtaining the flux upper limit and the corresponding $<\sigma v>$ limits as a function of $m_{DM}$, we have used our estimated J-factors from Table~5. In Fig.~3(a,b,c), we show the flux upper limit at 95$\%$ C.L. of the LSB galaxies as a function of $m_{DM}$ for three particular pair annihilation final states: $100\%$ $b\overline{b}$, $100\%$ $\rm{\tau^{+} \tau^{-}}$ and $100\%$ $\rm{\mu^{+} \mu^{-}}$, respectively. In Fig.~4(a,b,c), we show instead the $<\sigma~v>$ upper limits at 95$\%$ C.L. for these three annihilation channels. In Figs.~3(d) and ~4(d), we display the flux upper limit and $<\sigma~v>$ of UGC 12632 for four annihilation final states: $100\%$ $b\overline{b}$, $100\%$ $\rm{\tau^{+} \tau^{-}}$, $100\%$ $\rm{\mu^{+} \mu^{-}}$ and $100\%$ $\rm{W^{+} W^{-}}$. Fig.~3(d) shows that $100\%$ $\rm{\mu^{+} \mu^{-}}$ and $100\%$ $\rm{\tau^{+} \tau^{-}}$ annihilation channels provide the best flux upper limits, especially at higher energies with less background diffusion. From fig.~3(d), we also observe that at $m_{DM} \sim 1$~TeV, the flux upper limit for all four annihilation final states varies within a factor of $2$ but for low mass WIMP, this variation is about factor of 4. We have checked that all of our LSB galaxies have showed the same signature for each annihilation channels. Hence in Figs.~3(d) and 4(d), we have only shown the flux upper limit and corresponding $<\sigma v>$ limits of UGC 12632 for four annihilation final states. Here we would like to mention that for Figs.~3 and 4, we have only considered the median value of J-factor (Table~5).\\

\begin{figure}
\begin{center}
\includegraphics[width=0.9\linewidth]{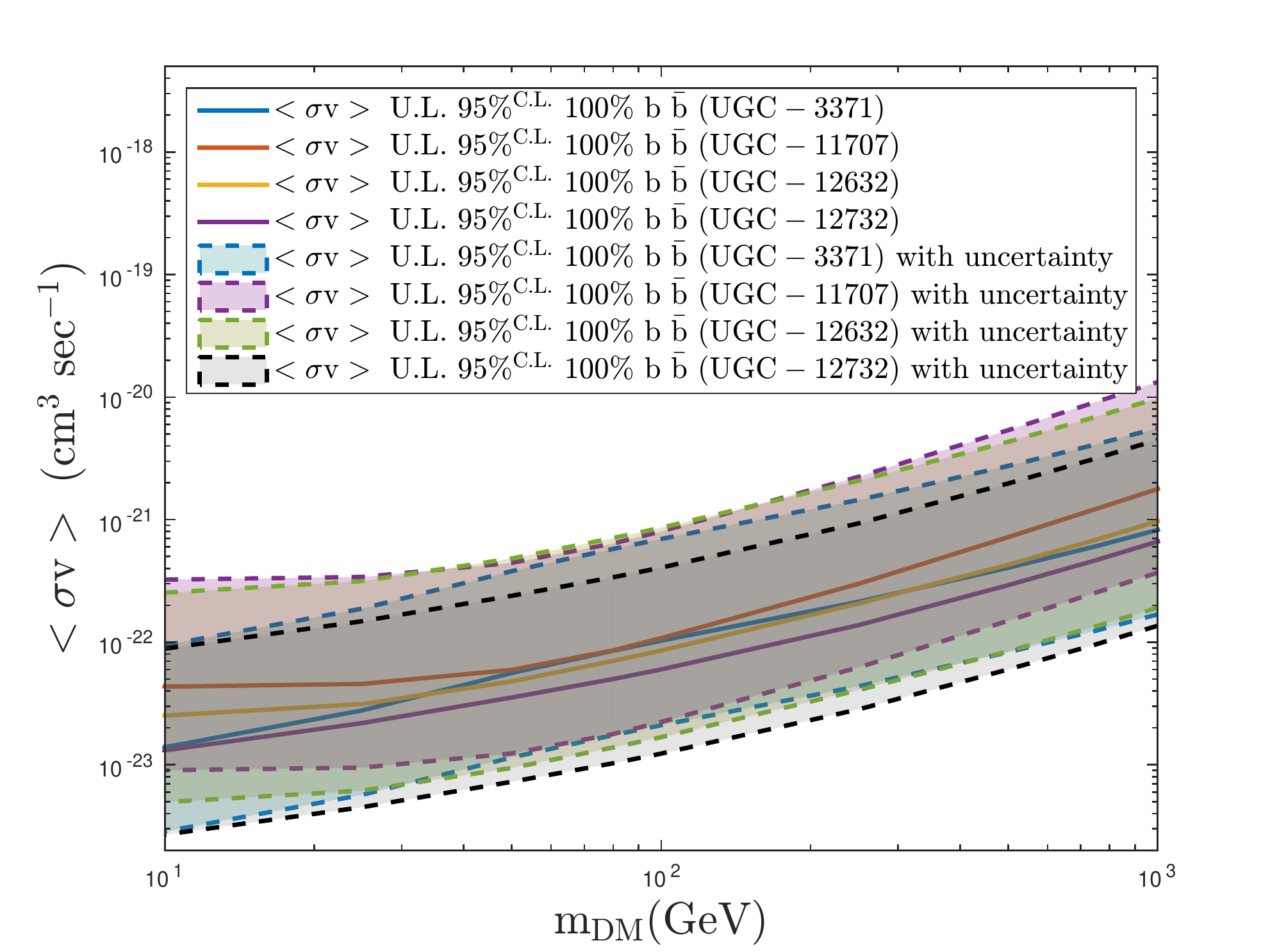}
\caption{95$\%$ C.L. upper limit of the WIMP pair-annihilation $<\sigma v>$ as a function of $m_{DM}$ for $b\overline{b}$ annihilation final states for the four selected LSB galaxies, using the median value of J-factor and taking into account its
uncertainties. The shaded band represents the uncertainty in J-factor for all four LSB galaxies.}
\end{center}
\end{figure}

\begin{figure}
\begin{center}
\includegraphics[width=0.8\linewidth]{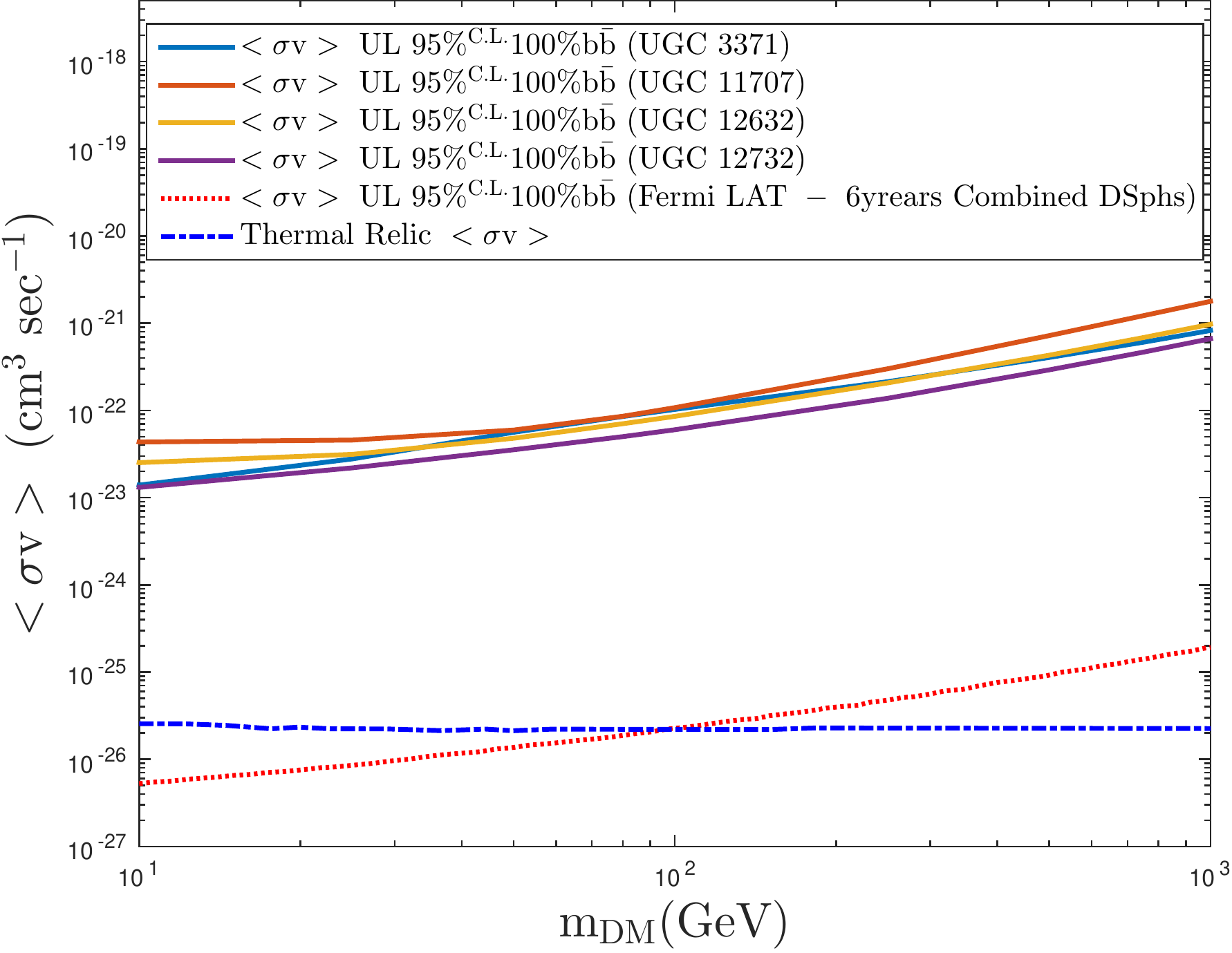}
\caption{95$\%$ C.L. upper limit of the WIMP pair-annihilation $<\sigma v>$ as a function of $m_{DM}$ for $b\overline{b}$ annihilation final states for the four selected LSB galaxies, using the median J-factor. We have also overplotted the relic abundance (or thermal) cross-section ($2.2~\times~10^{-26}~cm^{3}~s^{-1}$) estimated by \citet{ste1} and the $<\sigma v>$ values obtained from a combined analysis of 15 dSphs by \citet{ack}.}
\end{center}
\end{figure}

\noindent In Fig.~5, we have displayed the 95$\%$ C.L. upper limits of the thermally averaged WIMP pair-annihilation cross-section, as a function of the DM mass (${m_{DM}}$) for its median J-factor and considering the J-factor uncertainties. Only the $<\sigma v>$ upper limits for 100$\%$ $b\overline{b}$ annihilation channel has been considered in this figure as from Fig.~4(d) we have obtained that for $\gamma$-ray flux, 100$\%$ $b\overline{b}$ channel can provide the most stringent limits. From, Fig.~5, we can observe that the LSB galaxies have large 2$\sigma$ uncertainties and their uncertainties bands are overlapping each other. So, at this moment we could not favour one particular LSB galaxies over the others. The large uncertainty bands in LSB galaxies come from our inadequate knowledge of its internal structure. In the case of LSB galaxies, their lack of star formation and nuclear activity is the main obstacle in understanding their DM distribution.\\

\noindent A comparative study of the upper limits of WIMP pair-annihilation $<\sigma v>$, as obtained in this paper, with the $<\sigma v>$ obtained from other studies of the dSphs/UFDs is displayed in Fig.~6. The limits obtained from \citet{ack} is considered as one of the standard limits for indirect DM detection (the red dotted line in Fig.~6). These limits include the results obtained from a combined analysis of 15 dSphs from six years of Fermi-LAT data. From Fig.~6, we can notice that the LAT sensitivities obtained from these four LSB galaxies are nearly 3 orders of magnitude weaker than the ones derived by~\cite{ack}.

\subsection{\textbf{Stacking Analysis}}
\label{Section4.3}
\noindent In this section, we have performed a joint likelihood analysis on the four selected LSB galaxies. From Fig.~6, we have observed that the $<\sigma v>$ limits obtained from the individual LSB galaxies are nearly 3 orders of magnitude weaker than the LAT sensitivity estimated from the combined analysis by~\cite{ack} and the cross-section for relic abundances \citep{ste1}. Hence, the joint likelihood analysis would be an ideal approach for this scenario because it is expected that the sensitivity of our analysis would increase after stacking. For this purpose, we have generated a joint likelihood function as a product of individual likelihood functions of LSB galaxies. This function has combined the WIMP annihilation $<\sigma v>$, J-factor, DM mass, branching ration and the $\gamma$-ray spectrum of all individual LSB galaxies:
\begin{equation}
\mathcal{\widetilde{L}}(\mu, \{\alpha_{i}\}|\mathcal{D})=\prod \limits_{i=1} \mathcal{\widetilde{L}}_{i}(\mu, \alpha_{i}|\mathcal{D})
\end{equation}
\noindent For the combined likelihood analysis, the individual likelihood function for each LSB galaxy is weighted with their respective J-factor. The combined analysis of LSBs also has not shown any $\gamma$-ray emission resulting from the WIMP annihilation, hence, we have derived the upper limit of  $\gamma$-ray flux and $<\sigma v>$ in 95$\%$ C.L. by the delta-likelihood method.\\

\begin{figure}
\begin{center}
\subfigure[]
 { \includegraphics[width=0.8\linewidth]{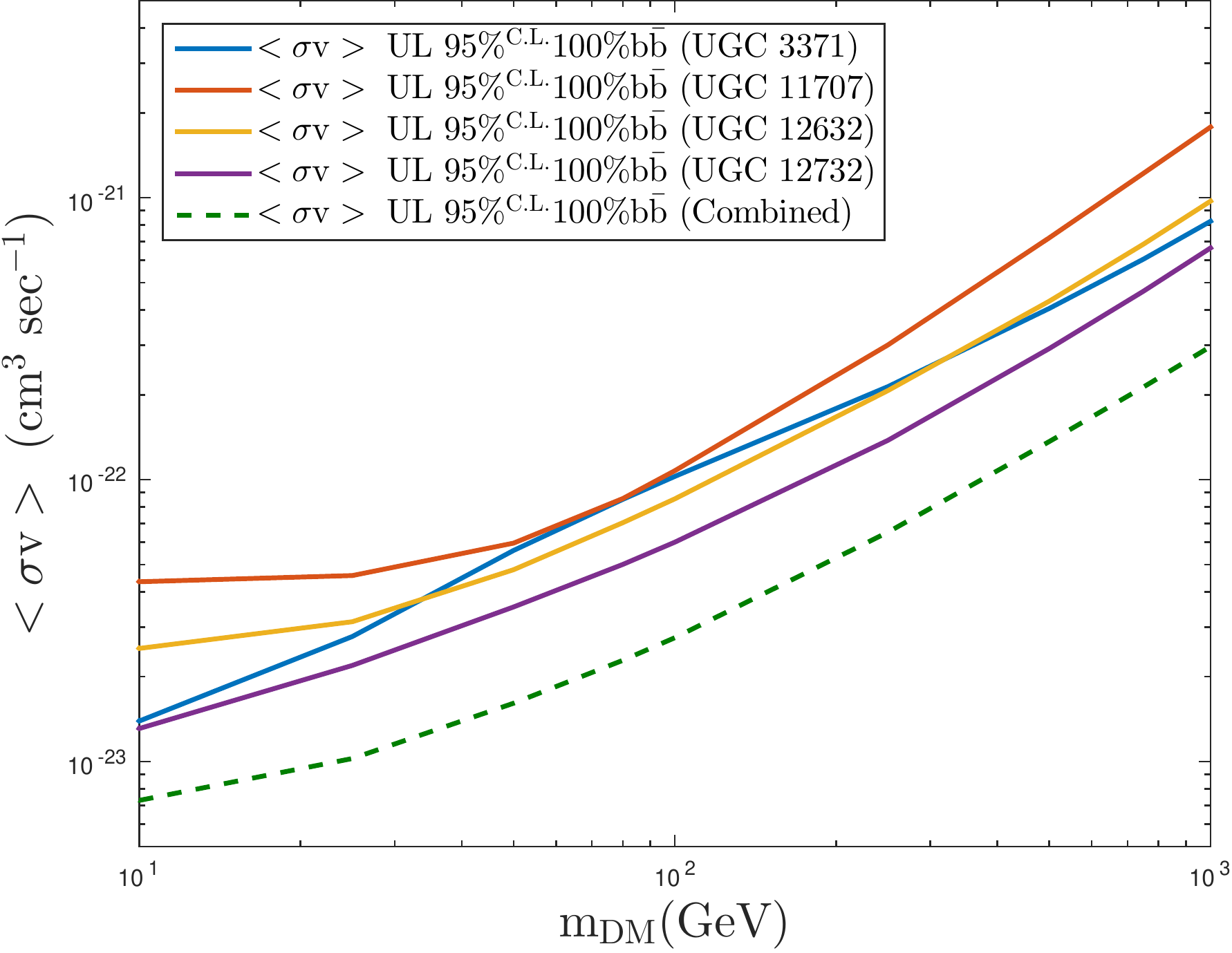}}
\subfigure[]
 { \includegraphics[width=0.8\linewidth]{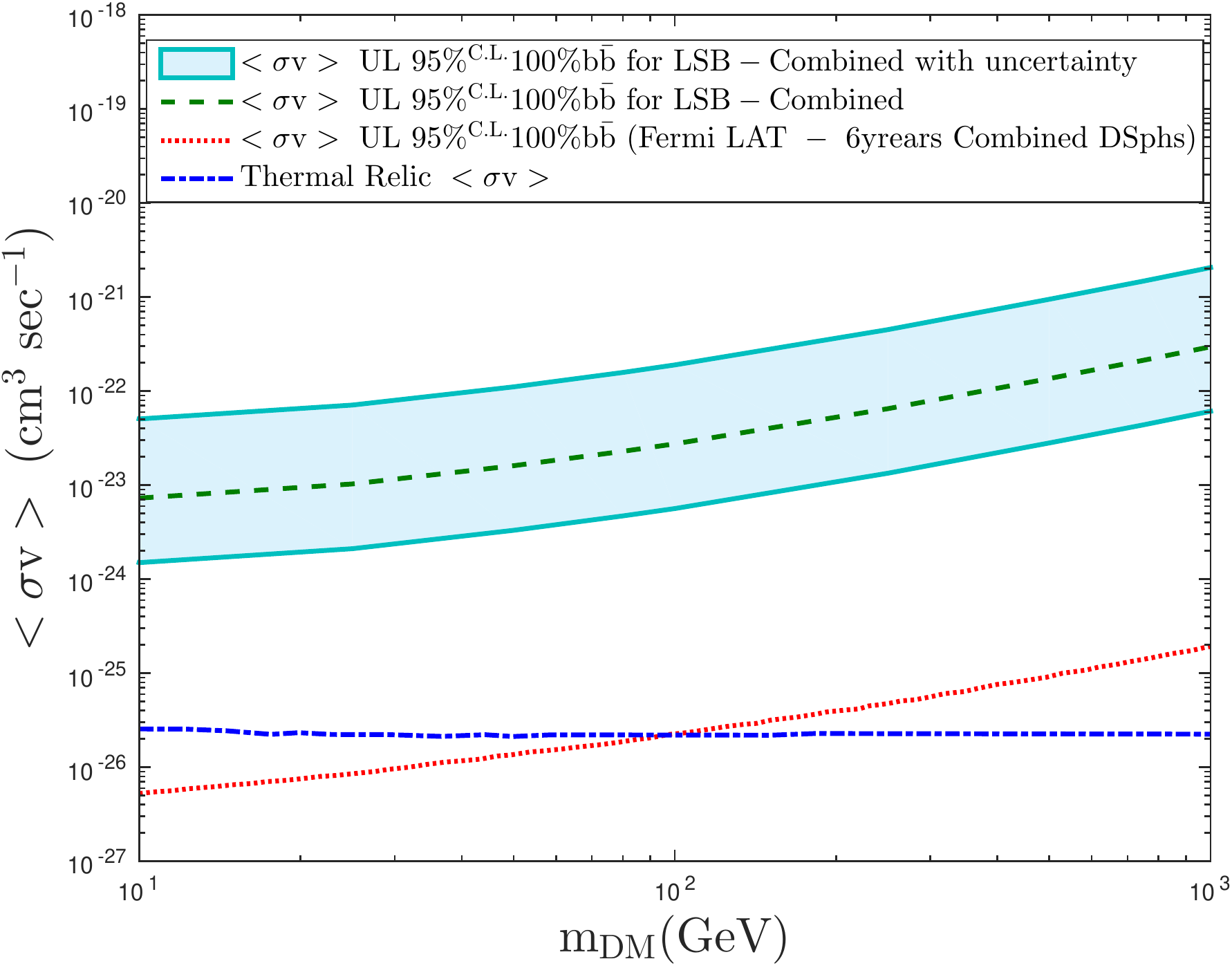}}
\caption{Comparison between the $<\sigma v>$ upper limits obtained from LSB stacking analysis with the limits from individual targets for $100\%$ $b\overline{b}$ pair annihilation final state, considering the median J-factor. b) Comparison between the $<\sigma v>$ upper limits obtained from LSB stacking analysis with the predictions from relic abundance (or thermal) cross-section estimated by \citet{ste1} and upper limit obtained by \citet{ack}. The shaded band represents the uncertainty of the stacking limits.}
\end{center}
\end{figure}

\noindent In Fig.~7(a), we show the upper limit of $<\sigma v>$ as a function of $m_{DM}$ from the combined analysis along with their individual limits for $100\%$ $b\overline{b}$ annihilation channel. In this figure, $<\sigma v>$ upper limits for each LSB galaxies are associated with their median J values. In Fig.~7(b), we have compared the stacking limits for LSB galaxies with the limits obtained from \cite{ack} and thermal abundances from \cite{ste1}. In Fig.~7(b), we have also shown the 2$\sigma$ uncertainty band for combined $<\sigma v>$ upper limits.\\

\noindent From Fig.~7, we find that the stacking analysis has improved the combined $<\sigma~v>$ limits obtained from individual LSB galaxies by a factor $\sim$4. This represents a marginal improvement, if compared to the limits obtained in \cite{ack}. Nevertheless, the constraints on DM models achievable from LSB galaxies may significantly improve in the future, as many new LSB galaxies are expected to be discovered with the future optical surveys.

\subsection{\textbf{Possible radio constraint obtained from LSB galaxies}}
\label{Section4.4}

\begin{figure}
\subfigure[UGC 3371]
 { \includegraphics[width=0.78\linewidth]{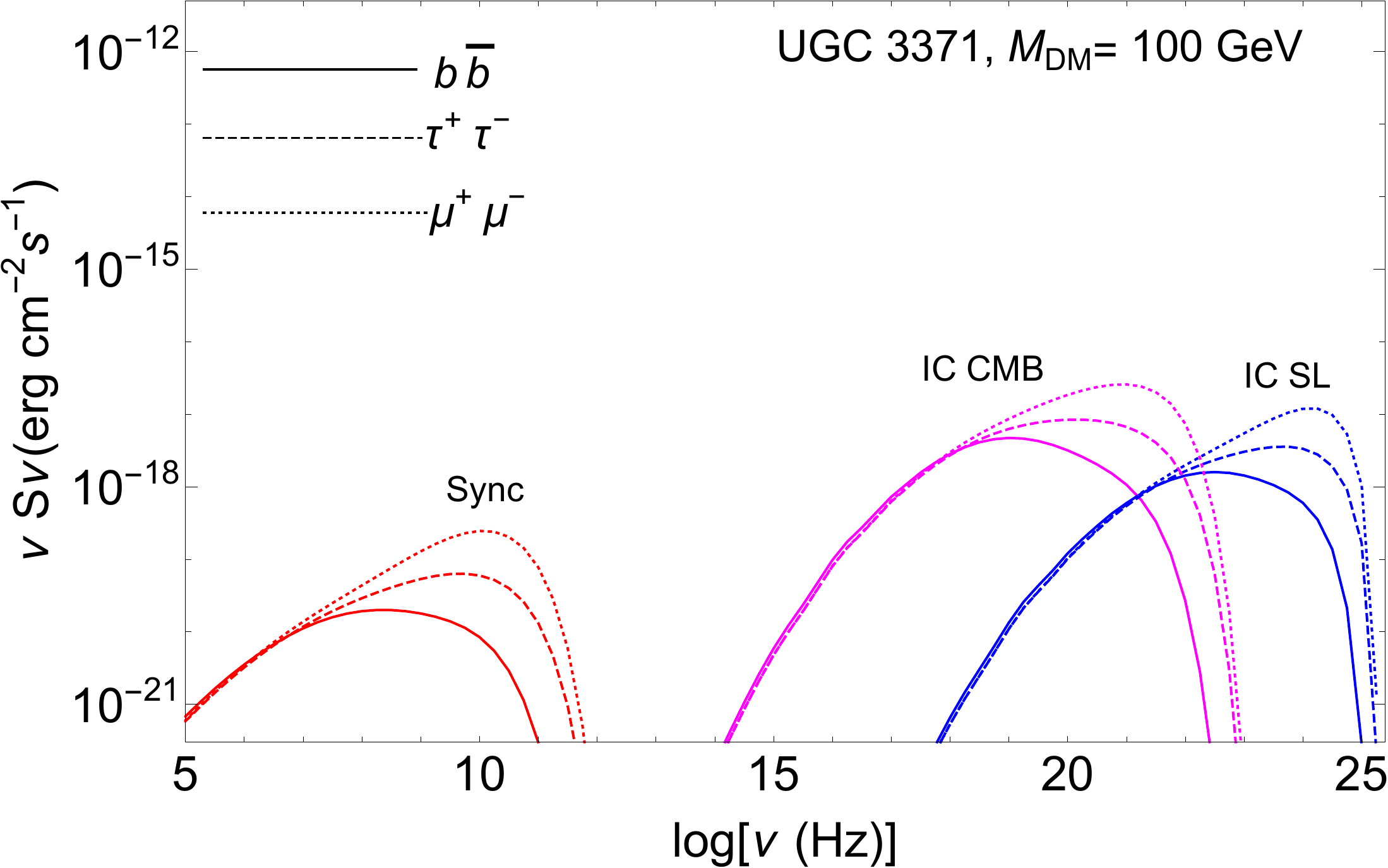}}
\subfigure[ UGC 11707]
 { \includegraphics[width=0.78\linewidth]{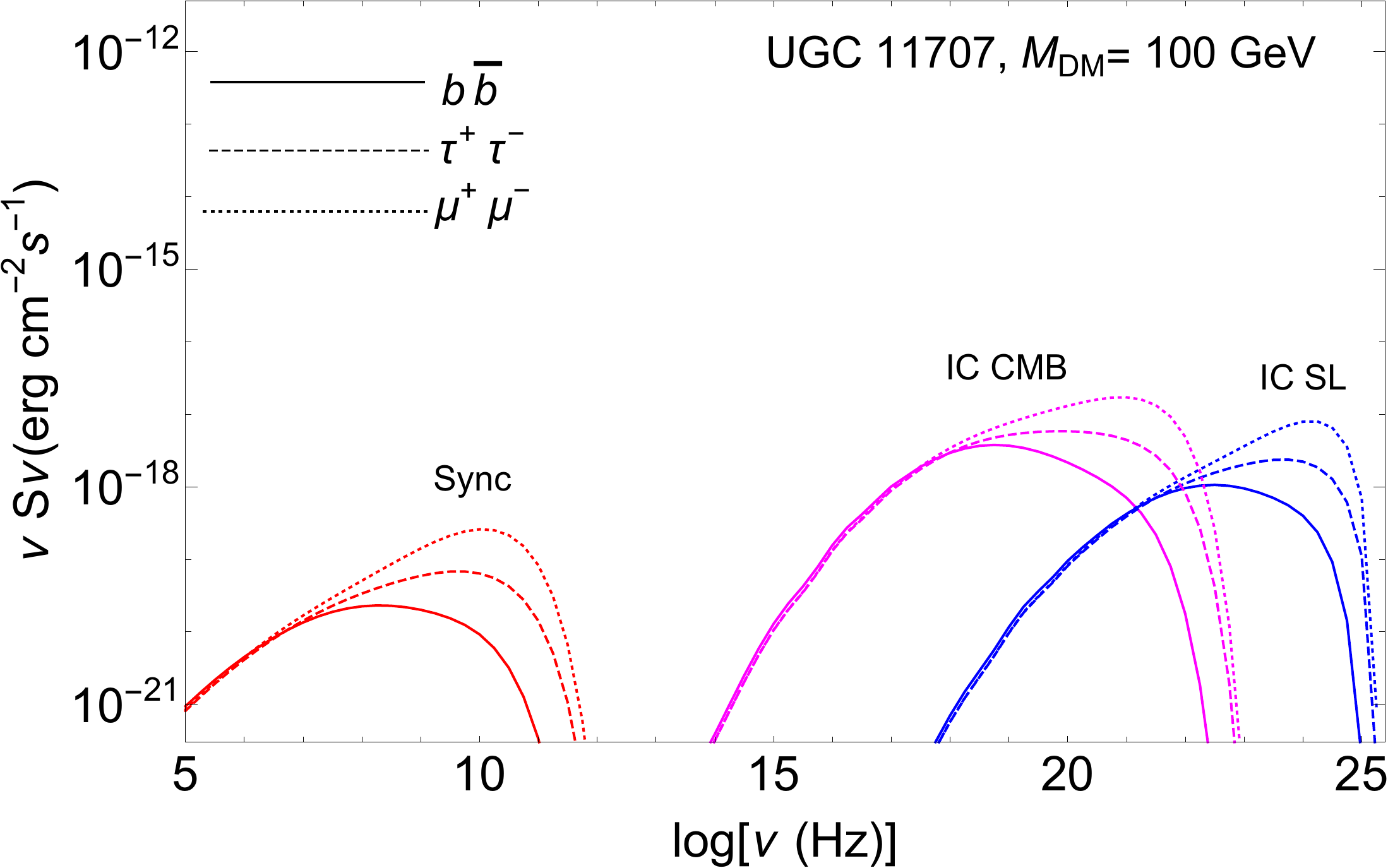}}
\subfigure[ UGC 12632]
 { \includegraphics[width=0.78\linewidth]{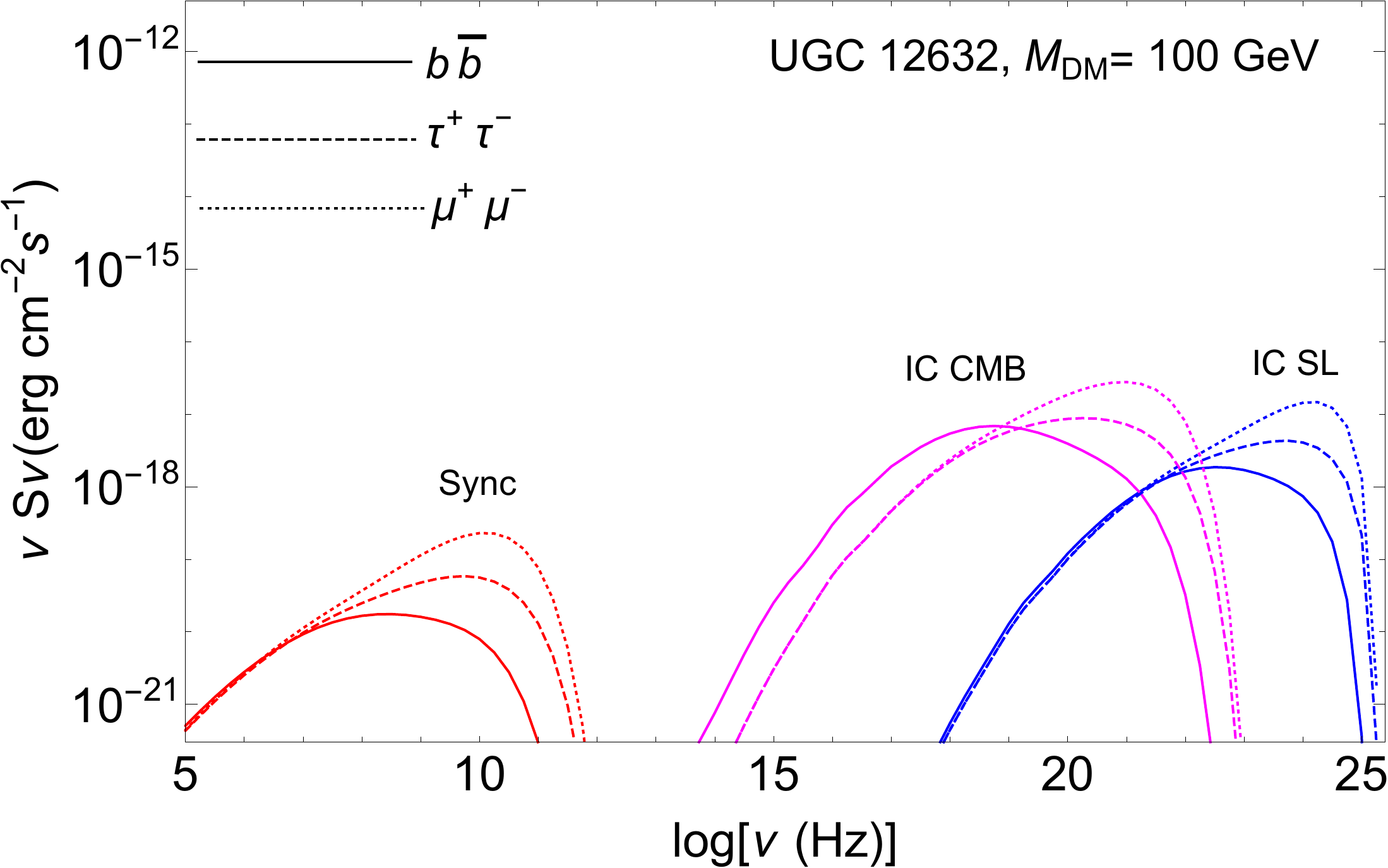}}
\subfigure[ UGC 12732]
 { \includegraphics[width=0.78\linewidth]{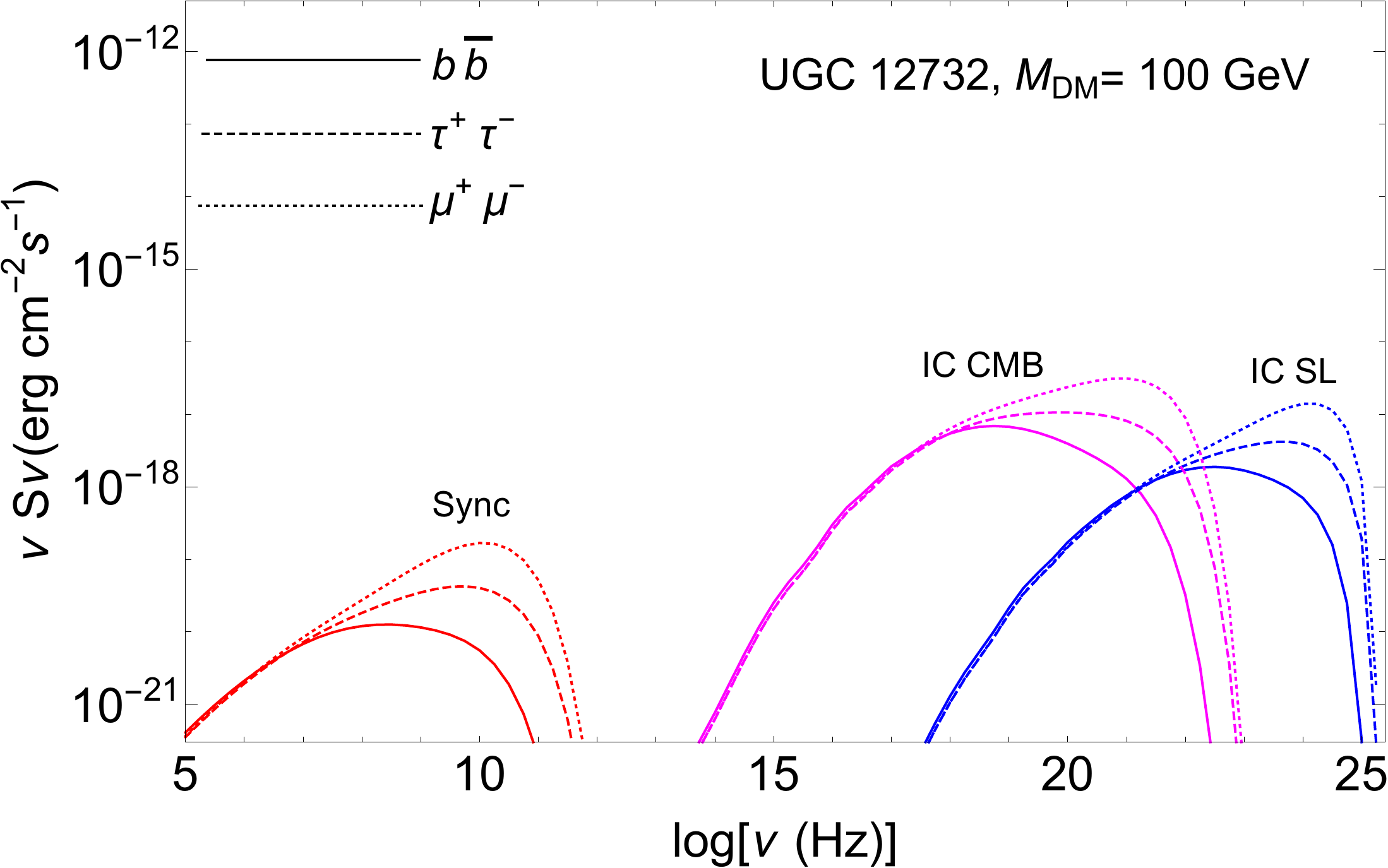}}
\caption{Multiwavelength SED of all the selected LSB galaxies for WIMP annihilating into $b\overline{b}$, $\tau^{+}\tau^{-}$ and $\mu^{+}\mu^{-}$ final states. We assume diffusion constant, $D_{0}$ = $3\times10^{28}~cm^{2}s^{-1}$ and $m_{DM}$=100 GeV.}
\end{figure}

\begin{figure}
\subfigure[Variation with $B_{0}$]
 { \includegraphics[width=0.8\linewidth]{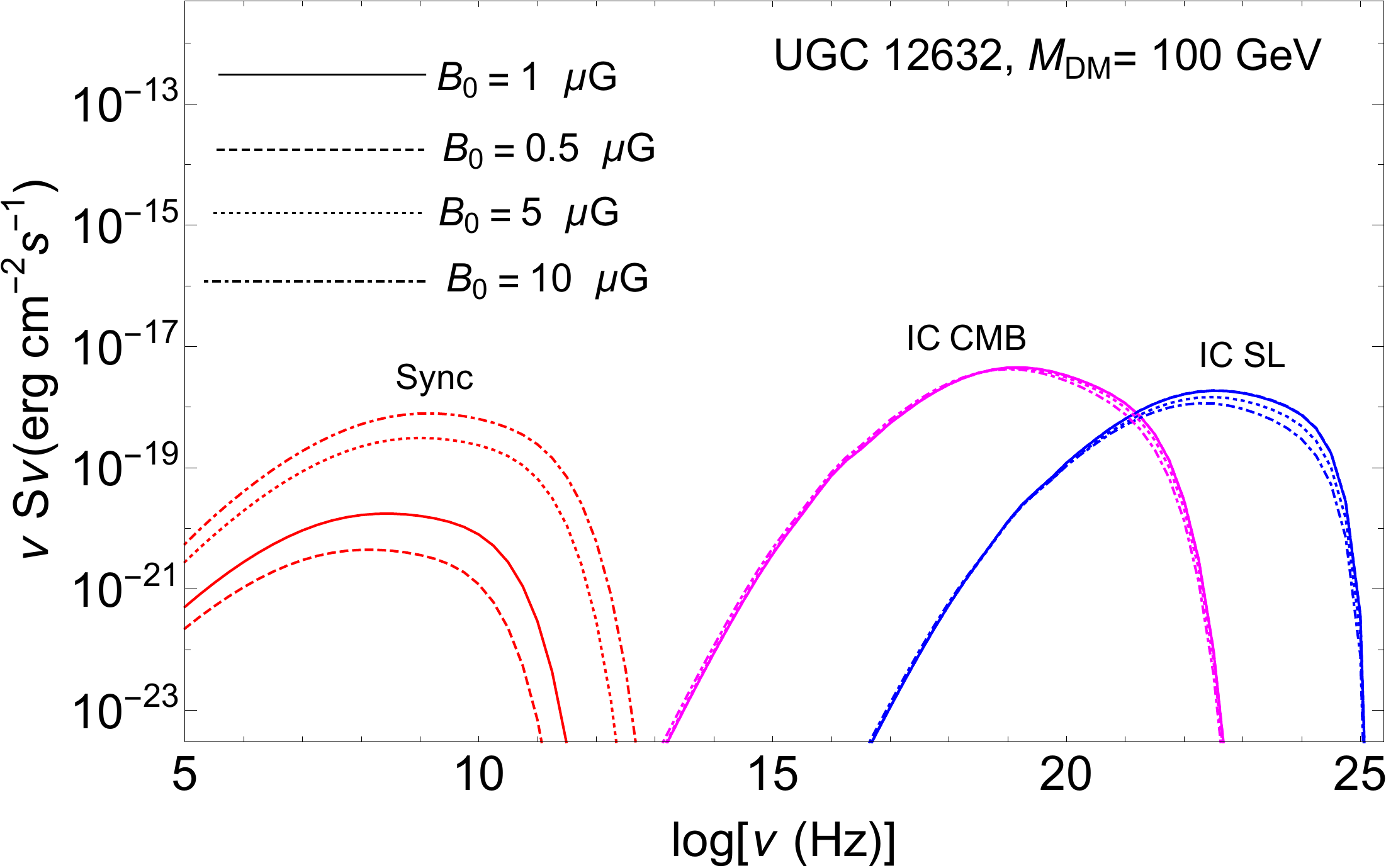}}
\subfigure[Variation with $D_{0}$]
 { \includegraphics[width=0.80\linewidth]{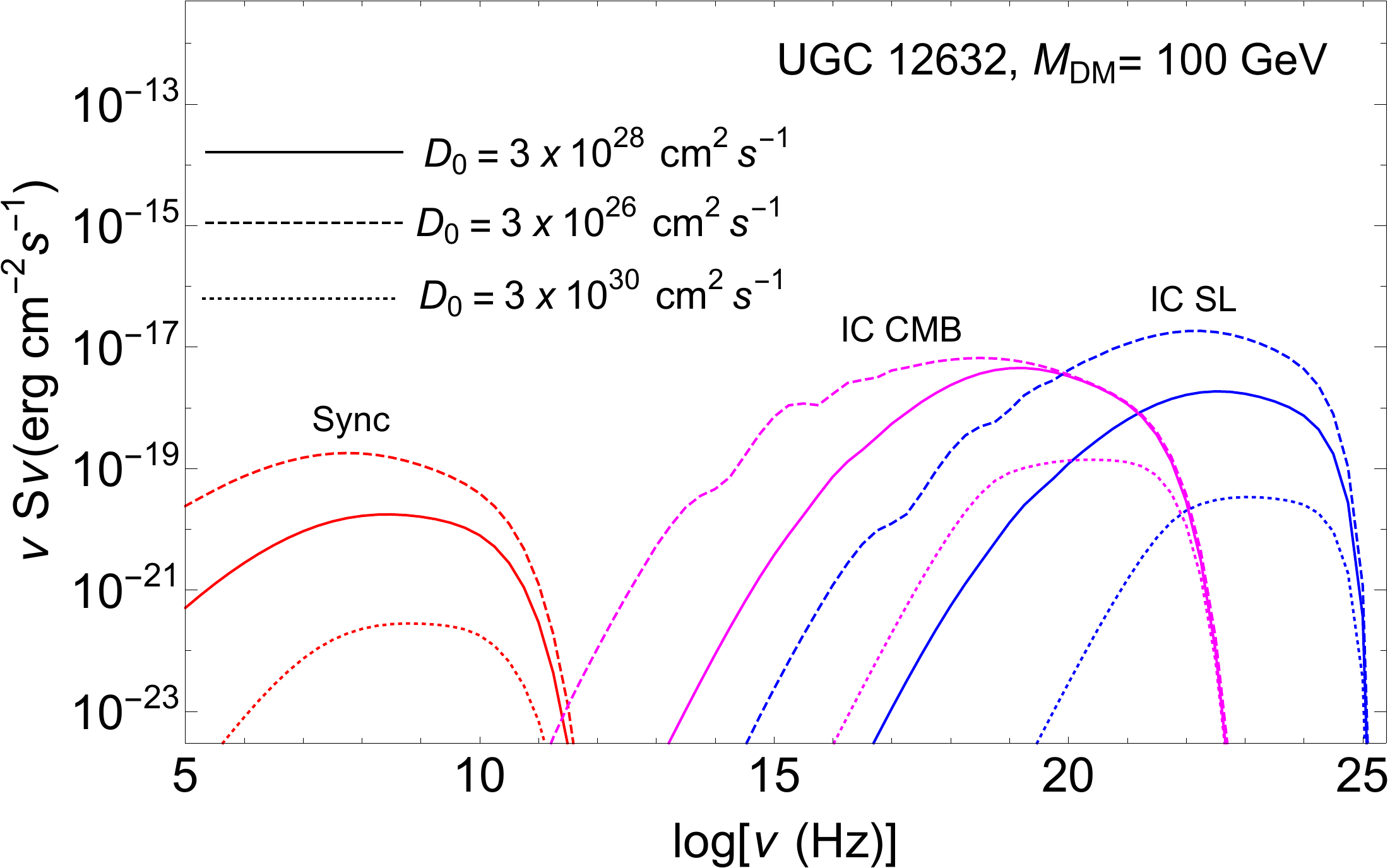}}
\subfigure[Variation with $\gamma$]
 { \includegraphics[width=0.8\linewidth]{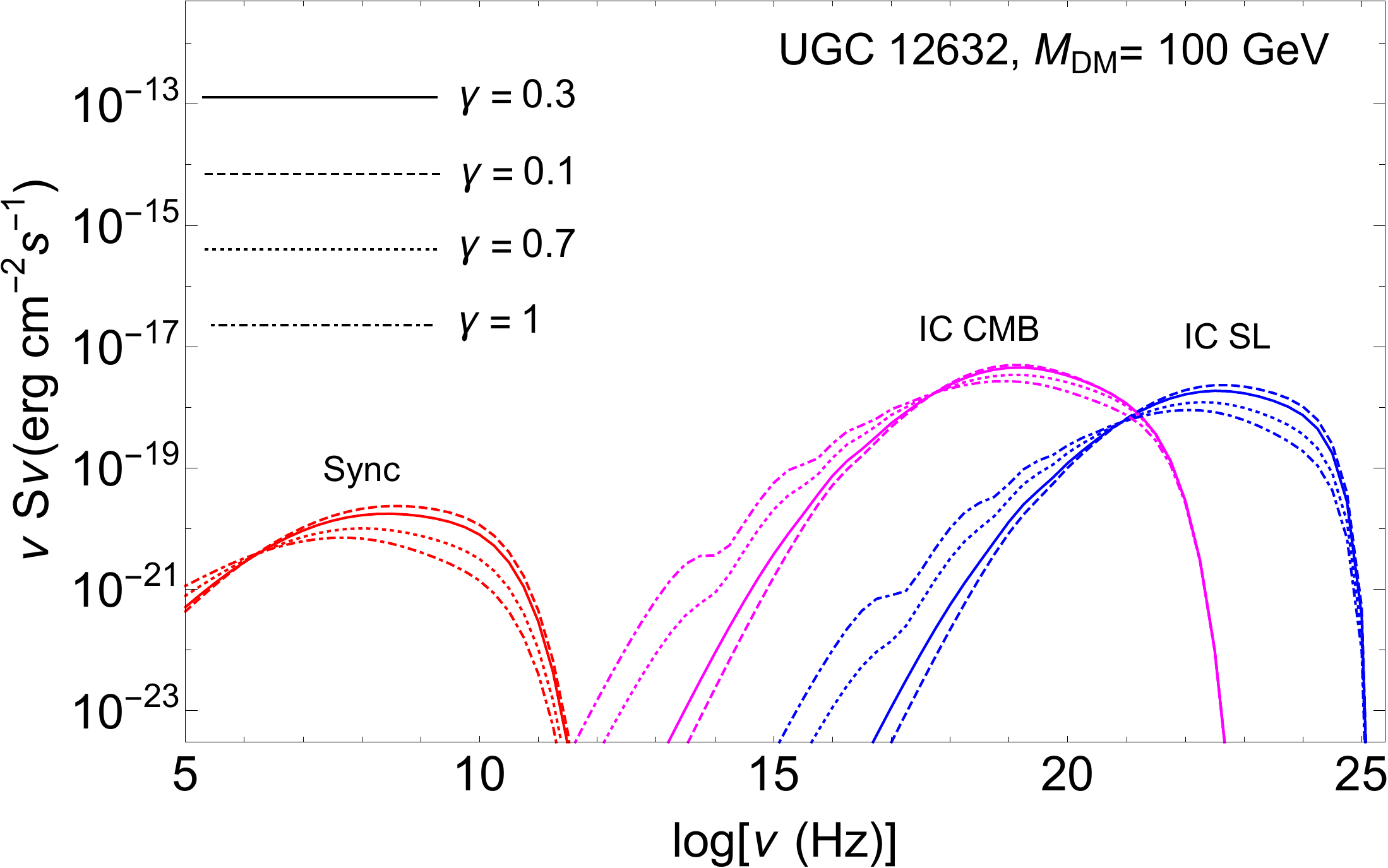}}
\caption{Variation of SED for UGC 12632 with (a) four possible values of $B_{0}$, (b) three possible values of $D_{0}$ and (c) four possible values of $\gamma$. All three results are shown for $m_{DM}$=100 GeV and the thermal averaged DM annihilation cross-section is fixed to the value to $<\sigma v> = 3 \times 10^{26}~cm^{3}s^{-1}$.}
\end{figure}

\noindent In this section, we examine the possible radio emission from LSB galaxies. The $\gamma$-ray analysis is one of the most popular ways to examine the indirect detection of DM signature \citep{ackermann2010, ack, abramowski2011, abramowski2014, charles2016, daylan2016} but a multiwavelength approach can also provide a complementary probe of $\gamma$-ray analysis \citep{strom2012, strom2017, regis2017}. For dSphs, several pieces of literature point out that the constraint obtained from the radio data can be competitive with $\gamma$-ray data.\\   

\noindent WIMP can self-annihilate to the standard model particles such as bosons, quarks, leptons and then they decay to the charged particles such as positrons, electrons, etc. Secondary charged particles are generated through the various mechanisms of generation and the decay of charged pions, i.e., $\pi^{\pm} \rightarrow \mu^{\pm} + \nu_{\mu}(\overline{\nu_{\mu}})$, with $\mu^{\pm} \rightarrow e^{\pm} + \overline{\nu_{\mu}}(\nu_{\mu})$. These secondary charged particles in the astrophysical system can lead to a generation of multiwavelength energy spectrum caused by several radiation mechanisms like synchrotron (sync), inverse Compton (IC), bremsstrahlung and Coulomb energy loss \citep{ginzburg, colafrancesco2006, longair}. At high energies, the synchrotron and the inverse Compton radiation mechanisms play a dominant role. In order to predict the radio or the X-ray emission through DM annihilation, we need to solve the diffusion equation of the secondary electron spectrum. The transport equation is \citep{colafrancesco2006, colafrancesco2007}:
\begin{equation}
\frac{\partial}{\partial t} \frac{\partial n_{e}}{\partial E} = \nabla \Big[D(E,r) \nabla \frac{\partial n_{e}}{\partial E} \Big] + \frac{\partial}{\partial E} \Big[b(E,r) \frac{\partial n_{e}}{\partial E} \Big] + Q_{e}(E,r)
\end{equation}
\begin{equation}
Q_{e}(E,r)~ = ~ \frac{<\sigma v>}{2~m^{2}_{DM}} \rho^{2} (r) \sum_{f} \frac{dN^{e}_{f}}{dE}B_{f}.\\
\end{equation}

\noindent Here, we have assumed that at high energy $e^{\pm}$ are in equilibrium i.e., $\frac{\partial}{\partial t} \frac{\partial n_{e}}{\partial E} \approx 0 $ in eq.~18. $Q_{e}(E,r)$ is the $e^{\pm}$ source term and $\frac{\partial n_{e}}{\partial E}$ is the equilibrium energy density of $e^{\pm}$. In Eq.~19, $\rho(r)$ is the DM mass density profile of the LSB galaxy, $\frac{dN^{e}_{f}}{dE}$ and $B_{f}$ denote the $e^\pm$ injection spectrum and branching fraction corresponding to the $f^{\rm{th}}$ final state of WIMP annihilation, respectively. $b(E,r)$ is the $e^{\pm}$ energy loss per unit time where, $b(E,r) = b_{inverse~Compton} + b_{synchrotron} + b_{bremsstrahlung} + b_{Coulomb}$ \citep{colafrancesco2006} and $D~(E,r)$ is the diffusion coefficient.\\

\noindent The synchrotron emission from secondary electrons is the result of ambient magnetic fields that accelerate the charged particles, causing them to emit radiation at radio wavelengths \citep{ginzburg, longair}. The IC radiation peaks at X-ray frequencies and is the result of photons from various radiation sources such as the cosmic microwave background (CMB) and starlight being up-scattered by the relativistic particles \citep{ginzburg, longair}. In this subsection, we would wish to estimate the radio signature from LSB galaxies and for such purpose, we have used a publicly available code, RX-DMFIT \citep{alex}. This code is an extension of the DMFit tool \citep{gondo, jel} which is used for $\gamma$-ray fitting in Fermi-LAT science tools. Like our gamma-ray analysis, here we have again used the NFW density profile to model the DM distribution in LSB galaxies and for calculating the source term ($Q_{e}$) from Eq.~19, RX-DMFIT code calls for the Fortran package DarkSUSY v5.1.2. DarkSUSY determines the electron/positron injection spectrum per DM annihilation event, $\sum_{f} \frac{dN^{e}_{f}}{dE}B_{f}$ which is dependent on the DM particle mass, annihilation channel, and the source energy, $E$. \\

\noindent RX-DMFIT includes a large range of astrophysical and particle parameters such as distance ($d$) to the DM source, DM density profile ($\rho$), diffusion constant ($D_{0}$), diffusion zone ($r_{h}$), magnetic field ($B_{0}$), core radius ($r_{c}$), etc. We can customize them according to our needs. This code \citep{alex} incorporates important astrophysical scenarios including diffusion of charged particles, relevant radiative energy losses, and magnetic field modelling. It uses two forms of diffusion coefficients and magnetic field models. Here we want to mention that there is no such detailed study on diffusion mechanism and magnetic field for our selected LSB galaxies. But like dSphs, LSB galaxies are considered as the clean sources (i.e. there is no presence of any significant emission due to astrophysical phenomenon) and the kinematics of LSB galaxies are not much different from the dSphs. Thus, for our LSB galaxies, we can adapt the values of astrophysical parameters that are generally used for the dSphs case. For our purpose, we have assumed the Kolmogorov form of diffusion coefficient, $D(E)$=$D_{0}~E^{\gamma}$ where $D_{0}$ is the diffusion constant and the diffusion mechanism would extend to the diffusion zone, $r_{h}$. Like dSphs, the magnetic field strength of LSBs is very weak and their spatial extension is not well defined. Hence, we have taken the exponential form of magnetic field, i.e. $B(r)$=$B_{0}~e^{-r/r_{c}}$, where, $r_{c}$ is the core radius. For LSB galaxies, we have taken $r_{h}$ = 2 $\times$ $R_{last}$ , whereas $r_{c}$ = $r_{d}$ (see Table~1). We have fixed the magnetic field of the LSB galaxies at 1$\mu G$. There are no observations of the magnetic field in LSB galaxies and hence we have assumed a value of 1$\mu G$ which is typical of late-type spiral galaxies \citep{fitt1993}. For galaxy
clusters, the $D_{0}$ lies between $10^{28}$-$10^{30}$ $cm^{2}s^{-1}$ \citep{natarajan2015, Jeltema2008}, whereas for Milky Way $D_{0}$ typically range between $10^{27}$-$10^{29}$ $cm^{2}s^{-1}$ \citep{Webber1992, baltz1998, maurin2011}.
Similarly, the expected range for $\gamma$ is assumed to lie between 0 to 1 \citep{Jeltema2008}. Thus, for our study, we have chosen the values which are close to their geometric means. For Diffusion coefficient ($D_{0}$) we have fixed its value at $3~\times~10^{-28}~cm^{-2}~s^{-1}$, while $\gamma$ is fixed at 0.3.
The thermally-averaged DM annihilation cross-section is fixed to the value $<\sigma v>$ $\approx$ $3\times10^{-26} cm^{3} s^{-1}$. The parameters that we have used as inputs of RX-DMFIT code are mentioned in Table~7.\\

\noindent With RX-DMFIT code, we can predict the properties of secondary emission generated from DM annihilation due to synchrotron and inverse Compton process. In Fig. 8, we have shown the multiwavelength spectral energy distribution (SED) of four LSB galaxies for three annihilation channels at $m_{DM}$=100 GeV. In this figure, `Sync' denotes SED for synchrotron emission while `IC CMB' and `IC SL' refer the inverse Compton emission resulting from the CMB photons and starlight photons, respectively. The `solid', `dashed' and `dotted' line style have been used for $b\overline{b}$, $\tau^{+}\tau^{-}$ and $\mu^{+}\mu^{-}$, respectively. From Fig. 8, we find that for radio and X-ray emission, $\tau^{+}\tau^{-}$ and $\mu^{+}\mu^{-}$ produce the harder spectrum than $b\overline{b}$ annihilation channel.\\

\noindent The SED plots displayed in Fig.~8 are directly dependent on the choice of different astrophysical parameters, mostly on the magnetic field ($B_{0}$) and diffusion constant ($D_{0}$). As we have already discussed, there is not ample observational study on LSB galaxies and thus the parameters that we have used for our analysis are not very precise. In Fig.~9, we have tried to check how does the SED of LSB galaxies would change if we vary $D_{0}$, $B_{0}$ and $\gamma$ within their approved ranges. For this study, we have only shown the result of UGC 12632 for the $b\overline{b}$ annihilation channel. In Fig~9(a), we have shown the SED for four values of $B_{0}$. From this plot, we can observe that the synchrotron emission is strongly dependent on the magnetic field and with increasing the magnetic field it would increase the emission. But the magnetic field does not have any strong influence on IC emission resulting from CMB photons and starlight photons. In Fig.~9(b), we have shown the SED for three $D_{0}$ values and from this figure, we can state that the SED of LSB galaxies strongly depends on the strength of the $D_{0}$. Finally, in Fig.~9(c), we have shown the variation of SED for four $\gamma$-values.\\

\begin{table}
\begin{center}
\caption{parameter set for RXDMFIT code.}
\begin{tabular}{|p{0.8cm}|p{1cm}|p{1cm}|p{0.5cm}|p{1cm}|p{0.4cm}|p{0.3cm}|p{0.5cm}|}
\hline 
\hline
Galaxy & $d$ & $c$ & $r_{h}$ & $D_{0}$ & $\gamma$ & $B_{0}$ & $r_{c}$ \\
$ $ & Mpc &  & Kpc & $cm^{2}s^{-1}$ & & $\mu G$ & Kpc \\
\hline
UGC 3371 & $12.73^{+0.90}_{-0.90}$ & $14.5^{+14.6}_{-10.2}$ & 20.4 & $3\times10^{28}$ & 0.3 & 1 & 3.09  \\
\hline
UGC 11707 & $14.95^{+1.05}_{-1.05}$ & $14.7^{+14.6}_{-10.3}$ & 30.0 & $3\times10^{28}$ & 0.3 & 1 & 4.30 \\
\hline
UGC 12632 & $8.36^{+0.60}_{-0.60}$ & $15.6^{+15.5}_{-10.9}$ & 17.06 & $3\times10^{28}$ & 0.3 & 1 & 2.57 \\
\hline
UGC 12732 & $12.38^{+0.87}_{-0.87}$ & $14.3^{+14.4}_{-10}$ & 30.8 & $3\times10^{28}$ & 0.3 & 1 & 2.21 \\
\hline
\hline
\end{tabular}
\end{center}
\end{table}

\begin{table}
\begin{center}
\caption{Radio flux-limit obtained from VLA radio telescope at 1.4 GHz.}
\begin{tabular}{|p{2cm}|p{5cm}|}
\hline 
\hline
Galaxy & Observed Flux density in mJy \\
\hline
UGC~3371 & $<$ 0.45 mJy \\
\hline
UGC~11707 & $1.17$ mJy \\
\hline
UGC~12632 & $<$ 0.45 mJy \\
\hline
UGC~12732 & $<$ 0.45 mJy \\
\hline
\hline
\end{tabular}
\end{center}
\end{table}

\noindent One of the important aspects of this code is that with its help we can predict the possible limits on the DM cross-section from the observed data of radio and X-ray emission. From the observed flux density data, one can estimate the limits for the ($<\sigma v>$ vs. $m_{DM}$) plane for several annihilation channels. LSB galaxies could be an ideal place for studying the diffuse radio signals obtained from DM pair annihilation, as their star formation rates are very low and active galactic nuclei are rare. This minimizes the uncertain contribution of astrophysical processes. For our study, we have predicted the DM constraints of four LSB galaxies using the radio data from the NVSS survey \citep{condon1998} which was an all-sky survey done using the Very Large Array (VLA) radio telescope at the frequency of ($\nu$)= 1.4 GHz. The VLA is a 27 element interferometric array that produces images of the radio sky over a wide range of frequencies and resolutions. The VLA is located at an elevation of 2100 meters on the Plains of San Agustin in southwestern New Mexico. The spatial size of the NVSS images around UGC 3371, UGC 11707, UGC 12632 and UGC 12732 were 185.40", 300.70", 66.00" and 307.70", respectively. For estimating the radio-limits on $<\sigma v>$ from DM annihilation, we have taken the observed radio data from NVSS images\footnote{\url{https://www.cv.nrao.edu/nvss/}} and except for UGC 11707, we have obtained the upper limits to the flux density (Table~8).\\

\begin{figure}
\subfigure[]
 { \includegraphics[width=0.8\linewidth]{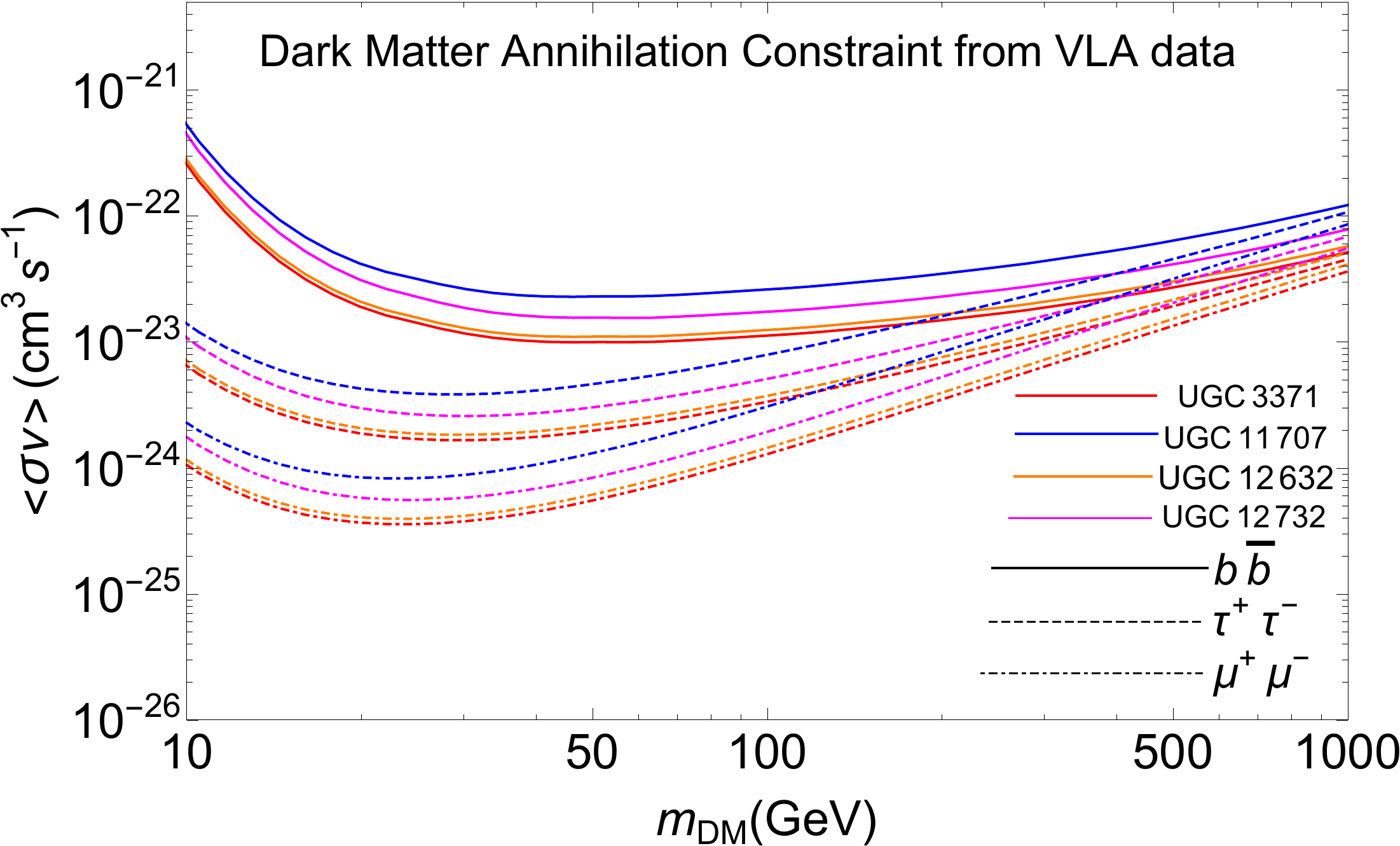}}
\subfigure[]
 { \includegraphics[width=0.8\linewidth]{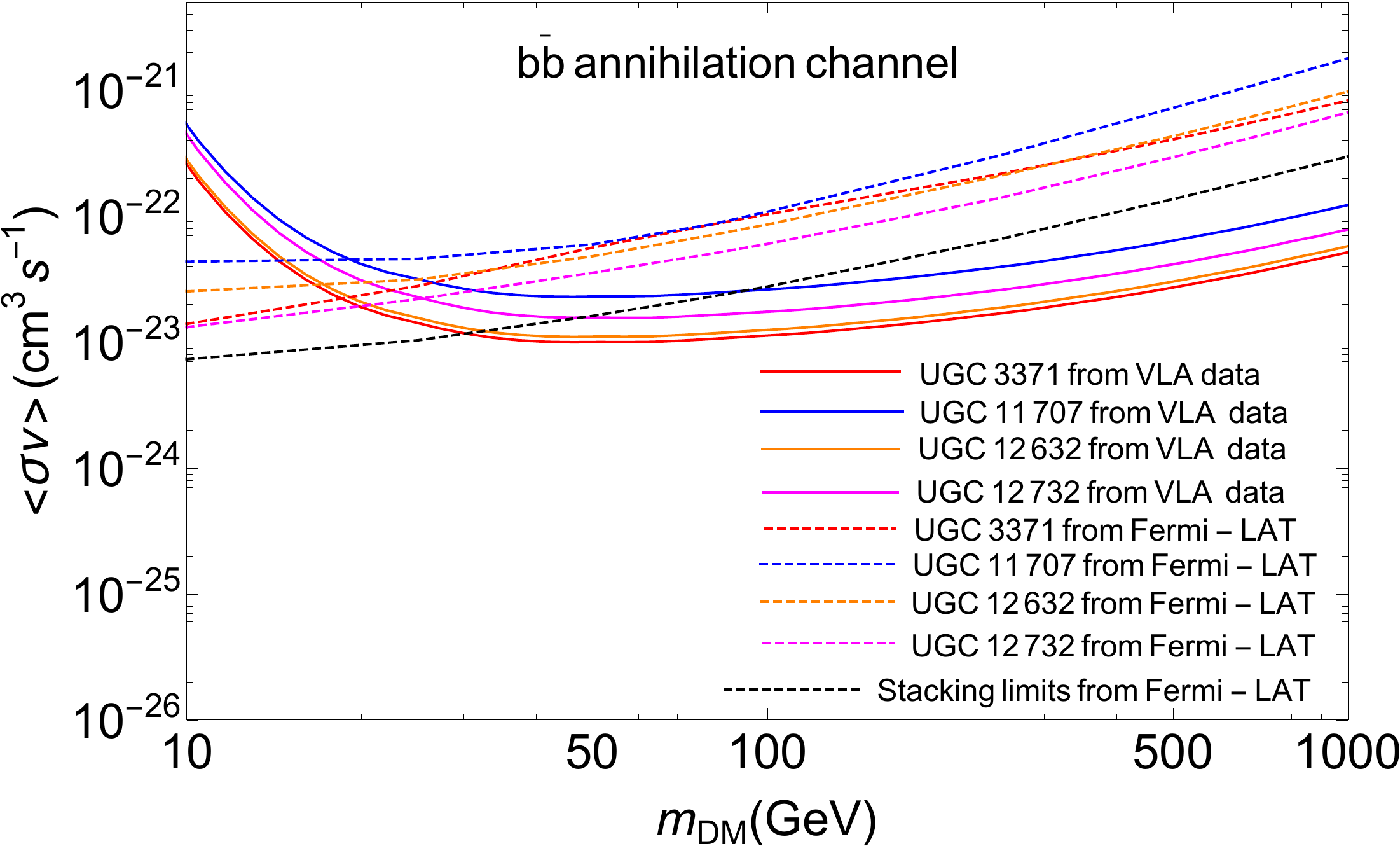}}
\subfigure[]
 { \includegraphics[width=0.8\linewidth]{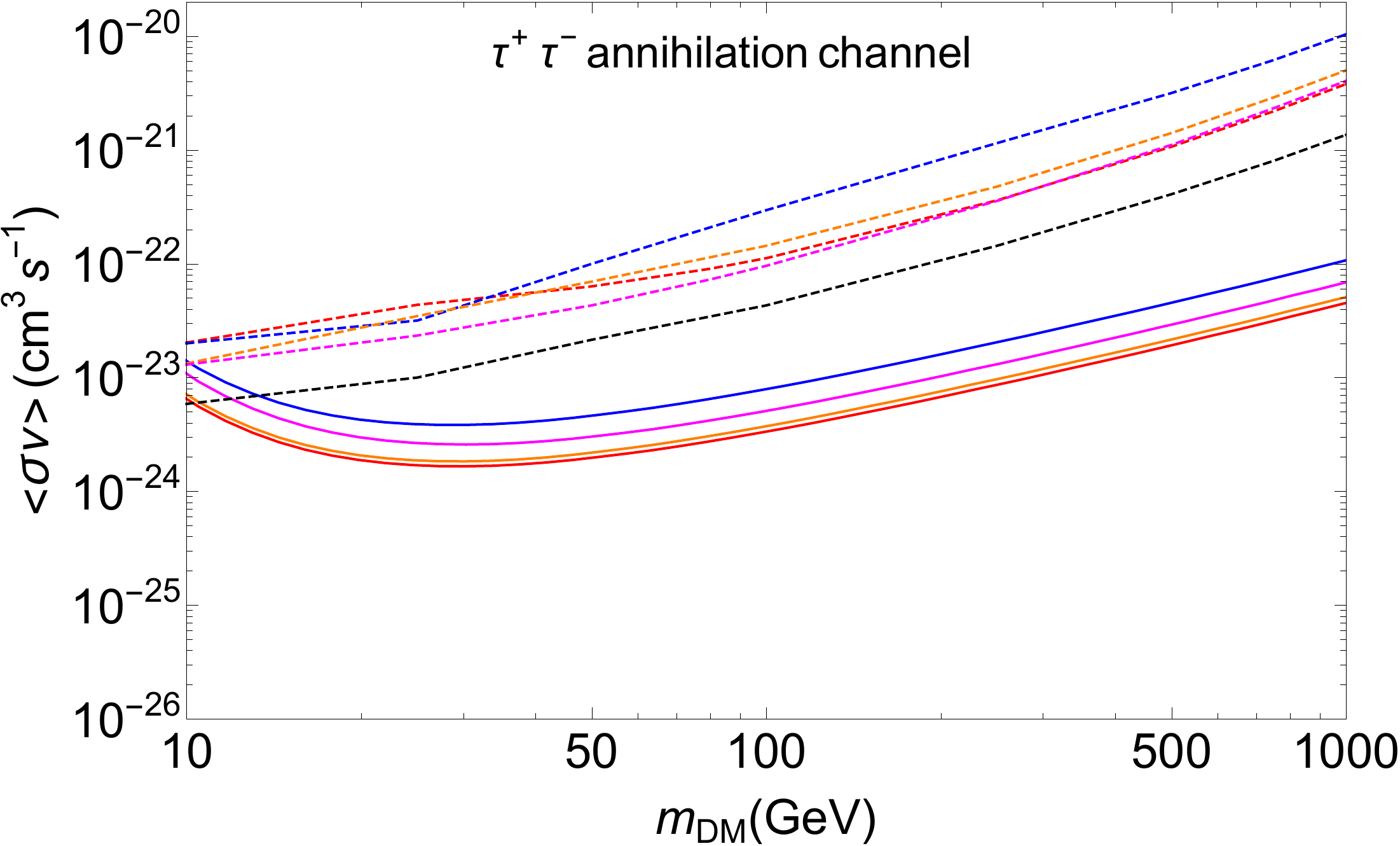}}
\subfigure[]
 { \includegraphics[width=0.8\linewidth]{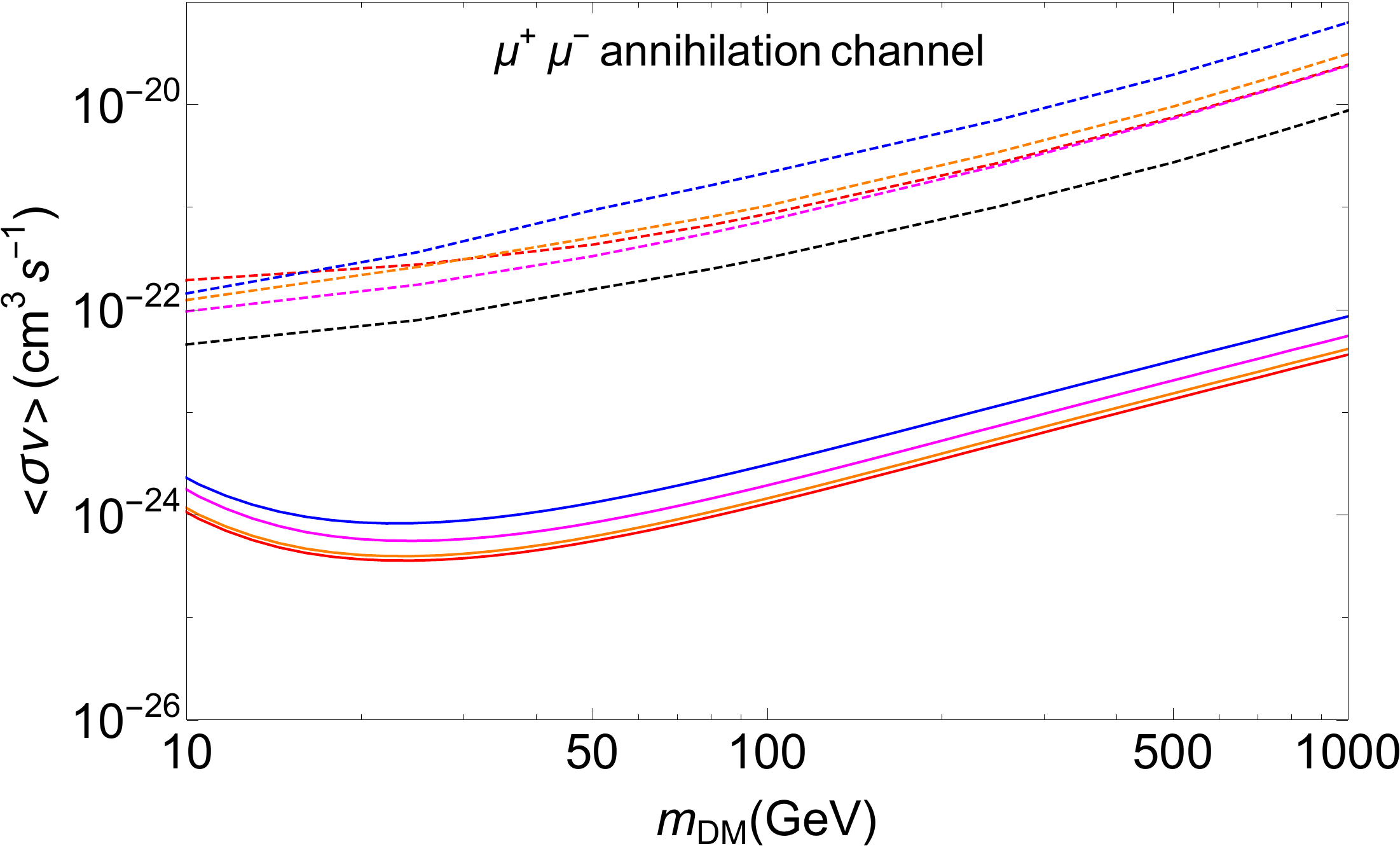}}
\caption{(a) Limits on the DM particle cross-section ($<\sigma v>$) from VLA-observed radio data for three annihilation final states. For (a), `solid', `dashed' and `dotdashed' line style have been used to denote $b\overline{b}$, $\tau^{+}\tau^{-}$ and $\mu^{+}\mu^{-}$, respectively. Comparison of the constraint on $<\sigma v>$ limits from radio data with individual and stacked limits obtained from Fermi gamma-ray data for (b) $b\overline{b}$, (c) $\tau^{+}\tau^{-}$ and (d) $\mu^{+}\mu^{-}$ final states. The halo profile is NFW, $B_{0}$=1 $\mu$G and the thermal averaged DM annihilation cross-section is fixed to the value to $<\sigma v> = 3 \times 10^{26}~cm^{3}s^{-1}$. We have used the same line styles for (b), (c) and (d),thus we have not separately added the legend for (c) $\tau^{+}\tau^{-}$ and (d) $\mu^{+}\mu^{-}$ final states.}
\end{figure}

\begin{figure}
\begin{center}
\subfigure[$b\overline{b}$]
 { \includegraphics[width=1.2\linewidth]{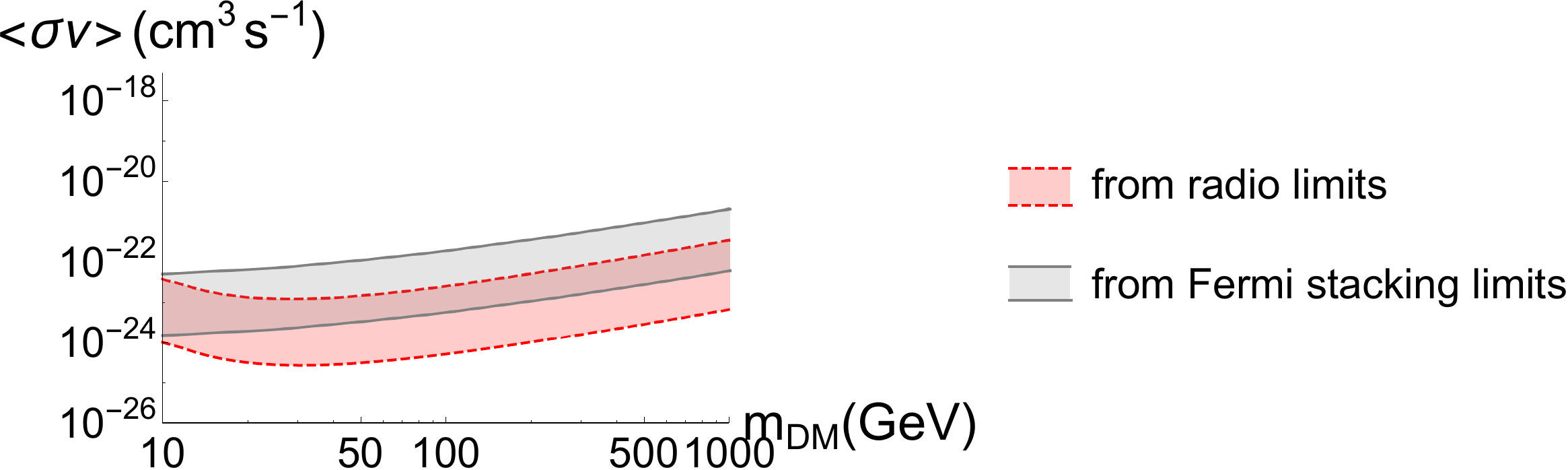}}
\subfigure[$\tau^{+}\tau^{-}$]
 { \includegraphics[width=1.2\linewidth]{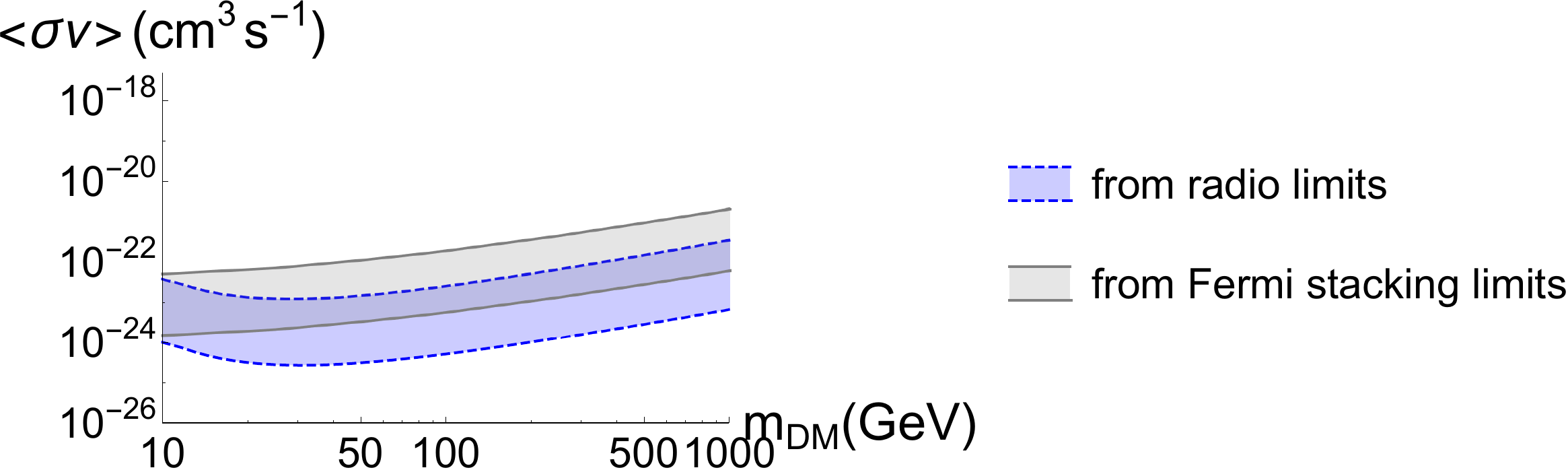}}
\subfigure[$\mu^{+}\mu^{-}$]
 { \includegraphics[width=1.2\linewidth]{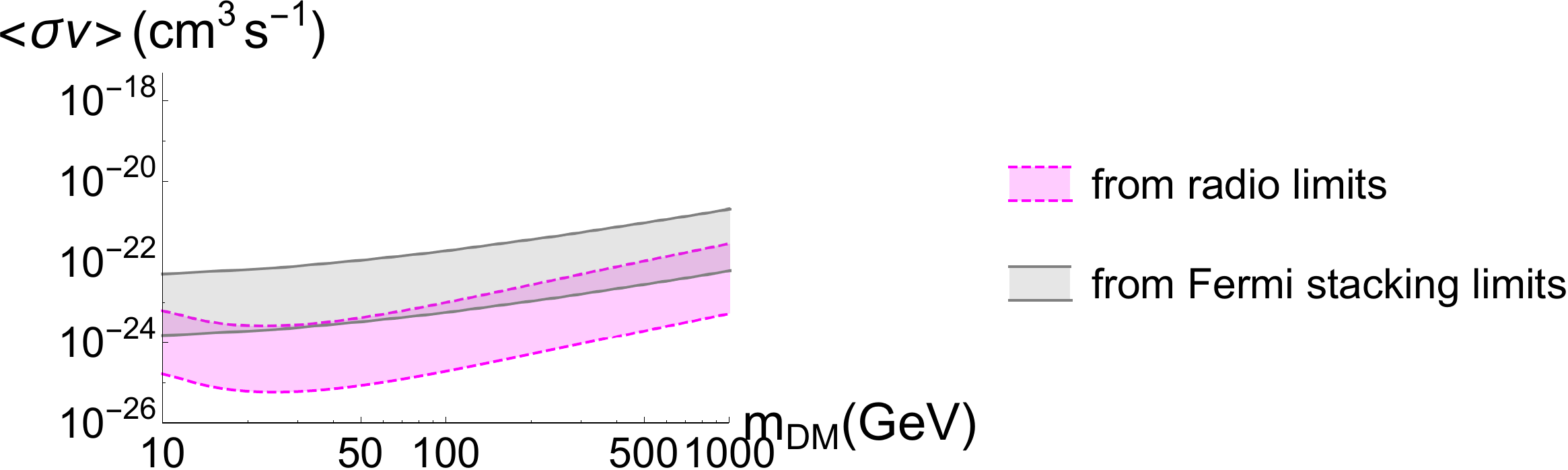}}
\caption{Uncertainties associated with the $<\sigma v>$ limits obtained from NVSS images for (a) $b\overline{b}$, (b) $\tau^{+}\tau^{-}$ and (c) $\mu^{+}\mu^{-}$ final states. The radio limits for each annihilation channels are compared with the uncertainty band associated with $\gamma$-ray stacking limits for $b\overline{b}$. The shaded region between dashed lines displays the uncertainty band for radio limits, while the shaded region between solid lines shows the uncertainty band for $\gamma$-ray stacking limits.}
\end{center}
\end{figure}

\begin{figure}
\subfigure[UGC 3371]
 { \includegraphics[width=0.8\linewidth]{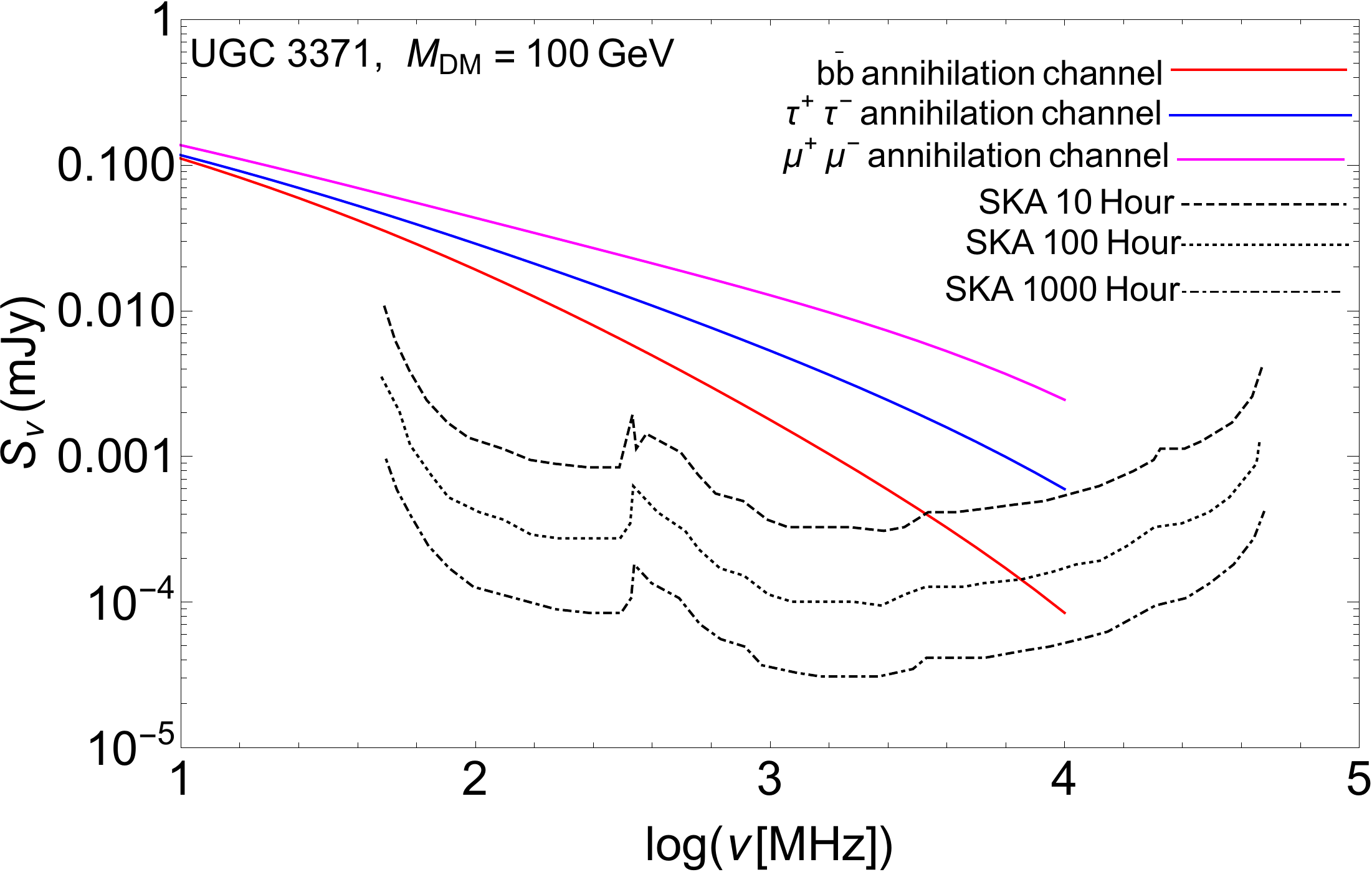}}
\subfigure[ UGC 11707]
 { \includegraphics[width=0.8\linewidth]{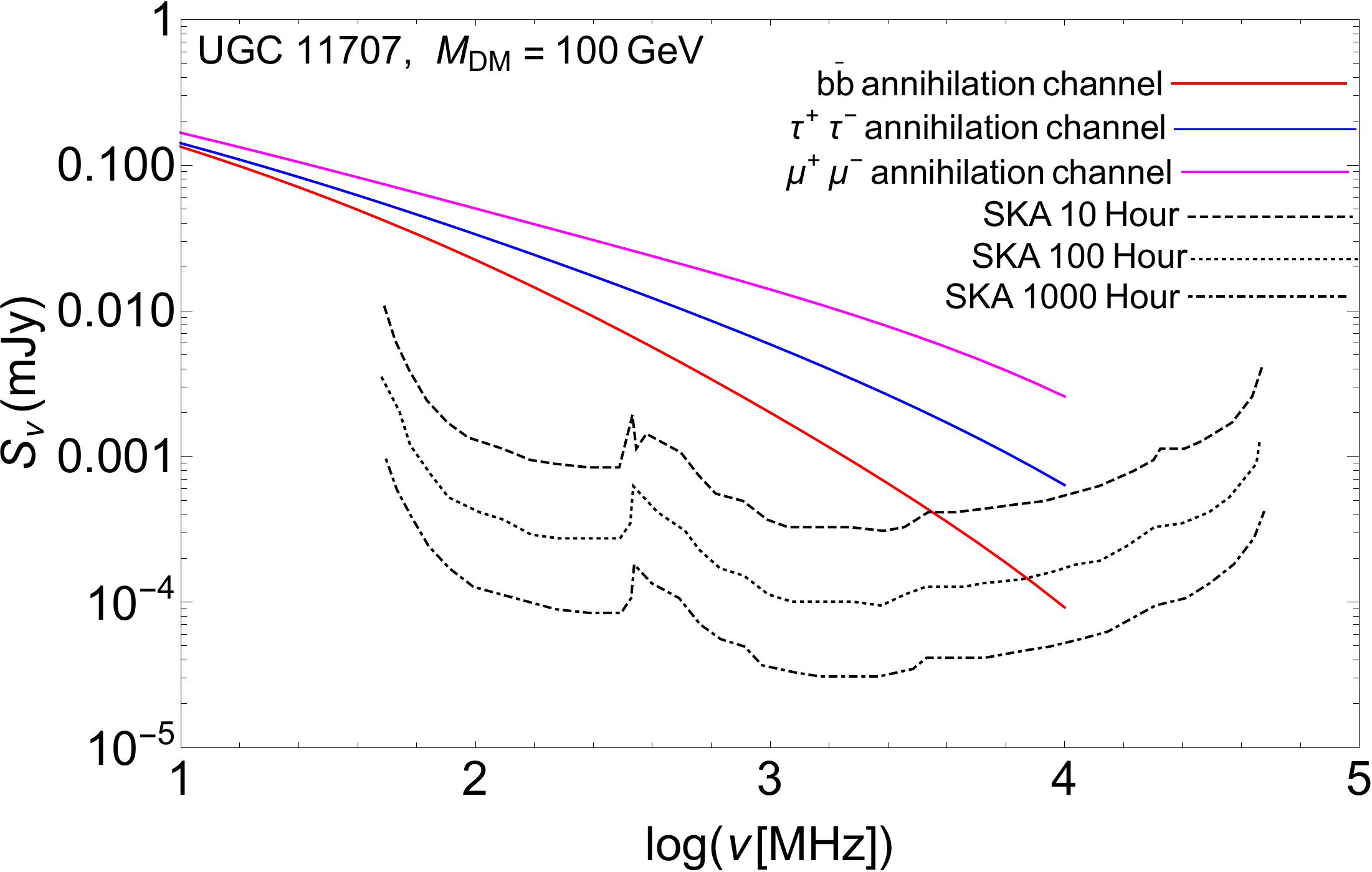}}
\subfigure[ UGC 12632]
 { \includegraphics[width=0.8\linewidth]{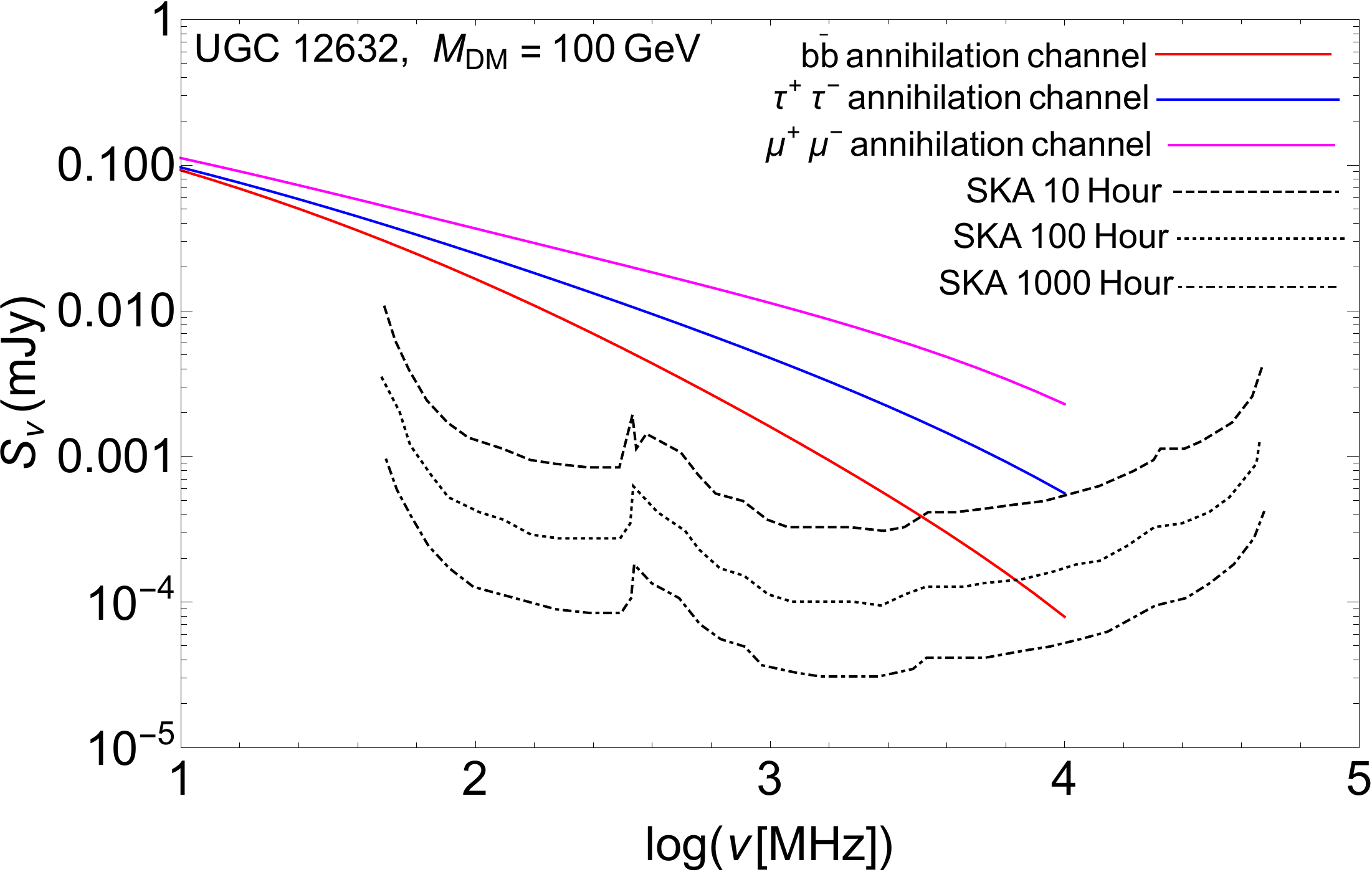}}
\subfigure[ UGC 12732]
 { \includegraphics[width=0.8\linewidth]{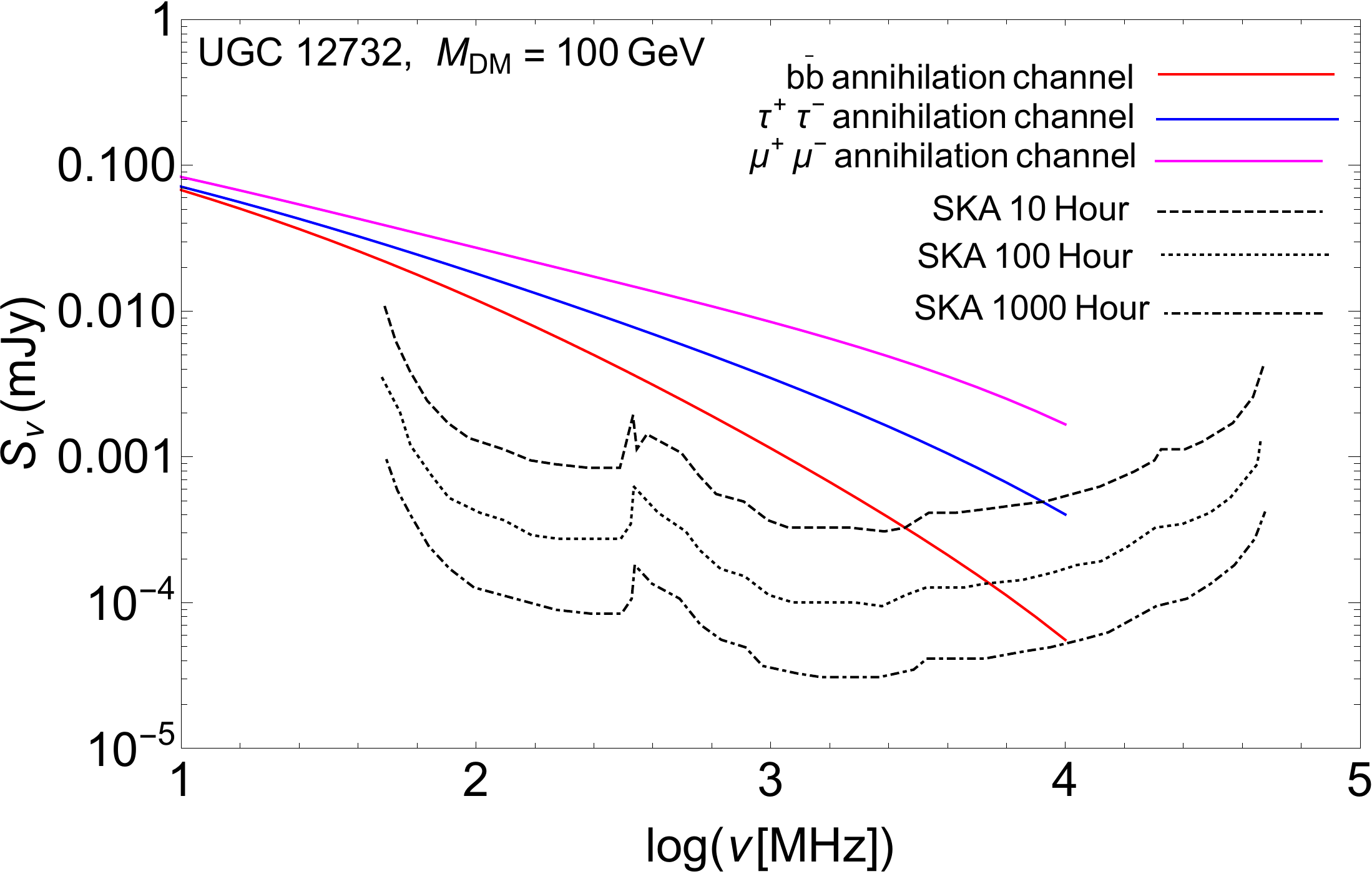}}
\caption{Flux densities of four LSB galaxies for DM annihilating into $b\overline{b}$, $\mu^{+}\mu^{-}$ and $\tau^{+}\tau^{-}$ final states. The halo profile is NFW, $B_{0}$=1 $\mu$G, $m_{DM}$ = 100 GeV and the thermal averaged DM annihilation cross-section is fixed to the value to $<\sigma v> = 3 \times 10^{-26}~cm^{3}s^{-1}$. The SKA sensitivity limits for integration times of 10, 100 and 1000 hours are overplotted as dashed, dotted and dot-dashed black lines, respectively.}
\end{figure}

\noindent Here we would like to mention that the emission detected from the UGC 11707 in the NVSS survey is less than 3$\sigma$ and such emission could be related to the fluctuation in astroparticle physics. But, it is likely to be real as the survey images are of high sensitivity. There is also a good match between the location of the emission and the optical position of the UGC 11707. The emission is diffuse in nature and could come from stars. However, there is no strong star formation going on in this galaxy, hence DM annihilation cannot be ruled out as an origin for the emission. But, further observations at 1.4GHz, which are more sensitive and deeper than the NVSS, are required to study the nature of the emission. Thus, in order to predict the radio $<\sigma v>$ limits from VLA data, we have used the same analysis method for all the four LSB galaxies. In Fig. 10 (a), we have shown the limits on the annihilation cross-section of all four LSB galaxies for three annihilation channels using VLA radio data with diffusion constant $D_{0}=3 \times 10^{28}~cm^{2}~s^{-1}$. From Fig.~10 (a), it is evident that $\mu^{+}\mu^{-}$ and $\tau^{+}\tau^{-}$ final states provide the strongest limit on DM cross-section. Here we would like to add that for gamma-ray data (Fig~4(d)) we have obtained that $b\overline{b}$ provides the most stringent limits, while for radio data we can find that $\mu^{+}\mu^{-}$ and $\tau^{+}\tau^{-}$ channels give a better limit than $b\overline{b}$. There is a high probability that the $b\overline{b}$ channel would annihilate to neutral pions and then neutral pions would decay the high energy photons, whereas the leptonic channel would mostly give the electron-positron as the end products of their annihilation channel. Thus, it is quite expected that $b\overline{b}$ channel would provide the best limits if we are dealing with the gamma-ray observation. But the scenario would be completely different for radio observation. The synchrotron emission from secondary electrons is the result of an ambient magnetic field that accelerates the charged particles and causing them to emit radiation at radio wavelengths. Hence, for radio emission, the leptonic channels would generally produce the more stringent limit than $b\overline{b}$ final states.\\

\noindent In Figs.~10 (b,c,d), we have compared the constraints on the DM particle cross-section from radio observations with the stacked and the individual limits obtained from Fermi-LAT data (sections~4.2 and 4.3) for a variety of annihilation channels. From Figs.~10 (b,c,d) we can find that the predicted radio constraint provides a more stringent constraint than the $\gamma$-ray data. Especially for the $\mu^{+}\mu^{-}$ annihilation channel (Fig.~10 (d)) at $m_{DM}$=100 GeV, the constraints on the DM annihilation cross-section obtained from the VLA radio data is $\approx$ 2 orders more stringent than the stacked limits obtained by Fermi-LAT in the $\gamma$-rays. These limits scale with the strength of the magnetic field. For our study, we have fixed the magnetic field of LSB galaxies at 1 $\mu$G. \\

\noindent From Figs.~10, we can expect that for LSB galaxies, radio emission might provide better limits on DM cross-section than $\gamma$-ray data observed by Fermi-LAT. But we should also keep in mind that, due to insufficient kinematics of LSB galaxies, our $<\sigma v>$ limits obtaining from NVSS images might come with large uncertainties. Thus, before investigating the uncertainty associated with the radio $<\sigma v>$ limits, we could not firmly state that
radio limits impose the strong limits on theoretical DM models. In Fig~11, we have shown the 2$\sigma$ uncertainty band associated with radio $<\sigma v>$ limits for three annihilation channels. For gamma-ray data, $b\overline{b}$ provides the most stringent limits, thus in Fig~11, we have compared the radio $<\sigma v>$ limits with the stacking limits for $b\overline{b}$ final states. For this study, we have only shown the result of UGC 12632. 
From Fig.~11, it is evident that for each annihilation final states, LSB galaxy would impose a large uncertainty band, nearly at the order of magnitude 2. But, it is also important to mention that from Fig.~11, we can observe an overlapping region between the $<\sigma v>$ limits obtained from $\gamma$-ray and radio data and that same nature is followed by all three annihilation channels. Hence, from Fig.~11, we can not strongly comment that radio limits would provide stronger limits than $\gamma$-ray but our analysis at best hints that the radio limits of $<\sigma v>$ are competitive with the results obtained from gamma-ray. With the more precise observation, in the future, we can expect to reduce such band to a possible single upper limit curve of $<\sigma v>$ and that
might improve the constraint limit on beyond Standard Model.\\

\noindent Next, we explore a more sensitive telescope-like Square Kilometre Array (SKA) for detecting the radio synchrotron signal \citep{colafrancesco2015} from LSB galaxies. The proposed SKA radio telescope will be one of the most sensitive radio telescopes in the next decade and is expected to explore many important questions in astrophysics and cosmology. Such as the fundamental physics aspects of dark energy, gravitation and magnetism. The exploration of the nature of DM is one of the most important additions to its scientific themes \citep{braun2015}.\\

\noindent We have predicted radio flux density $S(\nu)$ for DM annihilating into a specific channel in the form of synchrotron emission. In Fig.~12, we plot the $S(\nu)$ of radiation as a function of $\nu$ for DM annihilating into $b\overline{b}$, $\mu^{+}\mu^{-}$ and $\tau^{+}\tau^{-}$ channels, using the benchmark of thermally averaged cross-section value $<\sigma v> = 3 \times 10^{26}~cm^{3}s^{-1}$ and DM mass ($m_{DM}$) 100 GeV. From, Fig.~12 (a,b,c,d), we can observe that for these annihilation channels, the radio emission from LSB galaxies is seemed to be detected with the SKA sensitivity curve with 10-100-1000 hours of observation. The $\mu^{+}\mu^{-}$ and $\tau^{+}\tau^{-}$ annihilation channels are favoured over $b\overline{b}$ final states and the SKA detection threshold with 1000 hours of observation has higher chances to detect the radiation from LSB galaxies. Here we would also like to mention that for deriving the Fig.~12, we have taken the parameter values from Table~7. But from Fig.~9, we have already shown that the SED from synchrotron emission is strongly dependent on the magnetic field and diffusion zone of the galaxy. Thus without having the precise knowledge on the astrophysical parameters of LSB galaxies, we can never state the SKA would positively detect any emission from LSB galaxies. Besides, a dedicated simulation study is also needed to perform to predict whether with SKA we can detect the positive DM annihilation signal from LSB galaxies. Thus, from our study (Fig.~12), we can at best hint that in the future SKA would play an important role to study these LSB galaxies and we can also hope that they might detect any radio emission from LSB galaxies.\\

\noindent Hence, in this section, we have investigated whether radio emission can produce a stronger constraint on the DM cross-section and we find that if we consider the uncertainties associated with radio and gamma-ray limits in parameter space of ($<\sigma v>$, $m_{DM}$), they would be competitive to each other. We can also expect that in the future it might be possible for SKA to detect the radio emission from LSB galaxies. \\

\subsection{\textbf{Applying the NFW, Burkert and Pseudo Isothermal density profiles to the LSB galaxies}}
\label{Section4.5}
\begin{table}
\centering
\caption{ J-factor for three density profiles ($h_{0}=0.75$).}
\label{table-1}
\begin{tabular}{|p{1cm}|p{2cm}|p{3.5cm}|}
\hline \hline
Galaxy name &  Density Profile & J-factor ($\rm{GeV^{2}/cm^{5}}$)\\
\hline \hline
UGC & NFW & $0.739^{+2.87}_{-0.63}\times10^{16}$ \\
3371 & ISO & $0.188^{+0.775}_{-0.169}\times10^{16}$  \\
& BURKERT & $0.385^{+1.594}_{-0.346}\times10^{16}$ \\
\hline \hline
UGC & NFW & $0.485^{+1.85}_{-0.42}\times10^{16}$  \\
11707 & ISO & $0.123^{+0.501}_{-0.110}\times10^{16}$  \\
& BURKERT & $0.253^{+1.03}_{-0.227}\times10^{16}$  \\
\hline \hline
UGC & NFW & $0.795^{+3.08}_{-0.68}\times10^{16}$  \\
12632 & IS0 & $0.202^{+0.835}_{-0.182}\times10^{16}$  \\
& BURKERT & $0.414^{+1.717}_{-0.373}\times10^{16}$  \\
\hline \hline
UGC & NFW & $0.880^{+3.40}_{-0.75}\times10^{16}$  \\
12732 & ISO & $0.223^{+0.919}_{-0.1997}\times10^{16}$  \\
& BURKERT & $0.459^{+1.888}_{-0.411}\times10^{16}$ \\
\hline \hline
\end{tabular}
\end{table}
 \begin{figure}
\begin{center}
 { \includegraphics[width=0.9\linewidth]{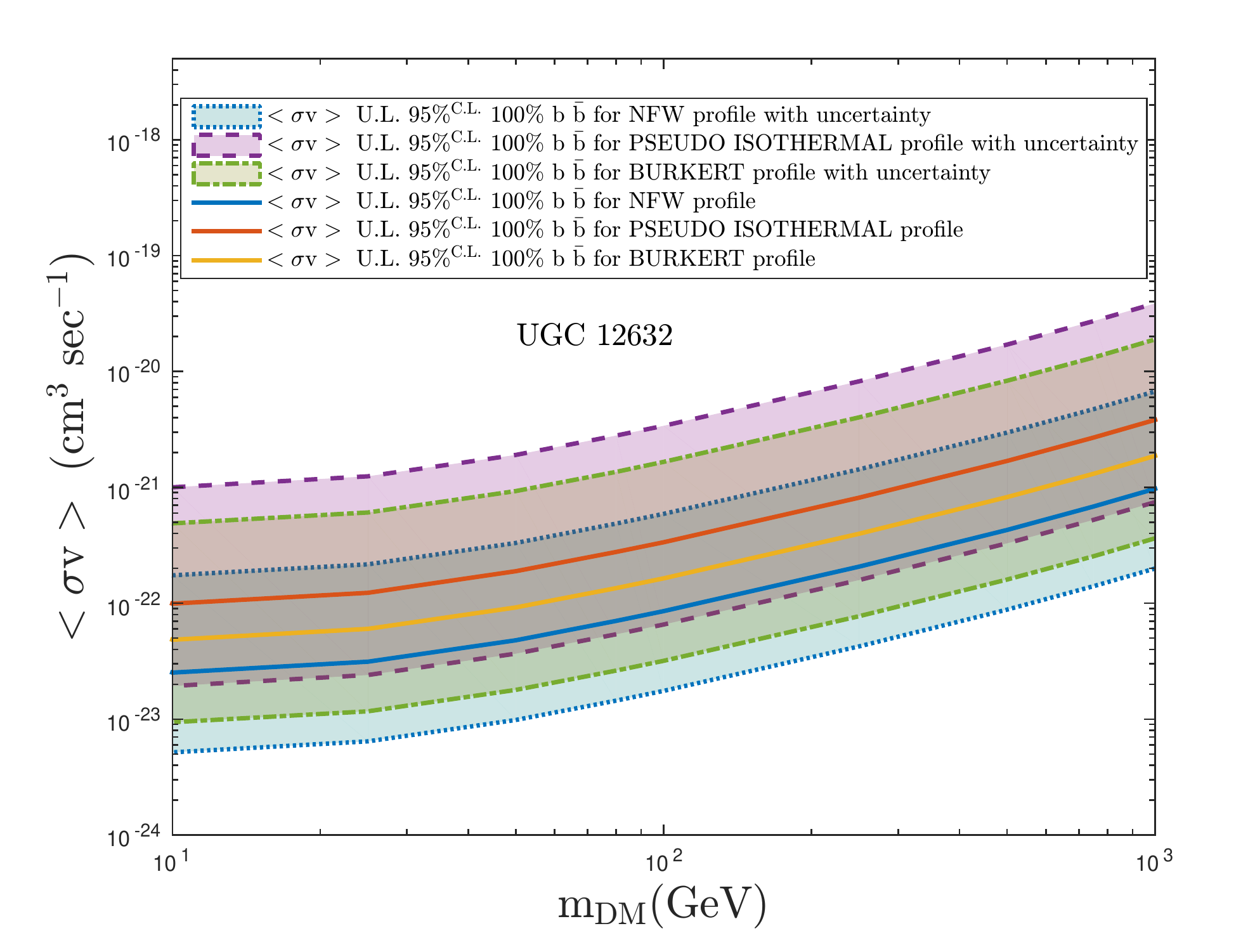}}
\caption{Comparison between the $<\sigma v>$ upper limits displayed in ($m_{DM}$, $<\sigma v>$) plane for J-factor with its uncertainties for three density profiles for $100\%$ $b\overline{b}$ final state. The shaded band represents the uncertainty in the DM density profiles in the LSB galaxies.}
\end{center}
\end{figure}

\noindent In several studies, it has been mentioned that two types of DM density profiles are popularly used to fit the observational data obtained from dSphs or LSB galaxies \citep{boyarsky2009}, which are the cusp-like profile such as NFW profile \citep{nav} and the cored profile such as Pseudo Isothermal (ISO) profile \citep{iso} or Burkert (BURK) profile \citep{burk, sal}. Unfortunately, the available rotation curves cannot constrain the particular type of DM density profile. So in this study, we have used the NFW density profile throughout our analysis. However, several studies in the literature also suggest that for some LSB galaxies the BURK profile can reproduce the rotation curves and for them, BURK profile is favoured over the NFW profile \citep{cadena2018, gammaldi2017}.\\

\noindent According to \citet{bosch}, the rotational curve of our selected LSB galaxies are consistent with cuspy profile \citep{bosch, bosch2001, swaters2003}. Even the observational study by \citet{bosch, bosch2001, swaters2003} cannot discriminate between dark halos with
constant density cores and $1/r$ cusps. Some recent studies also stress that for some LSB galaxies the cuspy profiles can produce a good fitting to their rotation curves \citep{nav, navarro2010, cadena2018}. \\

\noindent In this section, we have compared the J-factor values obtained from NFW with ISO and BURK density profiles.
\noindent The mathematical form of three distribution profiles are the following:\\
\begin{eqnarray}
\rho_{\rm NFW}(r)&=&\frac{\rho_{s}r_{s}^{3}}{r(r_{s} + r)^{2}}\\
\rho_{\rm BURK}(r)&=&\frac{\rho_{B}r_{B}^{3}}{(r_{B}+r)(r_{B}^{2} + r^{2})} \\
\rho_{\rm ISO}(r)&=&\frac{\rho_{c}}{(1+\frac{r^{2}}{r_{c}^{2}})} 
\end{eqnarray}
\noindent where, $\rho_{s}$ is the characteristic density of NFW profile, $\rho_{B}$ and $\rho_c$  are
central densities for BURK and ISO profiles, respectively. $r_{s}$ is the scale radius of NFW and  
$r_{B}$ and $r_{c}$ are core radius for BURK and ISO profiles, respectively. 
The parameters of these three distribution profiles of DM are related to each other through some scaling relation based on the study of DM dominated
galaxies by the authors of \cite{boyarsky2009}. The length scale parameters of 
these distributions are related as $r_s \simeq 6.1 r_c $ and 
$r_s \simeq 1.6 r_B $, while the relationship between the 
central densities are $\rho_s \simeq 0.11 \rho_c $ and $\rho_s \simeq 0.37 \rho_B$ 
respectively. We have derived the J-factor for three density profiles from Eq.~4. In Table~9, we present the J-factor of UGC 12632 obtained from NFW, ISO and BURK profiles. From this table, we can find out that the NFW density profile gives the largest values of J-factor and, hence, the NFW should produce a more stringent limit on DM particle models than ISO and BURK density profile.\\

\noindent We also have estimated the upper limits on annihilation $<\sigma v>$ of UGC 12632 for each of the density profiles for $\rm{100\%~b\overline{b}~channel}$. In Fig.~13, we show a comparison between three density profiles in the parameter space of ($<\sigma v>$, $m_{DM}$). Here we have considered the uncertainty associated with the density profiles and from Fig.~13, we could distinctly observe an overlapping region in parameter space of ($<\sigma v>$, $m_{DM}$). So, in view of the indirect DM search, it is not possible to favour one particular density profile but from Fig. 13, although it is also evident that for the median J value, the NFW profile gives more stringent $<\sigma v>$ upper limits than the other two density profiles.\\

\section{\textbf{The Future of LSB galaxies for Dark Matter Searches and the impact of the CTA}}
\label{Section5}
\noindent In the next decade, the Cherenkov Telescope Array (CTA) will be the most sensitive instrument to study high-energy $\gamma$-rays. It can detect the $\gamma$-rays from very faint and distant sources over a very large range of energies, and approximately between 20 GeV to 300 TeV. It has a large field of view, $\sim$ 8-10 degrees for medium and small-sized of telescopes. The large angular resolution of the CTA is about 2 arcmin and its energy resolution is well below $\sim$10$\%$. The effective area of the CTA increases with increasing energies, such as $5\times10^{4}~m^{2}$ at 50 GeV, $10^{6}~ m^{2}$ at 1 TeV, and $5\times10^{6}~m^{2}$ at 10 TeV. Hence, the CTA is capable of reaching higher sensitivities compared to other ground-based and space-based $\gamma$-ray instruments. All of these qualities make CTA an important facility to search for signatures from DM particle self-annihilation or decay.\\

\noindent In this section, we consider the sensitivity of the CTA and Fermi-LAT for $\gamma$-ray detection and try to assess whether the CTA can detect any $\gamma$-ray signal from LSB galaxies. Concerning energy resolution, angular resolution, effective area and any other key features, there are several differences between space-based and ground-based telescopes and for our study, this comparison is especially important. We have adopted the CTA differential flux sensitivities curve from \citep{maier2017}, whereas for Fermi-LAT we have taken the sensitivity curve for 10 years of observations ($P8R3\_V2~IRF$) of point-like and high-Galactic latitude sources\footnote{\url{http://www.slac.stanford.edu/exp/glast/groups/canda/lat_Performance.htm}}. The CTA sensitivity curve is estimated for the point sources with power-law modelling and also with the detection significance of 5$\sigma$ for 50 hours of observation \citep{maier2017}. For the Fermi-LAT instrument, that sensitivity curve is drawn with a similar approach \footnote{\url{http://www.slac.stanford.edu/exp/glast/groups/canda/lat_Performance.htm}}.\\

\noindent In Fig.~14, we show the flux upper limits achieved from our Fermi-LAT analysis (see Section 3.1) of all LSB galaxies and compare them with Fermi-LAT and CTA sensitivity. From Fig.~14 we can observe that within the 100 GeV to 1 TeV energy range, it might be possible for CTA to detect in 50 hours of observation a $\gamma$-ray signal from LSB galaxies, or to improve the flux upper limits. Either DM annihilation/decay or any astrophysical scenario could be the source of such emission. Without a dedicated simulation study, we could not strongly comment on that. But, the detailed simulation for CTA is beyond the scope of our current analysis. Thus from Fig.~14, we can, at best, conclude there is a possibility that in the future, CTA will be a very suitable instrument for investigating the gamma-ray signal from LSB galaxies.

\begin{figure}
\begin{center}
 { \includegraphics[width=0.9\linewidth]{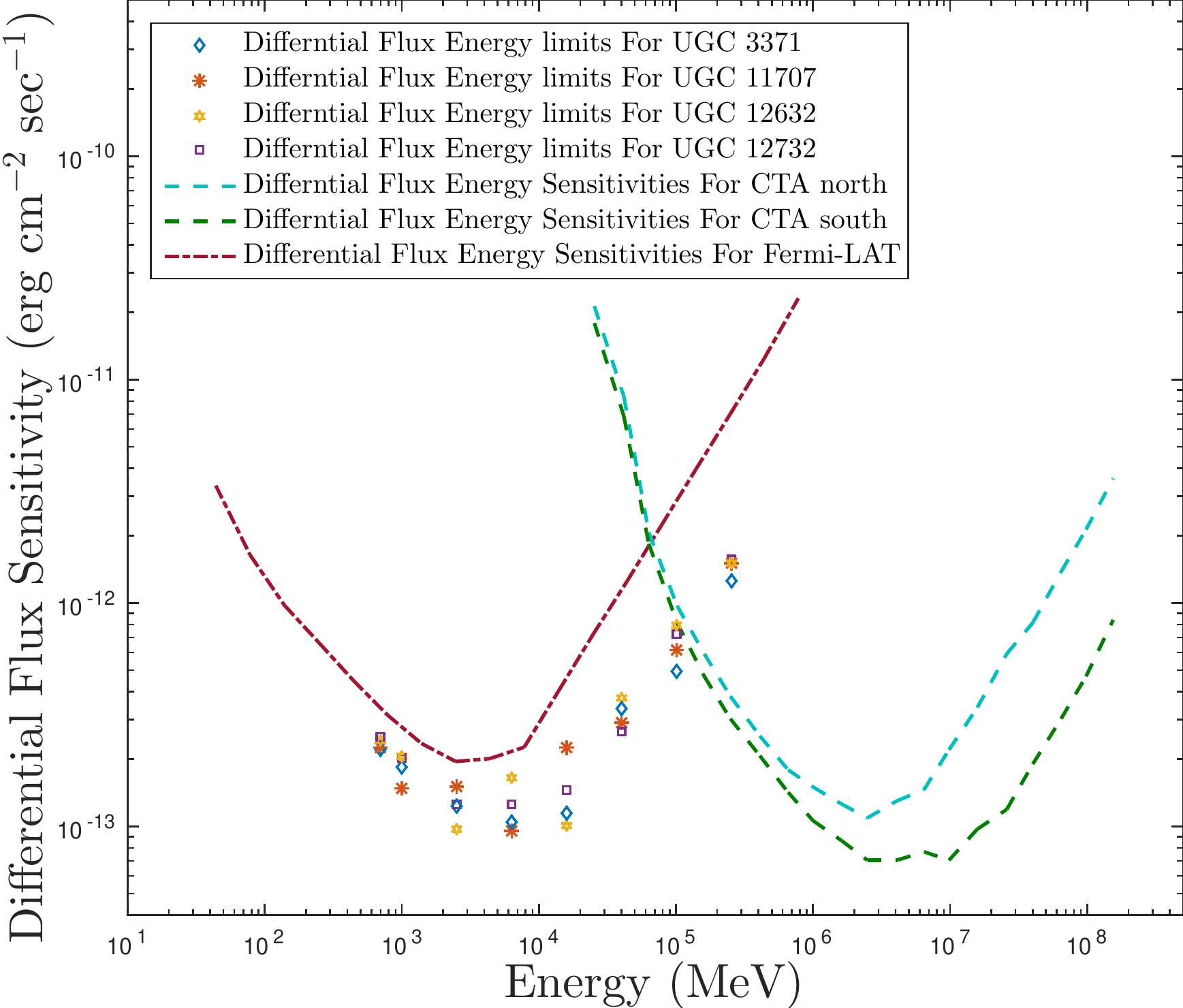}}
\caption{Comparison of differential energy flux of LSB galaxies with CTA and Fermi-LAT sensitivity curves.}
\end{center}
\end{figure}

\section{Discussions \& Conclusions}
\label{Section6}
\noindent In our paper, we have analyzed nearly nine years of Fermi-LAT $\gamma$-ray data along the direction of four LSB galaxies and for each case, no excess $\gamma$-ray emission was detected along the line of sight. We have then determined the flux upper limit for those LSB galaxies. With the $\gamma$-ray spectrum from DM annihilation (i.e. with DMFit function), we have calculated the flux and $<\sigma v>$ upper limit for several pair-annihilation final states. Our estimates show that for LSB galaxies, the $<\sigma v>$ upper limit for $100\%$ $b\overline{b}$ annihilation could not impose a strong constraint on any theoretical model. Even though LSB galaxies are considered as one of the possible DM dominated galaxies, their individual $<\sigma v>$ upper limits are almost 3 orders of magnitude weaker than the ones obtained from Milky Way dSphs at the same DM mass \citep{ack}. The available rotation curves of our LSB galaxies cannot favour any particular type of DM density profile. The observational study shows that the rotational curve of our selected LSB galaxies is consistent $\Lambda$CDM and the cuspy profile can also fit the DM distribution of their internal core. Thus throughout our analysis, we have considered only the standard NFW density profile. We have also shown a comparison between three popular density profiles for DM distribution (Fig.~13). If we consider the uncertainty associated with three density profiles, from Fig.~13 we could distinctly observe an overlapping region among three density profiles in parameter space of ($<\sigma v>$, $m_{DM}$). So, in view of the indirect DM search, it is not possible to favour one particular density profile but from figure 13, it is also evident that for median J value, NFW profile would provide the strongest limits on ($<\sigma v>$, $m_{DM}$) space.\\

\noindent We have also performed a joint likelihood analysis on these four LSB galaxies. The joint likelihood would be an ideal approach to obtain more stringent limits from LSB galaxies because through joint likelihood we have combined individual likelihood function of each LSB galaxy. Hence, it is expected that such an approach would increase the sensitivity of the analysis. After performing the combined likelihood, we found that the stacking analysis has improved the sensitivity of LSB limits but the combined limits are still nearly 2 orders of magnitude weaker than the limits achieved with the observation of the Milky Way dSphs by Fermi-LAT. New-generation optical facilities are expected to increase the number of known LSB galaxies. This will improve the prospects for indirect DM searches with this class of targets in the near future.\\

\noindent For the indirect detection of DM signature, the $\gamma$-ray analysis is believed to be one of the most popular methods to examine the DM signature but a multiwavelength approach can also provide a complementary probe of $\gamma$-ray analysis. LSB galaxies have very low star formation rates and this might minimize the uncertain contribution of astrophysical processes. Hence, LSB galaxies could be an ideal place for studying the diffuse radio signals obtained from DM pair annihilation. We have considered the multiwavelength approach and attempted to predict the possible radio emission from LSB galaxies. For this purpose, we have used a publicly available code, RX-DMFIT, an extension of the DMFit tool. We have predicted the SED plots of our selected LSB galaxies for several possible conditions (Fig.~8 $\&$ 9). We have also taken the radio data observed by VLA for all four LSB galaxies but except for UGC 11707, we have only obtained the upper limit of flux density for the other three LSB galaxies. Using the observational radio data, we have estimated the predicted radio constraint in parameter space of ($<\sigma v>$, $m_{DM}$). From our study, we find that the uncertainty band associated with radio $<\sigma v>$ limits are overlapped with the uncertainty band associated with the Fermi-LAT stacking limits (Fig.~11). The same nature is followed by all three annihilation channels. Thus, we could not firmly comment that radio data would provide stronger limits than gamma-ray data. But from our study, at best, we can conclude that multiwavelength analysis is competitive with the results obtained from gamma-ray. In the future with more precise analysis, we can expect them to provide strong limits on theoretical DM models. \\

\noindent We have tried to find out whether in the future SKA and CTA can detect any emission from LSB galaxies. From our study (Fig.~12 (a,b,c,d)) we have observed that radio emission from LSB galaxies could be detected with the SKA already in 10 hours. But to comment on anything precisely, we need to follow a detailed simulation study. Besides, we also should have accurate knowledge of the magnetic field, diffusion zone, DM distribution, etc. of LSB galaxies. Thus, our study only indicates the possibility of detecting radio emission by SKA. From Fig.~14, we also find that between 100 GeV to 1 TeV energy range, CTA might be able to detect a gamma-ray signal from LSB galaxies within 50 hours observations. But just like SKA, a dedicated simulation study is necessary to investigate whether CTA would positively detect the DM annihilation/decay signal from LSB galaxies. Our results, at best, hint that in the future SKA telescope and CTA telescope may play an important role in detecting the radio and gamma signal, respectively from LSB galaxies. 

\section*{Acknowledgement}
\noindent We would like to thank Dr Stefano Profumo and Dr Tesla Jeltema of the University of California, Santa Cruz, U.S.A., for providing useful guidance in Fermi-LAT data analysis. We are particularly grateful to Alex McDaniel of the University of California for providing us with the RX-DMFIT code and all the necessary suggestions to run it. PB is grateful to DST INSPIRE Fellowship Scheme for providing support for her research. PB is thankful to Prateek Chawla for his help regarding uncertainty calculation. We are particularly grateful to the Fermi Science Tools for allowing us to freely access the Fermi-LAT data and providing all the needed guidance for our analysis.

\section*{Data Availability}
\noindent The Fermi-LAT data using for this paper are available at \url{https://fermi.gsfc.nasa.gov/ssc/data/access/lat/}.
The NVSS images that we accessed for our study are available at \url{https://www.cv.nrao.edu/nvss/}.

\label{lastpage}
\end{document}